\documentclass[preprint,10pt,authoryear]{elsarticle}

\usepackage{amssymb}
\usepackage{amsmath}
\usepackage{tikzfill}
\usepackage{soul} % for the highlighting command \hl
\usepackage[margin=1in]{geometry}
\usepackage{mathtools}
\usepackage{amsthm}
\usepackage{amssymb}
\usepackage{tikz}
\usepackage{pgfplots}
\pgfplotsset{compat=1.18} % Compatibility level for pgfplots
\usepackage{caption}
\usepackage{subcaption}
\usepackage{comment}
\usepackage{algorithm}
\usepgfplotslibrary{groupplots}
\usepackage[]{algpseudocode}
\usepgfplotslibrary{fillbetween}
\usepackage{tabularx} % Include this in the preamble
\usetikzlibrary{decorations.markings}
\usetikzlibrary{arrows.meta}
\definecolor{dark green}{rgb}{0.0, 0.4, 0.0}

\usepackage{algorithm}
\usepackage{graphics}
\graphicspath{{fig/}}

\newtheorem{theorem}{Theorem}
\newtheorem{lemma}{Lemma}
\theoremstyle{definition}
\newtheorem{definition}{Definition}
\newtheorem{proposition}{Proposition}
\theoremstyle{remark}
\newtheorem{remark}{Remark}
\newcommand{\bbxi}{\boldsymbol{\xi}}
\newcommand{\cc}{\mathbf{c}}
\newcommand{\dd}{\mathbf{d}}
\newcommand{\ee}{\mathbf{e}}
\newcommand{\xx}{\mathbf{x}}
\newcommand{\yy}{\mathbf{y}}
\newcommand{\rr}{\mathbf{r}}

\renewcommand{\ss}{\boldsymbol{\sigma}}

\newcommand{\bb}{\mathbf{b}}

\newcommand{\uu}{\mathbf{u}}
\newcommand{\UU}{\mathbf{U}}
\newcommand{\vv}{\mathbf{v}}
\newcommand{\ww}{\mathbf{w}}
\newcommand{\zz}{\mathbf{z}}

\newcommand{\HH}{\mathbf{H}}

\newcommand{\EE}{\mathbf{E}}
\newcommand{\bS}{\mathbf{S}}

\newcommand{\LL}{\mathcal{L}}
\newcommand{\bzz}{{\boldsymbol{\zeta}}}
\renewcommand{\HH}{\mathcal{H}}
\numberwithin{equation}{section}
\journal{arXiv}
\begin{document}

\begin{frontmatter}



\title{Frame Based Inelasticity for Fracture in Anisotropic Media}
%\title{Frame Based Peridynamics for Strength and Fracture in Anisotropic Media}

%\author[1]{Rober}
%\author[1]{Semsi Coskun}
%\author[1]{Davood Damircheli}
\author[1]{Robert Lipton }
\author[1]{Prakash K.C.}

\affiliation[1]{organization={Department of Mathematics and Center for Computation and Technology, Louisiana State University},
            addressline={Lockett Hall}, 
            city={Baton Rouge},
            postcode={70803-4918}, 
            state={LA},
            country={USA}}

%% Abstract
\begin{abstract}
A field theory is presented for predicting damage and fracture in brittle   materials.  Four  concepts are used: a new  nonlocal deformation gradient for dynamic stress analysis in anisotropic media, a new type of phase field coming from a constitutive law, the evolution of material displacement  from Newton's second law, and a new incorporation of  scaled strength surfaces into constitutive laws for  predicting dynamic and quasi-static fracture patterns. The elastic softening is governed by the strain dynamics through the constitutive law and  in this way we provide the two way coupling  between  strain and fracture. Simulations validating the approach for both  dynamic and quasi-static fracture analysis are compared to experimental results and other simulation methods. The model  is capable of predicting material strength in titania and the competition between  strength and energy in soda lime glass and titana and captures dynamic crack branching in soda lime glass and concrete.

%%%%%%%%%%%%%%%%%%%
%%%%%%%%%%%%%%%%%

\end{abstract}

\end{frontmatter}



\section{Introduction}
\label{sec.intro}

A field theory is constructed for modeling fracture  in elastically anisotropic  materials allowing for both differentiable and discontinuous deformations.  The material can change topology as part of the evolution. Within this context a suitable notion of nonlocal gradient and divergence are developed that are simple to use for  stress calculation. 
 The field theory presented in this paper  is new and different than existing nonlocal free fracture formulations  including phase field, see   \cite{Bourdin-Francfort-EtAl-2000,Aranson,Karma,BourdinLarsenRichardson},  peridynamics, see \cite{silling2000reformulation,silling2007peridynamic}, cohesive element see, \cite{HillerborgPetersson1976,XuNeedeleman1994},  eigen deformation, see  \cite{Schmidt}, and combined methods \cite{Partmann}. Here the new field theory is distinct from earlier ones and is based on the fact that traction forces acting on surfaces are inherently  nonlocal and can in principle allow for discontinuities in the underlying deformation field.  The crack set is  defined as the internal boundary where traction force on one side of the crack does not  influence the motion of material  points on the other side.  This method  is  naturally driven by the need to understand  the effects of inelastic fields in otherwise linearly elastic  anisotropic media. The material sample can be elastically  heterogeneous and of composite or polycrystalline form. The theory is also driven by the need to find a computationally less expensive way to model brittle fracture in an otherwise anisotropic elastic media. The field theory is shown to autonomously interpolate between inelastic and linear elastic behavior, see Section \ref{Interpolation}.
 
The idea is to develop an inelastic field theory based on nonlocal stress and strain pairs such that the nonlocal stress and strain rate are work conjugate fields.  We relate nonlocal stress to nonlocal strain through  constitutive laws that can depend on space and time. We start with a suitable nonlocal gradient that is easy to compute and use it to describe the nonlocal strain and an appropriate  divergence operator. The nonlocal gradient defined here is a matrix of directional difference quotients of the material displacement associated with a local coordinate frame. This is called a frame gradient.  Here the trace of the frame gradient operator applied to the stress  is seen to define the divergence operator. The theory is described  in Sections \ref{sec:nonlocal:model} through \ref{sec.Constiutivelaws BVP} where suitable boundary conditions are introduced and the initial boundary value problem (IBVP) is described. Existence and uniqueness of the IBVP is established with the aid of the Banach fixed point theorem in Section \ref{wellposedness}. The nonlocal stress and strain rate are shown to be work conjugate fields in Proposition \ref{conjugate} of Section \ref{powerbalanceprocesszone2}. These observations can be pushed further and we recover power balance \eqref{eq:processzonepowerbalance} for the fracture evolution.

\newpage \noindent Power balance is articulated as the damage condition:

\bigskip
{\em If the rate of energy put into the test specimen exceeds the material's capacity to generate kinetic and elastic energy through  displacement and velocity, then damage occurs.}
\bigskip

To develop a constitutive law for  the evolution of brittle fracture within an otherwise elastic medium we  consider damaging material characterized by softening elastic properties. Here a phase field is introduced as a damage field and depends on the stress and  the prescribed strength surface, eg., Von Mises, Drucker Prager, Mohr Coulomb, etc.  To fix ideas we use the Drucker Prager strength surface and in the numerical section we apply the frame gradient method to titania and numerically corroborate the experimental results of \cite{ely1972strength} describing the onset of fracture.

Energy balance equations are derived from the evolution equation involving both elastic energies  and  dissipation energies. Integration by parts shows that the dissipation energy of a flat crack is the product of the fracture toughness and the surface area of the crack as given by Griffith. What is important is that this is done independently of the length scale $\epsilon$. Hence no convergence $\Gamma$ or otherwise  is necessary as $\epsilon \rightarrow 0$ for computing the energy dissipation of flat cracks. 
Here the formula for the fracture toughness is given in terms of model parameters through multiplying the evolution equation by the velocity field integrating and finding the stress work necessary to fail a frame. It  does not involve the length scale of nonlocality $\epsilon$. The length scale independence is accomplished through a dimensionless scale factor introduced  in the definition of the phase field, see Sections \ref{scalefactor} and \ref{criticalenergyrelease}.  %Here the phase field takes values between one and zero with one in undamaged material and zero in fully damaged material.  
In this respect the frame gradient method is regarded as a nonlocal evolution over a mesoscopic continuum consistent with an idealized macroscopic  evolution obtained by sending $\epsilon$ to zero. Consequently the calibration of the  model to material parameters  will not involve the length scale of non locality $\epsilon$. 
This  type of scaling is recently introduced in the blended constitutive models developed for bond based peridynamics, see  \cite{coskun2025quasi}, \cite{bromleylipton}. This interpretation of nonlocal dynamics is motivated by the theoretical investigations  given by $\Gamma$ limits and evolution equations satisfied by the limit displacement and crack patterns reported in \cite{Liptonmeso}, \cite{lipton2014dynamic}, and \cite{lipton2016cohesive}. These theoretical results while worked out for simpler models show that the inelasticity is localized to surfaces, i.e., cracks and that the bulk of the material is linear elastic as $\epsilon$ tends to zero.

The computational complexity of the frame gradient method is the same order as bond based peridynamics \cite{silling2000reformulation} as only one quadrature is necessary to compute internal forces at each point in the domain.\footnote{The bond based scheme is a central force model theoretically valid for isotropic elastic materials with fixed Poisson ratio $\nu=1/3(1/4)$ depending on the dimension $d=2(3)$.}% But  give reliable results (within experimental tolerances) for materials with Poisson ratios close to $1/3(1/4)$.}.
Hence the computational complexity of this method is an order of magnitude less than the state based peridynamics. (In that method a second quadrature for all points neighboring a material point is required, \cite{silling2007peridynamic}.) When the frame gradient method is  compared to the phase field method one observes that only one equation for the evolution of the displacement needs to be considered for the frame gradient method and no evolution equation for the phase field needs to be computed. In the frame gradient context the fracture evolution is seen as a multiscale problem \cite{Liptonmeso} and in addition to a point $\xx$ in the domain $\Omega$  we have small scale degrees of freedom associated with $\xx$ described by points of the form $(\xx,\bbxi)$ where $\bbxi$ is in a ball of radius $\epsilon$ centered at the origin. Both $\xx$ and $\bbxi$ are three dimensional vectors (two dimensional vectors for plane stress loading). The constitutive law used in  frame gradient inelasticity  involves the softening of elastic moduli through a phase field viewed as a damage field that depends on the point $(\xx,\bbxi)$  and nonlocal strain history, see Section \ref{sec.Constiutivelaws BVP}.  The crack set is  defined as the internal boundary where traction force on one side of the crack does not influence the motion of material points on the other side.

The paper is organized as follows: frame gradients are defined in Section \ref{sec:nonlocal:model}. The constitutive law and stress tensor necessary for the dynamic initial value problem for linear elastic media is given in Section \ref{Linear Elastic Materials}. The inelastic space-time constitutive law and stress tensor necessary for the initial value problem for fracture is given in Section \ref{sec.Constiutivelaws BVP}.  %The frame gradient peridynamics is established as a new correspondence model in Section \ref. 
The well posedness, power balance, and thermodynamic consistency of the dynamic initial value problem  are established   in Sections \ref{wellposedness} and   \ref{powerbalanceprocesszone2}.  Energy balance is established in Section \ref{energybalanceprocesszone2}.  
The calibration of model parameters to the material properties given by fracture toughness and elasticity tensor are given in Section \ref{sec.Straight Cracks}.
We verify and validate frame gradient inelasticity with experiment and existing phase field methods in Section \ref{sec.numerics}.  In this section we  validate and compute the strength curves necessary for nucleation   and compare with the strength data provided by the experiments of \cite{ely1972strength}. We go further and analyze  the competition between strength and fracture energy on fracture nucleation. The studies are carried out for and verified for soda lime glass, \cite{kamarei2026nine}  and validated for polycrystalline rutile titania. We  then compare the frame gradient method to phase field methods for dynamic fracture propagation as introduced in \cite{borden2012phase}. Last we validate the model comparing numerical simulations to the dynamic fracture experiments of \cite{ovzbolt2011dynamic} carried out on notched concrete specimens. Concluding remarks are provided in Section \ref{sec:Conclusion}.

\section{Frames and gradient}
\label{sec:nonlocal:model}

The body containing the  material $\Omega$ is a bounded domain in two or three dimensions. Nonlocal force interactions are characterized by a length scale $\epsilon$. We introduce a ball  centered at the origin denoted by $\HH_\epsilon(0)$. Any coordinate frame of $d$ orthogonal vectors  $\bbxi^1,\ldots,\bbxi^d$, $d=2,3$  of common length less than $\epsilon$  are contained in the ball. The frame is chosen to form a right handed system. For simplicity we pick a vector $\bbxi$ in the ball and label it $\bbxi^1$.  
 Here $V^\epsilon_d=\omega_d\epsilon^d$ is the $d$-dimensional volume of the ball $\HH_\epsilon(0)$. 
  The displacement of a point $\xx$ inside the material at time $t$ is $\uu(t,\xx)$ and is defined for $0\leq t\leq T$ and $\xx$ in $\Omega$.

We describe the kinematics. Our approach is naturally motivated by the directional derivative of a vector valued function $\vv$.  The orthogonal basis for $\mathbb{R}^d$; $\ee^1,\ee^2,\ldots\ee^d$ defines a coordinate frame. The gradient written in terms of directional derivatives is
 \begin{align}\label{strain}
    \nabla \vv(t,\xx,\bbxi)=\lim_{h\rightarrow 0^+} \sum_{m=1}^d\frac{(\vv(t,\xx+h\ee^m)-\vv(t,\xx))}{h}\otimes\ee^m.
\end{align}
Motivated by this we introduce a {\em frame gradient}. Given a vector $\bbxi$ in $\HH_\epsilon(0)$ we construct  a local cartesian  coordinate frame based at $\xx$ generated by $\bbxi$ in $\HH_{\epsilon}(0)$.

\begin{definition}[{\bf The  frame generated by $\bbxi$ based at $\xx$}]
\label{local frame at x}
{\em The  frame generated by $\bbxi$ based at $\xx$,} is defined to be the right handed orthonormal basis given by  $\ee^1=\bbxi/|\bbxi|$, $\ee^2=\bbxi^\perp/|\bbxi|$, $\ee^3=\ee^1\times\ee^2$. Set $\bbxi^m=|\bbxi|\ee^m$ for $m=1,2,\ldots d$ and the frame legs are the vectors $\xx+\bbxi^m$. The frame length is denoted by $|\bbxi|$.
\end{definition}

\begin{definition}[\bf Frame gradient]
\label{First}

The frame gradient $\nabla^{\bbxi}\vv(t,\xx)$ based at $\xx$ {\em generated} by $\bbxi$ is defined by
\begin{align}\label{strainf3}
    \nabla^{\bbxi} \vv(t,\xx)=\sum_{m=1}^3\frac{(\vv(t,\xx+\bbxi^m)-\vv(t,\xx))}{|\bbxi|}\otimes\ee^m
\end{align}
When $\vv(t,\xx)$ is smooth  the frame gradient  recovers the gradient in the limit $|\bbxi|\rightarrow 0$.
For plane stress loading the state of strain is two dimensional, $H_\epsilon(0)$ is a disk and  the orthonormal basis reduces to $\ee^1=\bbxi/|\bbxi|$, $\ee^2=\bbxi^\perp/|\bbxi|$. Setting  $\bbxi^m=|\bbxi|\ee^m$ for $m=1,2$ and the strain gradient based at $\xx$ generated by $\bbxi$ is:
 \begin{align}\label{strainf2}
    \nabla^{\bbxi} \vv(t,\xx)=\sum_{m=1}^2\frac{(\vv(t,\xx+\bbxi^m)-\vv(t,\xx))}{|\bbxi|}\otimes\ee^m
\end{align}
\end{definition}
For linear displacements $\vv(t,\xx)=M\xx$  and recalling  $I=\sum_{n=1}^d\ee^n\otimes\,\ee^n$  we have   $ \nabla^f \vv(t,\xx,\bbxi)=\sum_{n=1}^dM\ee^n\ee^n=MI=M$.

\section{Linearly elastic materials}
\label{Linear Elastic Materials}
For $\xx$  in $\Omega$ and  $\bbxi$ in $\HH_{\epsilon}(0)$ the associated strain is the symmetrization of the frame gradient given by $\EE^{\bbxi}(\uu(t,\xx))=(  \nabla^{\bbxi}\uu(t,\xx)+  (\nabla^{\bbxi}\uu(t,\xx))^T)/2$. 
The strain vanishes for rigid motions and the full null space of the strain consists of only the rigid motions. This is established in the following:

\begin{theorem}
\label{rigidrot}
The null space of $\EE^{\bbxi}(\uu(t,\xx))$  are the rigid displacements $\Pi$  defined by
\begin{align}
\label{space}
\Pi=\{\uu:\,\uu=\mathbb{Q}\xx+\dd,\,\mathbb{Q}\hbox{ in }\mathbb{R}^{d\times d},\,  \mathbb{Q}=-\mathbb{Q}^T,\,\dd\hbox{ in }\mathbb{R}^d\}.
\end{align}
\end{theorem}
\begin{proof}
Calculation shows any  rigid motion belongs to the null space of the strain.  We illustrate that rigid motions in fact comprise all motions belonging to the null set of the strain.  This is established for any open set $\Omega$ that is arc-wise connected and continuous displacement field  $\uu(t,\xx)$  with respect to $\xx$ for fixed $t$.  For all frames for which $\xx$ and $\xx+\bbxi^n$, $n=1,\dots,d$ belong to $\Omega$ suppose that $\EE^{\bbxi}(\uu(t,\xx)=0$, then we consider the quadratic form $\EE^{\bbxi}(\uu(t,\xx))\bbxi^n\cdot\bbxi^n=2\nabla^{\bbxi}\uu(t,\xx)\bbxi^n\cdot\bbxi^n$ to find
\begin{align}
\label{inner product}
\EE^{\bbxi}(\uu(t,\xx))\bbxi^n\cdot\bbxi^n=2\nabla^{\bbxi}\uu(t,\xx)\bbxi^n\cdot\bbxi^n=2(\uu(\xx+\bbxi^n)-\uu(\xx))\cdot(\bbxi^n)=0.
\end{align}
Choose $n=1$,  $\bbxi^1=\bbxi$ and select $\yy$ according to $\yy=\xx+\bbxi$ we see that for any such pair $\xx$, $\yy$ in $\Omega$ that  $(\uu(\yy)-\uu(\xx))\cdot(\yy-\xx)=0$. The condition  $(\uu(\yy)-\uu(\xx))\cdot(\yy-\xx)=0$ for all  $\yy=\xx+\bbxi$ in $\Omega$ is  consistent with the hypothesis of Theorem  $5$ of \cite{ZhouDuGunLeh}.  Application of the steps used to prove the theorem shows the  displacement is of the form $\uu(\xx)=\mathbb{Q}\xx+\cc$  in $\Omega^\ast$, where $\mathbb{Q}^T=-\mathbb{Q}$ and $\cc$ is a constant vector. (The theorem holds more generally for $\uu(t,\xx)$  Lesbegue measurable in $\xx$, see Theorem 5 of \cite{ZhouDuGunLeh}, \cite{DuMrngesha}).
\end{proof}

Denote the elastic tensor of  the material by $\overline{\mathbb{C}}$.  In general the elasticity tensor  $\overline{\mathbb{C}}$ of the material is a positive definite symmetric linear map on symmetric $d\times d$ tensors and is characterized by $21$ independent parameters for $d=3$. The elastic material can be inhomogeneous and depend on the location $\xx$ within the sample of material $\Omega$ and we write $\overline{\mathbb{C}}(\xx)$. In this context the elastic tensor can be piece wise continuous and represent a polycrystalline  or a multiphase composite medium.  The frame stiffness is given by the normalized elasticity tensor $\mathbb{C}(\xx)$ used to calculate the stress tensor of a frame and is proportional to the elasticity tensor of the material and $\overline{\mathbb{C}}(\xx)=p\mathbb{C}(\xx)$  where  %$p=(1/(2\omega_d))\int_{\HH_1(0)}|\bbxi|J(|\bbxi|)\,d\bbxi$
the formula for the constant of proportionality  is given by \eqref{elasticpart3} of section \ref{elasticproperties}.  

Introducing standard conventions we recall that $\overline{\mathbb{C}}(\xx) \boldsymbol{\eta}=\overline{\mathbb{C}}_{ijkl}(\xx) \boldsymbol{\eta}_{kl}$ where repeated indices indicate summation and the  inner product of two matrices $A, B$ is denoted by $A:B=tr\{AB^T\}$ and the Hilbert Schmidt norm of a matrix is $|A|_H=\sqrt{A:A}$. 
To show existence of solution the elasticity tensor is  chosen coercive and bounded, 
\begin{align}
\label{coersivebounded}
0<\underline{\Lambda} |\boldsymbol{\eta}|_H^2\leq\overline{\mathbb{C}}(\xx)\boldsymbol{\eta}:\boldsymbol{\eta}\leq\overline{\Lambda} |\boldsymbol{\eta}|_H^2. 
\end{align}
For showing existence in Section \ref{wellposedness} we work out the general case and suppose the component functions of the elasticity tensor  $\overline{\mathbb{C}}_{ijkl}(\xx)$ belong to $L^\infty(\Omega,\mathbb{R}^d)$. (This generality allows for nonlocal study of  cermets, laminates, functionally graded materials, alloys, additively manufactured materials, etc.).
In the frame based formulation we introduce a stress tensor associated with linear elastic behavior $\ss$ associated with a frame at $\xx$ and generated by $\bbxi$ through the linear constitutive law
\begin{align}
\label{str}
\ss(t,\xx,\bbxi)=&{\mathbb{C}(\xx)\EE^{\bbxi}(\uu(t,\xx))}.
\end{align}

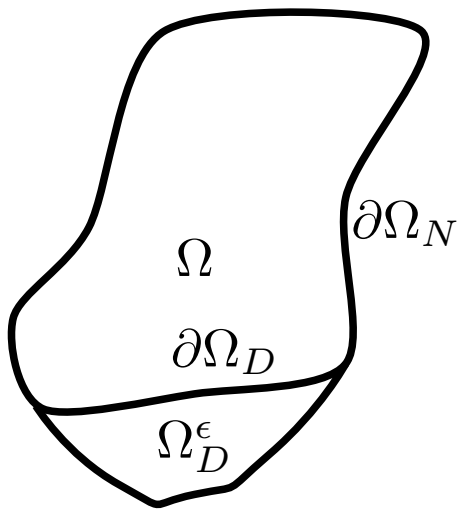
\begin{figure}
    \centering
    
      \begin{tikzpicture}[>=latex]
 \draw[line width=.5mm] plot [smooth cycle] coordinates {(0,0) (1,0.1) (2,0.3) (2,1.4) (2.5,2.5) (0.8,2.5) (0.3,1.2) (-0.2,0.6) } node at (1,1) {$\Omega$};
 \node [below] at (2.4,1.5) {$\partial\Omega_N$};
  \node [below] at (1.2,0.65) {$\partial\Omega_D$};
%\draw[line width=.5mm] plot [smooth cycle] coordinates {(0,0) (1,0.1) (2,0.3) (2,1.4) (1.5,2.5) (0.8,2.5) (0.3,1.2) (-0.2,0.6) } node at (1,1) {$\Omega$};
 %\draw[line width=.5mm] plot [smooth,tension=1.2] coordinates {(-0.05,0.02) (0.5,-0.5) (1.0,-0.565) (1.5,-0.295) (2,0.3)} node at (1.0,-0.25) {$\Omega_D^\epsilon$};
 \draw[line width=.5mm] plot [smooth,tension=1.2] coordinates {(-0.05,0.02) (0.5,-0.5) (1.0,-0.565) (1.5,-0.295) (2,0.3)} node at (1.0,-0.25) {$\Omega_D^\epsilon$};
\end{tikzpicture}
    \caption{ Domain $\Omega$ with prescribed Dirichlet data on $\Omega_D^\epsilon$. The union is denoted by $\Omega^*$}
    \label{Dirichlet}
\end{figure}

We start by describing the boundary conditions for frame based mechanics in anticipation of describing the Initial Boundary Value Problems (IBVP) for solving for the displacement.
Displacement loads are prescribed on the boundary of the specimen $\Omega$. For the nonlocal model, the displacement $\UU(t,\xx)$ is prescribed on a layer $\Omega_D^\epsilon$ surrounding part of the specimen, We write $\Omega^*=\Omega\cup\Omega_D^\epsilon$ where the layer is of minimum thickness greater than or equal to $\epsilon$, see Figure \ref{Dirichlet} . With this in mind, we introduce the non local kernel given by
\begin{equation}\label{scale2}
\rho^{ \ast,\epsilon}(\xx,\bbxi)=\chi_{\Omega}(\xx)\frac{\Pi_{n=1}^d\chi_{\Omega^\ast}(\xx+\bbxi^n)J^\epsilon(|\bbxi|)}{ \epsilon V^\epsilon_d},
\end{equation}
where $\chi_{\Omega}$  is the characteristic function of $\Omega$ and $\chi_{\Omega^\ast}$ is the characteristic function of $\Omega^\ast$,  $J^\epsilon(|\bbxi|)$ is a bounded dimension free influence function that is, a positive function of $|\bbxi|$ on the ball of radius $\epsilon$ centered at $0$ and is radially decreasing to $0$ inside the ball  and $0$  for $|\bbxi|\geq\epsilon$. This ball is denoted by $\mathcal{H}_\epsilon(0)$ gives all admissible frame legs. Stress interactions occur on a ball of radius $\sqrt{\epsilon}$ centered at the origin.
Note that the kernel \eqref{scale2} reduces to $J^\epsilon(|\bbxi|)\epsilon^{-1}(V_d^{\epsilon})^{-1}$ inside $\Omega$ for points inside $\Omega$ a distance $\epsilon$ away from the boundary $\partial\Omega^\ast$.
We also introduce the adjusted non local  kernel given by 
\begin{equation}\label{scale33}
\rho^{ \epsilon}(\xx,\bbxi)=\chi_{\Omega}(\xx)\frac{\Pi_{n=1}^d\chi_{\Omega}(\xx+\bbxi^n)J^\epsilon(|\bbxi|)}{ \epsilon V^\epsilon_d},
\end{equation}
where we have replaced all appearances of  $\chi_{\Omega^\ast}$ by  $\chi_\Omega$ in \eqref{scale2}.

 We now explain the details of Dirichlet and homogeneous Neumann boundary conditions for  frame based inelasticity
The frame admissibility conditions that deliver the boundary conditions are listed below, see Figure \ref{Dirichlet}.

\begin{definition}
\label{Boundary conditions}
{\bf Boundary Conditions}\\

The boundary components of $\Omega$ associated with Neumann boundary conditions and Dirichlet boundary conditions are denoted by $\partial\Omega_N$ and  $\partial\Omega_D$ respectively.
 \begin{enumerate} 
 \item Homogeneous Neumann Boundary Conditions: Only frames based at points $\xx$ inside $\Omega$ are admissible and $\partial\Omega=\partial\Omega_N$. 
 \begin{itemize}
 \item If any leg of a frame based at $\xx$  crosses $\partial\Omega_N$ then the corresponding frame gradient is set to zero.
 \end{itemize}
 \item Dirichlet and Homogeneous Neumann Boundary Conditions: Only frames based at points $\xx$ inside $\Omega$ are admissible and $\partial\Omega=\partial\Omega_N\cup\partial\Omega_D$:
 \begin{itemize}
 \item If any legs of the frame intersect $\partial\Omega_N$  then the frame gradient is set to zero.
 \item If any legs of the frame intersect $\Omega_D$ and the rest lie in $\Omega$ then the legs $\xx+\bbxi^m$ that belong to $\Omega_D$ satisfy $\uu(t,\xx+\bbxi^m)=\UU(t,\xx+\bbxi^m)$.
 \item If any  leg of the frame only intersects $\partial\Omega_D$ but lies outside $\Omega_D$ then the frame is inadmissible.
 \end{itemize}
 \end{enumerate}
 \end{definition}
 \noindent The last condition in $2$ can occur in a neighborhood of the junction where the Neumann and Dirichlet boundary conditions meet.

To introduce the ideas we describe the IBVP for linear elastic problems with homogeneous Neumann boundary conditions. We apply the more general boundary conditions for the fracture problem in the next section.
We suppose that the elasticity tensor is both positive definite and bounded and define the nonlocal elastic energy
\begin{align}\label{energyy}
\mathcal{E}(t,\uu)=\frac{1}{2}\int_{\Omega}\int_{\Omega}\,\rho^{\epsilon}(\xx,\bbxi)|\bbxi|\,\ss_{ij}(t,\xx,\bbxi)\EE_{ij}^{\bbxi}(\uu(t,\xx))\,d\bbxi\,d\xx.
\end{align}

\begin{figure}
    \centering
    
\begin{tikzpicture}

    % Define square coordinates (centered at origin)
    \coordinate (A) at (-.1,-.1);
    \coordinate (B) at (.1,-.1);
    \coordinate (C) at (.1,.1);
    \coordinate (D) at (-.1,.1);

    % Draw the square
  %  \draw[line width=1pt] (A) -- (B) -- (C) -- (D) -- cycle;

    % Calculate the center points of each face
    \coordinate (BottomCenter) at (0,-.1);
    \coordinate (RightCenter)  at (.1,0);
    \coordinate (TopCenter)    at (0,.1);
    \coordinate (LeftCenter)   at (-.1,0);

    % Draw normal force arrows exiting from the face centers
    \draw[-{[scale=1.2]}, blue, thick,<-] (TopCenter) -- (1,3.0) 
        node[above] {$\ss(t,\xx,\bbxi)(-\ee^2)$};
        
         \draw[-{[scale=1.2]}, black, thick, ->] (0,0) -- (0,2) 
        node[above] {$\ee^2$};
        
    \draw[-{[scale=1.2]}, blue, thick,<-] (RightCenter) -- (3.8,1) 
        node[right] {$\ss(t,\xx,\bbxi)(-\ee^1)$};
        
        \draw[-{[scale=1.2]}, black, thick,->] (0,0) -- (2,0) 
        node[right] {$\ee^1$};
        
    \draw[-{[scale=1.2]}, blue,thick,<-] (BottomCenter) -- (0.5,-3.0) 
        node[below] {$\ss(t,\xx-\bbxi^2,\bbxi)\ee^2$};
        
    \draw[-{[scale=1.2]}, blue, thick,<-] (LeftCenter) -- (-2.5,-0.3) 
        node[left] {$\ss(t,\xx-\bbxi^1,\bbxi)\ee^1$};

    % Optional: Mark the geometric center
   \filldraw[black]  (0,0) circle (2pt) node[ below left] {$\xx$};

\end{tikzpicture}

 \caption{ Tractions acting on $\xx$  in a plane stress state ($\ss_{33}=0$) by neighboring frames generated by the choice  $\bbxi=(\xi,0)$ in $\HH_\epsilon(0)$  for $d=2$. (For three dimensional stress states the tractions acting on  $\xx$ are given by \eqref{eq: force20} with $d=3$.)} 
    \label{normal stress}
\end{figure}
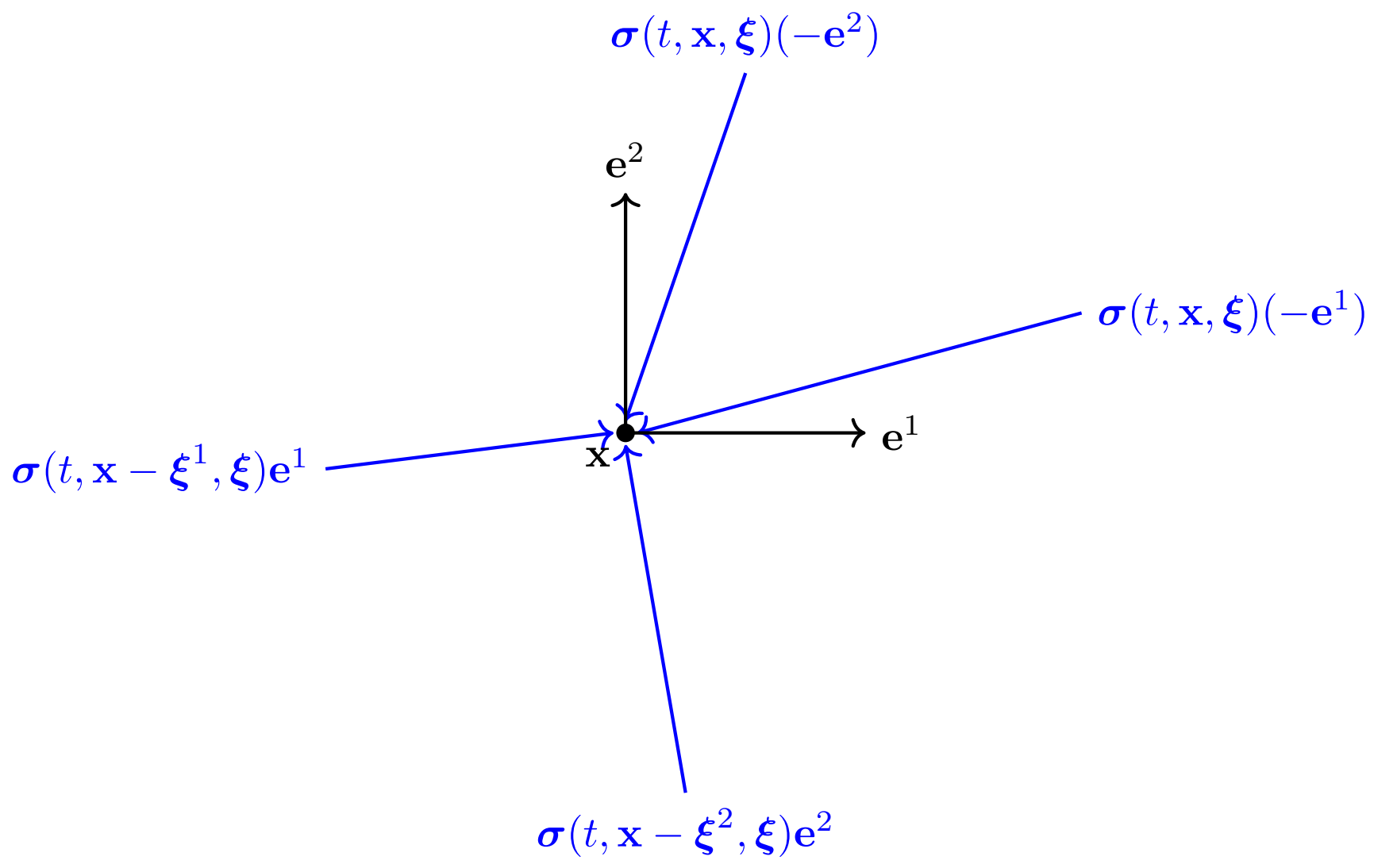

The traction force on the plane transverse to  $\xx$ in the direction $\ee^n$ due to the frame located at $\xx$ generated by $\bbxi$ is
 \begin{align}\label{eq: force2}
 -\ss_{ij}(t,\xx,\bbxi)\ee_j^n=-\mathbb{C}_{ijkl}(\xx)\EE_{kl}^{\bbxi}(\uu(t,\xx))\ee^n_j,
%  \LL^\epsilon [\uu](t,\xx) =-\int_{\Omega^*}\rho^{\ast, \epsilon}(\xx,\bbxi)\sum_{n=1}^d \ss\ee^nd\bbxi.
\end{align}
and the traction force per unit area on the plane transverse to $\xx$  due to  the neighboring frame on the $\ee^n$ axis located  at  the shifted point $\xx-\bbxi^n$ generated by $\bbxi$ is given by
 \begin{align}\label{eq: force3}
 \ss_{ij}(t,\xx-\bbxi^n,\bbxi)\ee_j^n=\mathbb{C}_{ijkl}(\xx-\bbxi^n)\EE_{kl}^{\bbxi}(\uu(t,\xx-\bbxi^n))\ee^n_j.
%  \LL^\epsilon [\uu](t,\xx) =-\int_{\Omega^*}\rho^{\ast, \epsilon}(\xx,\bbxi)\sum_{n=1}^d \ss\ee^nd\bbxi.
\end{align}
Summing over all $d$ directions $\ee^n$ and in view of the  factor $\epsilon V^\epsilon_d$ appearing in the denominator of the kernels shows that the total traction force  per unit volume$^2$  at $\xx$, corresponding to fixed $\bbxi$  is  given by the sum 
\begin{align}
      \label{eq: force20}
- \sum_{n=1}^d \left[\rho^{\epsilon}(\xx,\bbxi)\, \ss_{ij}(t,\xx,\bbxi)\ee_j^n -\rho^{\epsilon}(\xx-\bbxi^n,\bbxi)\,\ss_{ij}(t,\xx-\bbxi^n,\bbxi)\ee_j^n\right],
 \end{align}
 see Figure \ref {normal stress}.
 We can write the total fraction force in terms of the divergence associated with the frame. Taking the frame gradient of $\rho^{\epsilon}(\xx,\bbxi)\ss(t,\xx,\bbxi)$ gives
 \begin{align}\label{stress3}
    \nabla^{-\bbxi}[\rho^{\epsilon}(\xx,\bbxi)\ss(t,\xx,\bbxi)]=\sum_{m=1}^d\frac{\rho^{\epsilon}(\xx-\bbxi^m,\bbxi)(\ss(t,\xx-\bbxi^m,\bbxi)-\rho^{\epsilon}(\xx,\bbxi)\ss(t,\xx,\bbxi))}{|\bbxi|}\otimes\ee^m,
\end{align}
 and taking its trace we define the frame divergence    as
 \begin{align}\label{stress4}
\nabla^{-\bbxi}\cdot[\rho^{\epsilon}(\xx,\bbxi)\ss(t,\xx,\bbxi)]&=  tr\{ \nabla^{-\bbxi}\rho^{\epsilon}(\xx,\bbxi)\ss(t,\xx,\bbxi)\}\nonumber\\
&=\sum_{m=1}^d\frac{(\rho^{\epsilon}(\xx-\bbxi^n,\bbxi)\ss_{ij}(t,\xx-\bbxi^m)-\rho^{\epsilon}(\xx,\bbxi)\ss_{ij}(t,\xx,\bbxi)}{|\bbxi|}\ee_j^m.
\end{align}
We see that \eqref {eq: force20} is represented simply  by the product of frame length times the frame divergence given by \eqref{stress4}.
 On taking the first variation of the elastic energy \eqref{energyy} and applying the homogeneous Neumann boundary conditions
   we recover the nonlocal force density $\LL^\epsilon[\uu](t,\xx)$ defined for all points $\xx$ in $\Omega$  given by
\begin{align}
      \label{eq: force2}
 \LL^\epsilon [\uu](t,\xx) = \int_{\Omega}\nabla^{-\bbxi}\cdot[\rho^{\epsilon}(\xx,\bbxi)\ss(t,\xx,\bbxi)]|\bbxi|\,d\bbxi,
 \end{align}
 where the integrand is total traction force given by \eqref{eq: force20}. Set $\Omega_\epsilon=\{\xx\in\Omega;\, dist(\xx,\partial\Omega)>\epsilon\}$ and note from the definition of $\rho^\epsilon(\xx,\bbxi)$ when $\xx$ and $\xx+\bbxi^n$, $n=1,2,\ldots,d$ lie inside $\Omega_\epsilon$ then
 \begin{align}
      \label{eq: force3}
 \LL^\epsilon [\uu](t,\xx) = \frac{1}{V_d}\int_{\Omega}J^\epsilon(|\bbxi|)|\bbxi|\epsilon^{-1}[\nabla^{-\bbxi}\cdot\ss(t,\xx,\bbxi)]\,d\bbxi.
 \end{align}

  The integration by parts methods of  Section \ref{powerbalanceprocesszone2} can be used with out change to obtain  the displacement  force $\LL^\epsilon [\uu](t,\xx)$ density from the first variation of the energy.  Notably only one spatial integration is necessary for computing  the displacement force density for the frame based method.

 The initial boundary value problem for elastic evolution is to
find a displacement $\uu(t,\xx)$ on $\Omega$ that satisfies the homogenous Neumann boundary conditions given by Definition \ref{Boundary conditions}, part 1,
and for $\xx$ in
$\Omega$  satisfies the evolution equation 
\begin{align}\label{eq: linearmomentumbal2}
\rho\ddot{\uu}(t,\xx)+  \LL^\epsilon [\uu](t,\xx) =\bb(t,\xx),
\end{align}
for $0<t<T$, and satisfies the initial conditions on the displacement and velocity given by
\begin{align}
\uu(0,\xx)=\uu_0(\xx),\nonumber\\
\dot{\uu}(0,\xx)=\vv_0(\xx).\label{initialconditions3}
\end{align}

Note that  that any  rigid rotation solves the homogeneous Neumann boundary condition. Thus as in the local theory to insure a well posed IBVP we must impose a solvability condition on the body force $\bb(t,\xx)$. \footnote{More precisely  we ask that $\bb(t,\xx)$ be chosen to lie in a closed subspace of $L^\infty(\Omega,\mathbb{R}^d)$ that doesnt intersect $\Pi$.}

%%%%%%%%%%%%
%
% Square and forces fig
%
%%%%%%%%%%

%%%%%%%%%%%%%%
%%%%%%%%%%%%%

\section{Frame based constitutive laws with damage and the initial boundary value problem for brittle fracture}
\label{sec.Constiutivelaws BVP}

In this section we address inelastic behavior given by brittle fracture. The inelastic material behavior is represented  by swapping out the elastic  constitutive law and replacing it with a  new  space-time constitutive law  relating stress to strain appropriate for fracture. Here damage is modeled through softening of elastic moduli. We begin by setting up the problem in this introduction and defining the scaled effective stress. The next subsection describes the method of softening elastic moduli using effective stress, Section \ref{scalefactor}. The material domain is first chosen as a subset of   $\mathbb{R}^3$ and  then we consider the plane stress state case. The deviatoric and hydrostatic parts of the stress are denoted by $\ss=\ss_D+\ss_h$.  We introduce the fourth order projections onto the deviatoric and the hydrostatic parts of the stress  given by
\begin{align}
\mathbb{K}_{ijkl}=\frac{1}{2}(\delta_{ik} \delta_{jl} +\delta_{il}\delta_{kj})-\frac{1}{3}\delta_{ij}\delta_{kl},
\label{tracefree}
\end{align}
\begin{align}
\mathbb{J}_{ijkl}=\frac{1}{3}\delta_{ij}\delta_{kl},
\label{identity}
\end{align}
with the identity on symmetric $3\times3$ matrices in three dimensions given by
\begin{align}
\mathbb{I}_{ijkl}=\frac{1}{2}(\delta_{ik} \delta_{jl} +\delta_{il}\delta_{kj}).
\label{fourthordedident}
\end{align}

In the present context it is possible to model the strength of the material using several  different theories, i.e., Von Mises, Drucker Prager, Mohr Coulomb, etc. To fix the discussion we consider Drucker Prager  surfaces in stress space and apply it to soften elastic moduli. For a given strain $\EE^{\bbxi}(\uu(t,\xx))$ the effective stress  associated with  the frame gradient based at $\xx$ generated by $\bbxi$ is given by $\tilde{\ss}=\tilde{\ss}(t,\xx,\bbxi):=\mathbb{C}(\xx)\EE^{\bbxi}(\uu(t,\xx))$ where $\mathbb{C}(\xx)$ is the undamaged elastic tensor field. The Drucker Prager function acts on the effective stress and is defined by
\begin{align}
\label{strength surface}
q=\mathcal{F}(\tilde{\ss}) +\beta_C I_1(\tilde{\ss}),
\end{align}
where $\beta_C$ is a constant that depends on the material. 
Here   $\mathcal{F}$ is a homogeneous function of degree one and is $\mathcal{F}(\tilde{\ss})=\sqrt{\mathcal{J}_2(\tilde{\ss}_D)}$ where $\mathcal{J}_2({\tilde{\ss}_D)}$ is the second invariant of $\tilde{\ss}_D$ given by
\begin{align}
\label{Jtwo}
\mathcal{J}_2(\tilde{\ss}_D)=\frac{1}{2}\tilde{\ss}_D:\tilde{\ss}_D=\frac{1}{6}[(\tilde{\sigma}_D^1-\tilde{\sigma}_D^2)^2+(\tilde{\sigma}_D^2-\tilde{\sigma}_D^3)^2+(\tilde{\sigma}_D^3-\tilde{\sigma}_D^1)^2].
\end{align}
where $\tilde{\sigma}_D^1,\tilde{\sigma}_D^2,\tilde{\sigma}_D^3$ are the principal values of $\tilde{\ss}_D$ and $I_1$ is the first invariant of $\tilde{\ss}$  and $I_1(\tilde{\ss})=tr(\tilde{\ss})$.

For each $\xx$ in $\Omega$ and  $\bbxi$ in $\HH_\epsilon(0)$ we introduce the function $q(t,\xx,\bbxi,\uu)$ given by
\begin{align}
\label{q}
q(t,\xx,\bbxi,\uu)=q(\tilde{\ss}(t,\xx,\bbxi))=\mathcal{F}(\tilde{\ss}((t,\xx,\bbxi)) +\beta_C I_1(\tilde{\ss}((t,\xx,\bbxi)).
\end{align}

For convenience we will write $q(t):=q(t,\xx,\bbxi,\uu)$ and note that $q(\tilde{\ss})$ is homogeneous of degree one in its argument. It is clear from the discussion that $q(t)$ is a function of the strain evolution associated with the evolving displacement field $\uu(t,\xx)$. 

To model softening elastic moduli and frame failure a scale factor $\sqrt{|\bbxi|/L}$ is introduced  where $L$ is a length scale characteristic of the motion and $|\bbxi|$ is the frame length where $\bbxi$ belongs  to $\HH_{\epsilon}(0)$.  The scaled strain is defined as : $\bS(t,\xx,\bbxi,\uu)=\sqrt{|\bbxi|/L}\ \EE^{\bbxi}(\uu(t,\xx))$.
Using homogeneity we write the scaled effective stress surfaces
\begin{align}
\label{rescale}
q(\mathbb{C}(\xx)\bS(t,\xx,\bbxi,\uu))=q(\sqrt{|\bbxi|/L}\ \tilde{\ss}(t,\xx,\bbxi))=\sqrt{|\bbxi|/L}\ q(t):=r(t),
\end{align}
where $r(t)=r(t,\xx,\bbxi,\uu)$. 

\begin{remark}
\em{Here the scaling $\sqrt{|\bbxi|/L}$ is chosen such  that the stress work  to fail a frame is the same for all frames and independent of frame length  $|\bbxi|$,  As a consequence this guarantees that the energy necessary to form a straight (flat) crack of length $\ell$ is independent of $\epsilon$, 
see Section \ref{criticalenergyrelease}.}
\end{remark}

\begin{figure}[H]
    \centering
    {\includegraphics[width=3in]{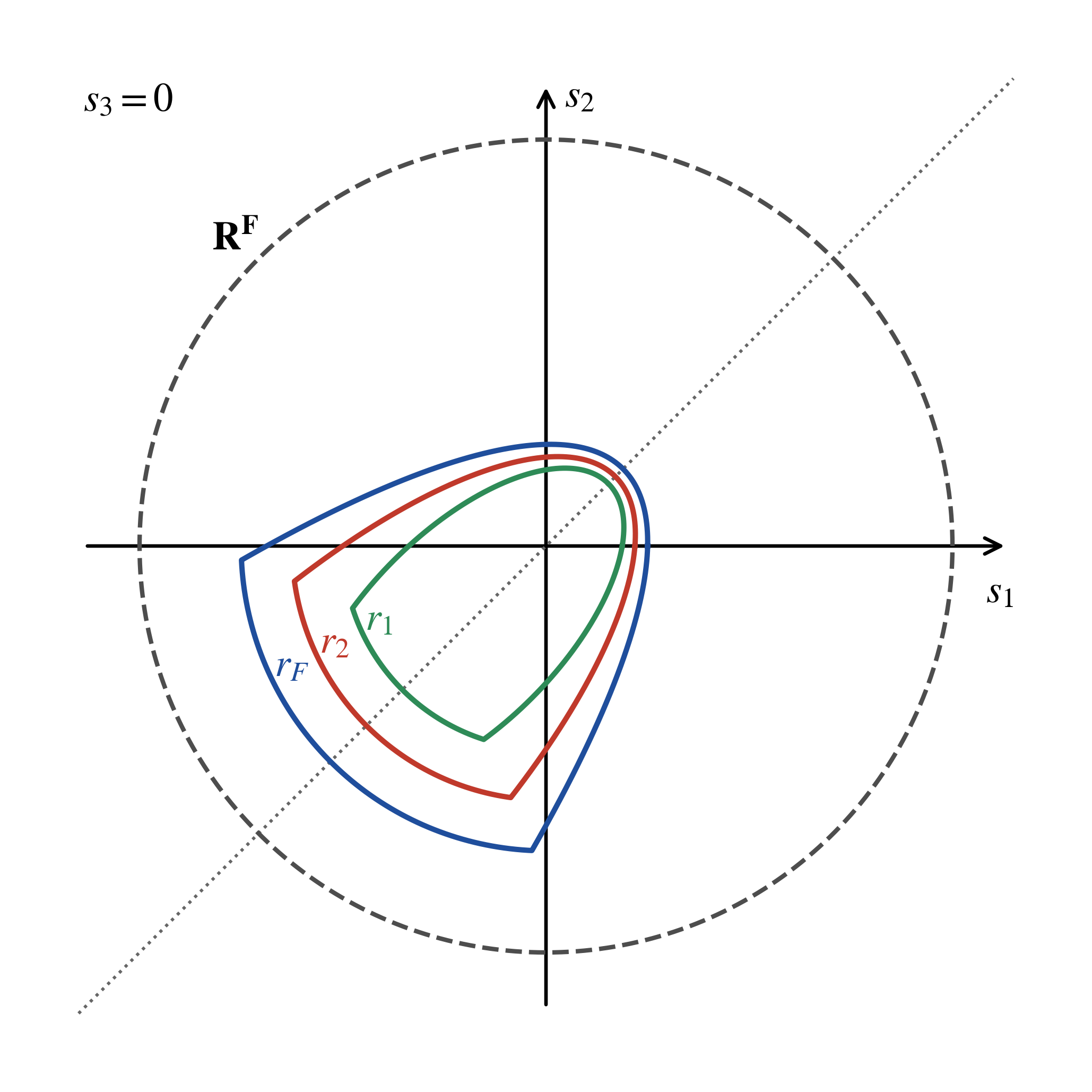}}
 \caption{%
    Closed scaled effective stress surfaces for plane stress loading. Scaled effective stress surfaces similarly closed in compressive octant for $3$ dimensional elasticity.}
  \label{iso}
\end{figure}

We introduce memory defined by
\begin{align}
\label{qEmem}
\\r^\ast(t)=r^\ast(t,\xx,\bbxi,\uu)=\max_{0\leq \tau\leq t}r(\tau,\xx,\bbxi,\uu),
\end{align}
where $r(t)\leq r^\ast(t)$ and $r(t)\leq r^\ast(t)=\max_{0\leq \tau\leq t}r(\tau,\xx,\bbxi,\uu)$.  It is stressed that the Drucker Prager  strength surface as classically defined is unbounded for triaxial  compressive stress. Motivated by \cite{lubliner} and \cite{Fenves}, with application to geology, granular materials  and concrete, we consider scaled equivalent stress surfaces  associated with $r$ in $[0,r^F]$  and define them in the compressive octant  so that the family of increasing iso-$r$ surfaces sweep out a finite volume in the scaled effective stress space that is contained inside a ball of radius $R^F$ centered at the origin. Hence  
\begin{align}
\label{bounding ball}
|\sqrt{|\bbxi|/L}\ \tilde{\ss}(t,\xx,\bbxi)|_H\leq R^F, 
\end{align}
see Figure \ref{iso}.
The quantity $r^\ast(t)$ is the maximum value of  $r(t)$ reached  during the evolution of $\rr(\tau,\xx,\bbxi,\uu)$ over the interval $0\leq \tau\leq t$ and $r^*(t)\leq r^F$.

\subsection{Residual strength, damage, loss of elasticity and fracture}
\label{scalefactor}
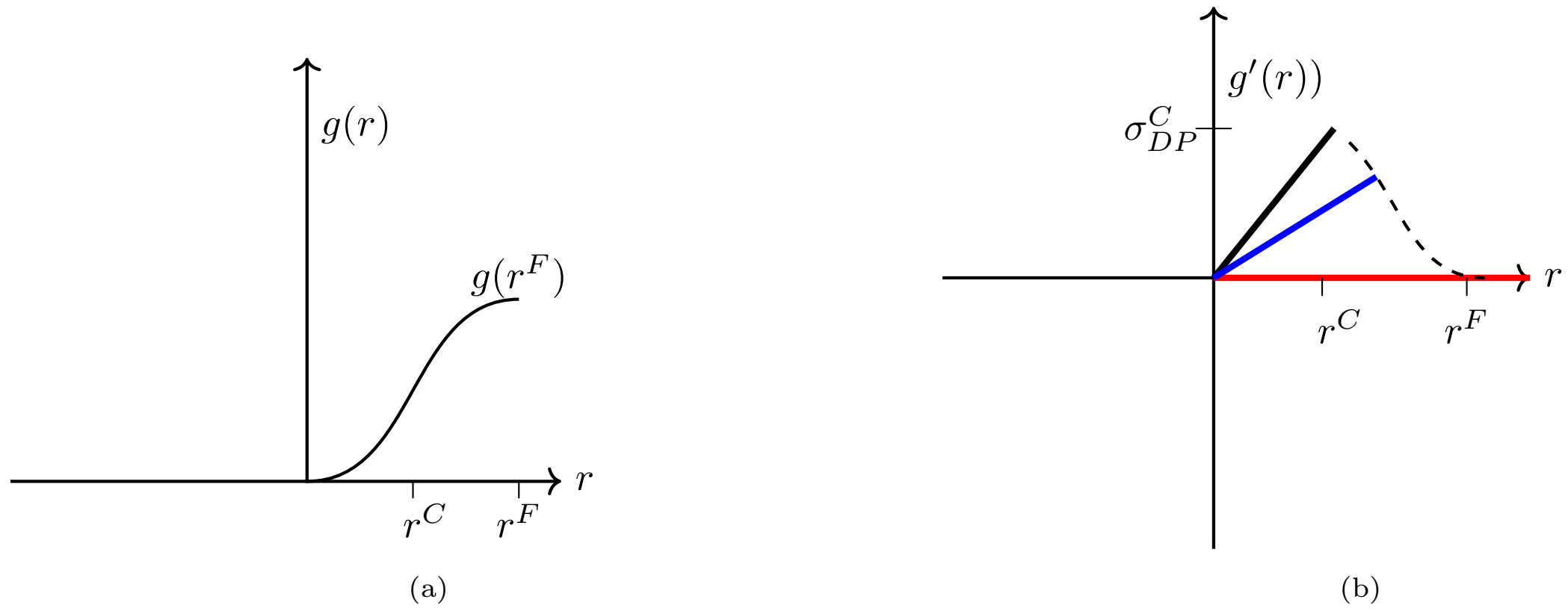
\begin{figure}
    \centering
    \begin{subfigure}{.45\linewidth}
        \begin{tikzpicture}[xscale=0.75,yscale=0.75]
		    \draw [<->,thick] (0,5) -- (0,0) -- (3.0,0);
			\draw [-,thick] (0,0) -- (-3.5,0);
		
			\draw [-,thick] (0,0) to [out=0,in=-180] (2.5,2.15);
			%\draw [-,thick] (-5.0,7.25) to [out=-5,in=180] (0,0);
			%\draw  [-,thick] (0,0) parabola (-2,4);
			
			\draw (2.5,-0.2) -- (2.5, 0.0);
			\node [above] at (2.5,2.0) {$g(r^F)$};
			\node [below] at (2.5,-0.1) {${r}^F$};

			\draw (1.25,-0.2) -- (1.25, 0.0);
			\node [below] at (1.4,-0.1) {${r}^C$};
			
			%\draw (0.8,-0.2) -- (0.8, 0.0);
			%\node [below] at (0.8,-0.1) {${r}^L$};

			\node [right] at (3,0) {$r$};
			
			\node [right] at (0,4.20) {$g(r)$};
			
		  \end{tikzpicture}
		  \caption{}
		  \label{ConvexConcavea}
    \end{subfigure}
    \hskip2em
    \begin{subfigure}{.45\linewidth}
        \begin{tikzpicture}[xscale=0.8,yscale=0.8]
		    \draw [<-,thick] (0,3) -- (0,-3);
			\draw [->,thick] (-3,0) -- (3.5,0);
            \draw [-,ultra thick,red] (0,0) -- (3.5,0);
			;
           % \draw [dashed,thick] (0,0) to [out=60,in=140] (1.5,1.5); 
            \draw [dashed,thick] (1.5,1.5) to [out=-45,in=180] (3,0.0);
						
			%%%%%\draw [-,ultra thick] (-1.735,-3.22214) -- (0,0); %Line
			
			\draw [-,ultra thick,black] (0,0) -- (1.332,1.653); %Secant Line 1
			%\draw [-,ultra thick] (0,0) -- (0.7,1.3); %Secant Line 0
			\draw [-,ultra thick,blue] (0,0) -- (1.8,1.119); %Secant Line 2
			\draw (2.8,-0.2) -- (2.8, 0.0);
			
			\draw (1.2,-0.2) -- (1.2, 0.0);
			%\draw (0.75,-0.2) -- (0.75, 0.0);
			\node [below] at (1.4,-0.2) {${r}^C$};
			%\node [below] at (0.75,-0.2) {${r}^L$};
			\node [below] at (2.8,-0.2) {$r^F$};
			
			\node [right] at (3.5,0) {$r$};
			\node [right] at (0,2.2) {$g'(r))$};
			
			\draw (-0.2,1.653) -- (0.2, 1.653);

			\node [left] at (0,1.653) {$\sigma_{DP}^C$};
			
		  \end{tikzpicture}
		   \caption{}
		   \label{ConvexConcaveb}
    \end{subfigure}
    \caption{  (a) The stress potential $g(r)$ used in characterizing the failure envelope for a frame of length $|\bbxi|$.  (b) The envelope of unloading laws  denoted by $\sigma_{DP}(r):=g'(r)$ is illustrated by the dashed curve for a frames of  length $|\bbxi|$. The frame stiffness becomes nonlinear at $r^C$ and goes smoothly to zero at $r^F$. Unloading laws are linear elastic with a softer stiffness controlled by the phase field and shown in blue. The frame offers zero  stiffness for $r^F<r ^\ast$.}
    \label{ConvexConcave}
\end{figure}

We introduce the history dependent constitutive law relating stress to strain. We start by introducing the parameterized family of failure surfaces for the effective stress $\tilde{\ss}(t,\xx,\bbxi)$ characterized by the Drucker-Prager strength function $\sigma_{DP}(r^\ast)$, see Figure \ref{ConvexConcaveb}.  The strength function of a frame of length $|\bbxi|$ is taken to be a derivative of a non negative convex-concave potential $g(r)$ defined for positive $r$ and  $\partial_r g(r)=\sigma_{DP} (r)$. For $0\leq r^\ast\leq r^C$ the function increases and $\sigma_{DP}(r^\ast)=r^\ast$. After it reaches the maximum value of  $\sigma_{DP}^C$ the function  $\sigma_{DP}(r^\ast)$ decreases continuously post peak to $0$ when $r^\ast=r^F$.

The Drucker Prager strength curve is the envelope of the unloading curves defined by lines connecting the origin $(0,0)$ to a point on the curve $(r^\ast,\sigma_{DP} (r^\ast))$. The curve is  given by $(r^\ast,\sigma_{DP}(r^\ast))$ for $0\leq r^\ast\leq r^F$, is also called the loading curve.
The phase field is the damage parameter $\gamma(t)=\gamma(t,\xx,\bbxi,\uu)$ given by the slope of the unloading curves 
\begin{align}
\label{sope}
\gamma=\frac{\sigma_{DP}(r^\ast)}{r^\ast},
\end{align}
For $r\leq r^\ast$the unloading curve is given by $\gamma r$. 
On the loading curve 
\begin{align}
\label{loadingunloading}
r(t,\xx,\bbxi,\uu)=r^\ast(t,\xx,\bbxi,\uu),
\end{align}
so $\gamma r=\gamma r^\ast=\sigma_{DP}(r^\ast)$ when the unloading curve intersects the loading curve, see Figure \ref{ConvexConcaveb}. Here $0\leq \gamma(t) \leq 1$ with $\gamma(t)=1$ indicating the  scaled strain is $\bS(t,\xx,\bbxi,\uu)$ has not reached the strength boundary given by the locus of all $\bS(t,\xx,\bbxi,\uu)=\sigma_{DP}^C$.  The condition  $\gamma(t)=0$ indicates the frame carries no stress.
Note since $\sqrt{\mathcal{J}_2}$ and $I_1$ are homogeneous of degree one in their arguments we can write the unloading curve for the damaging material as
\begin{align}
\label{homogeneity}
\gamma(t)r(t)=\mathcal{F}((\xx)\sqrt{|\bbxi|/L}\gamma(t)\mathbb{C}\EE^{\bbxi}(\uu(t, \xx))) +\beta_C  I_1(\sqrt{|\bbxi|/L}\gamma(t)\mathbb{C}(\xx)\EE^{\bbxi}(\uu(t,\xx)),
\end{align}
Inspection of \eqref{homogeneity}  reveals that the stress strain relation with elastic softening at the frame at $\xx$ generated by $\bbxi$   is defined to be
\begin{align}
\label{softsigma}
\ss(t,\xx,\bbxi)&=\gamma(t)\mathbb{C}(\xx)\EE^{\bbxi}(\uu(t,\xx),
\end{align}
here the damaged elastic tensor associated with the frame at $\xx$ generated by $\bbxi$  is
\begin{align}
\label{softening3d}
\gamma(t)\mathbb{C}(\xx)=\gamma(t,\xx,\bbxi,\uu)\mathbb{C}(\xx).
\end{align}
The shifted stress is due to the frame at $\xx-\bbxi^n$ generated by $\bbxi$ is written as
\begin{align}\label{histconstlaw2bis}
\ss(t,\xx-\bbxi^n,\bbxi)&=\gamma(t,\xx-\bbxi^n,\bbxi,\uu)\mathbb{C}(\xx-\bbxi^n)\EE^{\bbxi}(\uu(t,\xx-\bbxi^n).
\end{align}
The total traction force  per unit volume squared  at $\xx$, corresponding to fixed $\bbxi$  is  given by
 \begin{align}\label{eq: stress3d}
\nabla^{-\bbxi}\cdot[\rho^{\ast,\epsilon}(\xx,\bbxi)\ss(t,\xx,\bbxi)]|\bbxi| = \sum_{n=1}^3\left[ \rho^{\ast, \epsilon}(\xx,\bbxi)\, \ss(t,\xx,\bbxi)\ee^n -\rho^{\ast, \epsilon}(\xx-\bbxi^n,\bbxi)\,\ss(t,\xx-\bbxi^n,\bbxi)\ee^n\right] ,
\end{align}
the analogous tractions for plane stress are shown  in Figure \ref{normal stress}. Consequently the total traction force is the internal force given by
\begin{align}\label{eq: tottractforce3}
\LL^\epsilon [\uu](t,\xx)=\int_\Omega \nabla^{-\bbxi}\cdot[\rho^{\ast,\epsilon}(\xx,\bbxi)\ss(t,\xx,\bbxi)]|\bbxi| \,d\bbxi,
\end{align}

The initial boundary value problem for  linear elastic brittle fracture evolution is to
find a displacement $\uu(t,\xx)$ on $\Omega^*$ that satisfies 
\begin{align}\label{eq: prescrib boundary displacement2}
\uu(t,\xx)=\UU(t,\xx)& \hbox{ for }\xx \hbox{ in } \Omega^\epsilon_D,
\end{align}
and for $\xx$ in
$\Omega$  satisfies the evolution equation 
\begin{align}\label{eq: linearmomentumbal3}
\rho\ddot{\uu}(t,\xx)+ \LL^\epsilon [\uu](t,\xx)=\bb(t,\xx),
\end{align}
for $0<t<T$, and satisfies the initial conditions on the displacement and velocity given by
\begin{align}
\uu(0,\xx)=\uu_0(\xx),\nonumber\\
\dot{\uu}(0,\xx)=\vv_0(\xx).\label{initialconditions2}
\end{align}

As before only one spatial integration is necessary for computing  the displacement force density and this provides an order of magnitude savings in computational work over a state based peridynamic method. This is because state based methods require a separate integration  over  $\HH_\epsilon(0)$ to compute the gradient and then a second integration over $\HH_{\epsilon}(0)$ to compute the  force density $\LL^\epsilon [\uu](t,\xx)$.

\subsection{Calibration of  strength surface.}
\label{sec.Calibration}

In this section we  calibrate  the  strength  surface $\sigma_{DP}^C$. The majority of strength data for materials are obtained from uniaxial experiments  and with this in mind   the uniaxial  strength in tension is denoted by $\sigma_{\rm tf}>0$ and uniaxial  strength in compression given by $\sigma_{\rm cf}<0$. It is noted that $|\sigma_{\rm cf}|>\sigma_{\rm tf}$ for refractory materials, ceramics, concrete and rocks.
The Drucker Prager strength surface is given by all $\tilde{\ss}$ for which
\begin{align}
\label{strength surface}
\sigma_{DP}^C=\mathcal{F}(\tilde{\ss}_D) +\beta^C I_1(\tilde{\ss}).
\end{align}
Using the uniaxial strength in tension and compression we obtain
\begin{align}
\label{qc}
\sigma_{DP}^C=\frac{\sigma_{\rm tf}}{\sqrt{3}}\left( \frac{2|\sigma_{\rm cf}|}{|
\sigma_{\rm cf}|+\sigma_{\rm tf}}\right),
\end{align}
and 
\begin{align}
\label{bc}
\beta^C=\left( \frac{|\sigma_{\rm cf}|-\sigma_{\rm tf}}{\sqrt{3}(|
\sigma_{\rm cf}|+\sigma_{\rm tf})}\right).
\end{align}
To derive formulas \eqref{qc} and \eqref {bc} above  note  that failure under uniaxial tension corresponds to $\tilde{\ss}=diag(\tilde{\sigma}_{\rm tf},0,0)$, and failure under compression corresponds to  $\tilde{\ss}=diag(\tilde{\sigma}_{\rm cf},0,0)$.  Substitution of these two stresses into  the equation for the failure surface deliver \eqref{qc} and \eqref{bc}.

\subsubsection{Residual strength, damage, loss of elasticity and fracture for plain stress}
In this section we consider homogeneous isotropic materials subject to plane stress. In this context $\ss_{13}=\ss_{23}=\ss_{33}=0$  and   $\Omega$ in $\mathbb{R}^2$ and the elastic tensor $\mathbb{C}$  is assumed  isotropic.  Here $\ss_{ij}$, $1\leq i\leq j \leq 2$,  are the independent nonzero components of the stress stress tensor  which we continue to denote as $\ss$ for plain stress loading.  One can think of the strain as a $2\times 2$ symmetric tensor with  components $\EE_{kl}^{\bbxi}(\uu(t,\xx))$, $1\leq i\leq j \leq 2$.  For this case the fourth order projections onto the deviatoric part  and the hydrostatic part of the stress  are given by
\begin{align}
\mathbb{K}^{2d}_{ijkl}=\frac{1}{2}(\delta_{ik} \delta_{jl} +\delta_{il}\delta_{kj})-\frac{1}{2}\delta_{ij}\delta_{kl},
\label{tracefree2}
\end{align}
\begin{align}
\mathbb{J}^{2d}_{ijkl}=\frac{1}{2}\delta_{ij}\delta_{kl}.
\label{identity2}
\end{align}
Bulk and shear moduli $\kappa$ and $\mu$ are prescribed and the isotropic elastic tensor is a linear map from the space of $2\times 2$ symmetric matrices into itself. Isotropic elastic tensors for plane stress are given by
\begin{align}
\mathbb{C}(\xx)=2\mu(\xx) \mathbb{K}^{2d} + 2\left( \frac{9\kappa(\xx)\mu(\xx)}{3\kappa(\xx)+4\mu(\xx)}\right)\mathbb{J}^{2d},
\label{elastictensor2}
\end{align}
and $\tilde{\ss}(t,\xx,\bbxi)=\mathbb{C}(\xx)\EE^{\bbxi}((\uu(t,\xx))$. The effective stress  is decomposed into deviatoric and hydrostatic parts $\tilde{\ss}=\tilde{\ss}_D+\tilde{\ss}_h$.  For shear and bulk moduli $(\mu(\xx),\kappa(\xx))$  the deviatoric effective  stress  is given by 
$\tilde{\ss}_D=2\mu(\xx) \EE^{\bbxi}_D(\uu(t,\xx))$ and the hydrostatic stress is $\tilde{\ss}_h=2(9\kappa(\xx)\mu(\xx)/ 3\kappa(\xx)+4\mu(\xx))\EE^{\bbxi}_h(\uu(t,\xx))$.  With $\EE^{\bbxi}_h(\uu(t,\xx))=\mathbb{J}^{2d}\EE^{\bbxi}((\uu(t,\xx,\bbxi))$ and $\EE^{\bbxi}_D(\uu(t,\xx))=\mathbb{K}^{2d}\EE^{\bbxi}(\uu(t,\xx))$.
For plane stress loading
\begin{align}
\label{psdp}
\mathcal{J}_2(\tilde{\ss}_D)&=\frac{1}{2}\tilde{\ss}_D:\tilde{\ss}_D=\frac{1}{3}\left(\tilde{\ss}^2_{11}+\tilde{\ss}^2_{22}-\tilde{\ss}_{11}\tilde{\ss}_{22}+3\tilde{\ss}_{12}^2\right)\nonumber\\ 
&=\frac{1}{3}\left((tr\{\tilde{\ss}\})^2-3\,det\{\tilde{\ss}\}\right)\nonumber\\
&=\frac{1}{3}\left((\tilde{\sigma}_1+\tilde{\sigma}_2)^2-3\tilde{\sigma}_1\tilde{\sigma}_2\right),
\end{align}
where ${\sigma}_1$ and ${\sigma}_2$ are the principal stresses of  the full effective stress tensor $\tilde{\ss}$ for plane stress.
Here $\tilde{\ss}_{ij}$, $1\leq i\leq j \leq 2$,  are the independent nonzero components of the effective stress tensor and $\tilde{\ss}$ is symmetric. The effective  stress is given in terms of the elasticity tensor and strain
\begin{align}
\label{stresscomponents2d}
\tilde{\ss}_{ij}=\mathbb{C}_{ijkl}(\xx)\EE_{kl}^{\bbxi}(\uu(t,\xx,\bbxi)).
\end{align}
Note that $\tilde{\ss}_D$ is the original $3\times 3$ symmetric tensor
used in evaluating the strength. (Recall  $\tilde{\ss}_D=\tilde{\ss}-(1/3)(\tilde{\ss}_{11}+\tilde{\ss}_{22})I^{3\times 3}$, where $I^{3\times3}$ is the identity, in $\mathbb{R}^{3}$ and $\ss$ also denotes the $3\times 3$ stress with $\tilde{\ss}_{13}=\tilde{\ss}_{23}=\tilde{\ss}_{33}=0$.)% see \cite{huhes}, pg. 92.

Set $\mathcal{F}^{2d}\tilde{(\ss)}=\sqrt{\mathcal{J}_2(\tilde{\ss})}$ and note from \eqref{psdp} it is homogeneous of degree one, i.e.,  for any positive scalar $s$ that 
$s\times\mathcal{F}^{2d}(\tilde{\ss})=\mathcal{F}^{2d}(s\times \tilde{\ss})$. As before we introduce the scaled strain $\bS(t,\xx,\bbxi,\uu):=\sqrt{|\bbxi|/L}\ \EE^{\bbxi}(\uu(t,\xx))$  and\eqref{q} and \eqref{qEmem} still holds but now for plane strain loading $r(t)$ is given by
\begin{align}
\label{q2}
r(t):=r(t,\xx,\bbxi,\uu)=\mathcal{F}^{2d}(\mathbb{C}(\xx)\bS(t,\xx,\bbxi,\uu)) +\beta^C I_1(\mathbb{C}(\xx)\bS(t,\xx,\bbxi,\uu)),
\end{align}
and the memory  is given  by
\begin{align}
\label{qEmem2}
r^\ast(t)=r^\ast(t,\xx,\bbxi,\uu)=\max_{0\leq \tau\leq t}r(\tau,\xx,\bbxi,\uu).
\end{align}

To illustrate the ideas we specify the following loading curve $\sigma_{DP}(r^\ast)$
\begin{equation}\label{bilinearloading}
\sigma_{DP}(r^\ast)=\left\{\begin{array}{lc} r^\ast ,& r^*< r^C,\\
r^C\frac{r^F-r^\ast}{r^F-r^C},&r^C\leq r^*\leq r^F,\\
0,&r^F<r^*.\end{array}\right.
\end{equation}

For $r(t)\leq r^\ast(t)$ the unloading curves are given by $\gamma(t) r(t)$ where $\gamma(t)=\sigma_{DP}(r^\ast(t))/r^\ast(t))$.
From the 1-homogeneity  of $q(t)$ we find that

\begin{align}
\label{softsigma2d}
\ss(t,\xx,\bbxi)=\ss_D(t,\xx,\bbxi)+\ss_h(t,\xx,\bbxi)=\gamma(t)\mu(\xx)\EE^{\bbxi}_D(\uu(t,\xx))+ \gamma(t)\left(\frac{9\mu(\xx)\kappa(\xx)}{3\kappa(\xx)+4\mu(\xx)}\right)\EE^{\bbxi}_h(\uu(t,\xx)).
\end{align}
So the damaged elastic tensor is
\begin{align}
\gamma(t,\xx,\bbxi,\uu)\mathbb{C}=\gamma(t,\xx,\bbxi,\uu)2\mu(\xx) \mathbb{K} ^{2d}+ 2\gamma(t,\xx,\bbxi,\uu)\left( \frac{9\kappa(\xx)\mu(\xx)}{3\kappa(\xx)+4\mu(\xx)}\right)\mathbb{J}^{2d}.
\label{elastictensor2damage}
\end{align}
As before one concludes that the  history dependent constitutive law  delivers the the force  density at $\xx$ and $t$ given by
 \begin{align}\label{eq: force42d}
  \LL^\epsilon [\uu](t,\xx) &= \int_\Omega \nabla^{-\bbxi}\cdot[\rho^{\ast,\epsilon}(\xx,\bbxi)\ss(t,\xx,\bbxi)]|\bbxi| \,d\bbxi\nonumber\\
 &= \int_{\Omega}\sum_{n=1}^2\rho^{\ast,\epsilon}(\xx,\bbxi) [-\ss(t,\xx,\bbxi)\ee^n]+\rho^{\ast,\epsilon}(\xx-\bbxi^n,\bbxi)\ss(t,\xx-\bbxi^n,\bbxi)\ee^n\,d\bbxi,
\end{align}
where
\begin{align}
\label{histconstlaw2}
\ss(t,\xx,\bbxi)&={\gamma(t,\xx,\bbxi,\uu) }\mathbb{C}(\xx)\EE^{\bbxi}(\uu(t,\xx)),
\end{align}
and the shifted stress is

\begin{align}\label{histconstlaw220}
\ss(t,\xx-\bbxi^n,\bbxi)&=\gamma(t,\xx-\bbxi^n,\bbxi,\uu) \mathbb{C}(\xx-\bbxi^n)\EE^{\bbxi}(\uu(t,\xx-\bbxi^n)).
\end{align}

The initial boundary value problem for  linear elastic brittle fracture evolution is to
find a displacement $\uu(t,\xx)$ on $\Omega^*$ that satisfies 
\begin{align}\label{eq: prescrib boundary displacement22d}
\uu(t,\xx)=\UU(t,\xx)& \hbox{ for }\xx \hbox{ in } \Omega^\epsilon_D,
\end{align}
and for $\xx$ in
$\Omega$  satisfies the evolution equation 
\begin{align}\label{eq: linearmomentumbal32d}
\rho\ddot{\uu}(t,\xx)+   \LL^\epsilon [\uu](t,\xx)=\bb(t,\xx),
\end{align}
for $0<t<T$, and satisfies the initial conditions on the displacement and velocity given by
\begin{align}
\uu(0,\xx)=\uu_0(\xx),\nonumber\\
\dot{\uu}(0,\xx)=\vv_0(\xx).\label{initialconditions22d}
\end{align}

\subsection{Autonomous interpolation between inelasticity and linear elastiicity. }
\label{Interpolation}

In this subsection we show to leading order how the internal force operator autonomously morphs into the classic local internal force associated with linear elasticity in quiescent regions away from inelastic regions. The inelastic region is characterized by a phase field that changes with time, i.e., $\partial_t\gamma(t,\xx,\bbxi,\uu)>0$. In  quiescent regions free from damage  $\partial_t\gamma(t,\xx,\bbxi,\uu)=0$, $\gamma(t,\xx,\bbxi,\uu)=1$  and the constitutive law is linear. It is reasonable to assume the  solution is represented uniformly by its second order Taylor polynomial at $\xx$ in the ball $\HH_\epsilon(\xx)$ and to fix ideas suppose $\mathbb{C}(\xx)$ is constant in $\HH_\epsilon(\xx)$. Then to leading order 
\begin{align}
      \label{eq: forcelinear elast}
 \LL^\epsilon [\uu](t,\xx) = \int_{\Omega}\nabla^{-\bbxi}\cdot[\rho^{\epsilon}(\xx,\bbxi)\ss(t,\xx,\bbxi)]|\bbxi|\,d\bbxi= -\nabla\cdot\overline{\mathbb{C}}(\xx)\EE(\uu(t,\xx)),
 \end{align}
and we recover the internal force inside a linear elastic material where ``$\nabla\cdot$'' is the usual divergence operator, $\EE(\uu(t,\xx))$ is the local strain and $\overline{\mathbb{C}}(\xx)$ is the elasticity of the material in the ball.

\subsection{Material damage failure and fracture }
\label{Damage and Failure}

We observe that nonlocal dynamics for $\uu(t)$ over $\Omega$ in $\mathbb{R}^d$ presents itself as dynamics for the frame gradient at $\xx$ generated by $\bbxi$ given by  $\EE^{\bbxi}(\uu(t,\xx,\bbxi))$ evolving over $$\{(\xx,\bbxi)\in\Omega\times\HH_\epsilon(0)\},$$ for times $0\leq t\leq T$.  
The frame gradient dynamics generates subsets of $\Omega\times\HH_\epsilon(0)$  where there is a two way coupling between the stress $\ss(t,\xx,\bbxi)$ and the phase field and this evolves the internal force operator on these sets.
The set of  undamaged material given by undamaged elasticity tensors at time $t$ is associated with the set
\begin{align}\label{undamage}
UZ^\epsilon(t) =\{(\xx,\bbxi)\in\,\Omega\cup\HH_\epsilon(0)\,; \, \gamma(t,\xx,\bbxi,\uu))=1\hbox{ and }\partial_t\gamma(t,\xx,\bbxi,\uu)=0\},
\end{align}
with indicator function
\begin{equation}\label{undamagechar}
\chi_{UZ^\epsilon(t)}(\xx,\bbxi)=\left\{\begin{array}{ll}1,&\hbox{if  $(\xx,\bbxi)$ in $UZ^\epsilon(t)$}\\
0,& \hbox{otherwise}.\end{array}\right.
\end{equation}
The set of pairs $(\xx,\bbxi)$ corresponding to damaged material associated with elasticity tensors that have softened but are currently are not softening at time $t$ are given by the set
\begin{align}\label{damaged}
DZ^\epsilon(t) &=\{(\xx,\bbxi)\in\,\Omega\cup\HH_{\epsilon} \,;  \, 1>\gamma(t,\xx,\bbxi,\uu)\geq 0\hbox{ and }\partial_t\gamma(t,\xx,\bbxi,\uu)=0,\}
\end{align}
with indicator function
\begin{equation}\label{damagedchar}
\chi_{DZ^\epsilon(t)}(\xx,\bbxi)=\left\{\begin{array}{ll}1,&\hbox{if  $(\xx,\bbxi)$ in $DZ^\epsilon(t)$}\\
0,& \hbox{otherwise}.\end{array}\right.
\end{equation}
The set of elasticity tensors undergoing damage at time $t$ are given by the process zone given by the set of frames
\begin{align}\label{damaging}
PZ^\epsilon(t) &=\{(\xx,\bbxi)\in\,\Omega\cup\HH_{\epsilon}\,; \, \partial_t\gamma(t,\xx,\bbxi,\uu)<0.\}
\end{align}
with indicator function
\begin{equation}\label{processchar}
\chi_{PZ^\epsilon(t)}(\xx,\bbxi)=\left\{\begin{array}{ll}1,&\hbox{if  $(\xx,\bbxi)$ in $PZ^\epsilon(t)$}\\
0,& \hbox{otherwise}.\end{array}\right.
\end{equation}
The failure set $\Gamma^\epsilon(t)$ is the collection of pairs $(\xx,\bbxi)$ in $\Omega\times\HH_{\epsilon}$ such that the elastic tensor has gone completely soft at some time $\tau$, $0<\tau\leq t$. 
The failure set is written 
\begin{align}\label{damageset}
\Gamma^\epsilon(t)=\{\xx,\bbxi)\hbox{ in }\Omega\times\HH_{\epsilon}\, ; \, \gamma(t,\xx,\bbxi,\uu)=0\, \hbox{and}\, \, \partial_t\gamma(t,\xx,\bbxi,\uu)=0\}
\end{align}
with indicator function
\begin{equation}\label{busted}
\chi_{\Gamma^\epsilon(t)}(\xx,\bbxi)=\left\{\begin{array}{ll}1,&\hbox{if  $(\xx,\bbxi)$ in $\Gamma^\epsilon(t)$}\\
0,& \hbox{otherwise}.\end{array}\right.
\end{equation}
The energy dissipation necessary to create a flat  crack of area $R^\epsilon_t$ for $d$ dimensional stress states follows directly from the dynamic evolution equations \eqref{eq: linearmomentumbal3} and is calculated in Section \ref{sec.Straight Cracks}. Here a crack defined as the internal boundary where traction force on one side of the crack does not influence the motion of material points on the other side.

%%%%%%%%%%%%%%%%%%%%%%%%%%%%%%%%%%%%%%%%%%%%%%%%%%%%%%%%%%%%%%%%%%%%%%%%%%%%%%%%%%%%%%%%%

%%%%%%%%%%%%%%%%%%%%%%%%%%%%%%%%%%%%%%%%%

%%%%%%%%%%%%%%%%%%%%%%%%%%%%%%%%%%%%%%%%%%
%%%%%%%%%%%%%%%%%%%%%%%%%%%%%%%%%%%%%%%%%

%\section{Finite  deformation evolution and correspondence modeling with frame gradients.}
%\label{Large deformation}

\section{Existence and uniqueness of quasi-brittle fracture evolution.}
\label{wellposedness}
In this section we state and prove the existence and uniqueness of solution for the initial boundary value problem. %Body forces are easily chosen for which one can find a unique solution $\uu(t,\xx)$ 
To expedite the presentation we denote the function space  $L^\infty(\Omega;\mathbb{R}^d)$  by $X$. The norm on $C^2([0,T];X)$ is given by,
\begin{align}\label{norms}
\Vert \uu(t,\xx)\Vert_{C^2([0,T]; X)}=\sup_{0\leq t\leq T}\{ \sum_{i=0}^2||\partial_t^i\uu(t,\xx)||_{X}\}.
\end{align}
To simplify the notation, we denote the space $C([0,T];X)$ as $CX$ with the corresponding norm $\Vert\cdot \Vert_{CX}$. Similarly, we write $C^2([0,T];X)$ as $C^2X$. The existence and uniqueness of the solution is stated below.
%---------------------------------------------------------

\begin{theorem}[{\bf Existence and Uniqueness of Solution of Damage and  Fracture Evolution}]
\label{exist}
 The initial value problem given by \eqref{eq: prescrib boundary displacement2} through \eqref{initialconditions2} with initial data in $X$ and $\bb(t,\xx)$ belonging to $CX$
has a unique solution $\uu(t,\xx)$ in $C^2X$ with two strong derivatives in time.
 \end{theorem}

%---------------------------------------------------------
The initial value problem delivers a unique displacement-damage set pair: ${(\uu^\epsilon(t), DZ^\epsilon(t)\cup PZ^\epsilon(t))}$.  
The key points are that the operator $\LL^\epsilon[\uu]$ is Lipschitz continuous in $\uu$ with respect to the $CX$ norm and is a map from $CX$ into itself,. The theorem then follows from the Banach fixed-point theorem. To establish these properties, we first show Lipschitz continuity of  the operator $\mathcal{L}^\epsilon[\uu]$ on the function space $X$.

%--------------------------------------------------------

\begin{lemma}[{\bf Lipschitz continuity of \( \mathcal{L}^\epsilon [\cdot]\)}]
\label{linftcnttheta} 
Given two functions $\uu$ and $\ww$  in $CX$ there exists a constant  $C$ independent of $\uu$ and $\ww$ such that
\begin{align}\label{lipshitzg}
\|\mathcal{L}^\epsilon[\mathbf{u}] - \mathcal{L}^\epsilon[\mathbf{w}]\|_{CX} \leq C \|\mathbf{u} - \mathbf{w}\|_{CX},
\end{align}
 \end{lemma}
 
 \begin{proof}
Form
 \begin{align}\label{eq: first}
  \LL^\epsilon [\uu](t,\xx) -  \LL^\epsilon [\ww](t,\xx) &=  \int_{\Omega}\sum_{n=1}^d\rho^{\ast,\epsilon}(\xx,\bbxi) [-{\gamma(t,\xx,\bbxi,\uu) }\mathbb{C}(\xx)\EE^{\bbxi}(\uu(t,\xx))\ee^n]\, d\bbxi\nonumber\\
  &+\int_{\Omega}\sum_{n=1}^d \rho^{\ast,\epsilon}(\xx-\bbxi^n,\bbxi)\gamma(t,\xx-\bbxi^n,\bbxi,\uu) \mathbb{C}(\xx-\bbxi^n)\EE^{\bbxi}(\uu(t,\xx-\bbxi^n))\ee^n\,d\bbxi,\nonumber\\
  &-  \int_{\Omega}\sum_{n=1}^d\rho^{\ast,\epsilon}(\xx,\bbxi) [-{\gamma(t,\xx,\bbxi,\ww) }\mathbb{C}(\xx)\EE^{\bbxi}(\ww(t,\xx))\ee^n]\, d\bbxi\nonumber\\
  &-\int_{\Omega}\sum_{n=1}^d \rho^{\ast,\epsilon}(\xx-\bbxi^n,\bbxi)\gamma(t,\xx-\bbxi^n,\bbxi,\ww) \mathbb{C}(\xx-\bbxi^n)\EE^{\bbxi}(\ww(t,\xx-\bbxi^n))\ee^n\,d\bbxi,
\end{align}
to see that the integrand is of the form
\begin{align}\label{eq: second}
  &\sum_{n=1}^d \rho^{\ast,\epsilon}(\xx,\bbxi) \left([-{\gamma(t,\xx,\bbxi,\uu) }\mathbb{C}(\xx)\EE^{\bbxi}(\uu(t,\xx))\ee^n] - [-{\gamma(t,\xx,\bbxi,\ww) }\mathbb{C}(\xx)\EE^{\bbxi}(\ww(t,\xx))\ee^n]\right)\nonumber\\
  &+\sum_{n=1}^d \rho^{\ast,\epsilon}(\xx-\bbxi^n,\bbxi)\gamma(t,\xx-\bbxi^n,\bbxi,\uu) \mathbb{C}(\xx-\bbxi^n)\EE^{\bbxi}(\uu(t,\xx-\bbxi^n))\ee^n\nonumber\\
  &-\sum_{n=1}^d \rho^{\ast,\epsilon}(\xx-\bbxi^n,\bbxi)\gamma(t,\xx-\bbxi^n,\bbxi,\ww) \mathbb{C}(\xx-\bbxi^n)\EE^{\bbxi}(\ww(t,\xx-\bbxi^n))\ee^n.
\end{align}
To fix ideas we choose a typical difference denoted by $I$  from the summation above to show  that there is a constant $C$  independent of $\uu$, $\ww$, $t$, $\xx$ and $\bbxi^n$, $n=1,2\ldots$ for which
\begin{align}\label{eq: third}
I=& |\gamma(t,\xx-\bbxi^n,\bbxi,\uu) \mathbb{C}(\xx-\bbxi^n)\EE^{\bbxi}(\uu(t,\xx-\bbxi^n))\ee^n-\gamma(t,\xx-\bbxi^n,\bbxi,\ww) \mathbb{C}(\xx-\bbxi^n)\EE^{\bbxi}(\ww(t,\xx-\bbxi^n))\ee^n|\nonumber\\
&\leq C\frac{\Vert\uu(t,\xx)-\ww(t,\xx)\Vert_{CX}}{|\bbxi|},
\end{align}
holds almost everywhere in $\Omega$ for $0\leq t\leq T$. In what follows we will write $C$ to represent any constant independent of  $\uu$, $\ww$, $t$, $\xx$ and $\bbxi^n$, $n=1,2\ldots$. It is first noted that for any symmetric matrix $\ss$  and unit vector $\ee$ that $|\ss\ee|\leq|\ss|_H$ where $|\cdot|_H$ denotes the Hilbert Schmidt norm, $|\cdot|$ is the Euclidean norm in $\mathbb{R}^k$, $k=1,2,3$. Thus we have the estimate
\begin{align*}
%\label{eq: fourth}
 I\leq & |\mathbb{C}(\xx-\bbxi^n)\left[\gamma(t,\xx-\bbxi^n,\bbxi,\uu) \EE^{\bbxi}(\uu(t,\xx-\bbxi^n))-\gamma(t,\xx-\bbxi^n,\bbxi,\ww)\EE^{\bbxi}(\ww(t,\xx-\bbxi^n))\right]|_H.
 \end{align*}
Algebraic manipulation and the triangle inequality applied to the right hand side gives $I\leq I_1+I_2$ where
\begin{align*}
 I_1=& |(\gamma(t,\xx-\bbxi^n,\bbxi,\uu) - \gamma(t,\xx-\bbxi^n,\bbxi,\ww))||\mathbb{C}(\xx-\bbxi^n)\EE^{\bbxi}(\uu(t,\xx-\bbxi^n))|_H\\
 I_2=& \gamma(t,\xx-\bbxi^n,\bbxi,\ww)|\mathbb{C}(\xx-\bbxi^n)(\EE^{\bbxi}(\uu(t,\xx-\bbxi^n))-\EE^{\bbxi}(\ww(t,\xx-\bbxi^n)))|_H.
 \end{align*}
 We estimate the first factor of $I_1$ noting $t$ and $\xx-\bbxi^n$  are fixed and we abbreviate and write $\gamma_{\uu}=\sigma_{DP}(r_{\uu}^\ast)/r_{\uu}^\ast$  and $\gamma_{\ww}=\sigma_{DP}(r_{\ww}^\ast)/r_{\ww}^\ast$. We assume first that $r^\ast_{\uu}\geq r^C$ to find

 \begin{align}
 \label{lip1}
  |(\gamma_{\uu} - \gamma_{\ww})|&\leq|\frac{\sigma_{DP}(r_{\uu}^\ast)}{r_{\uu}^\ast}-\frac{\sigma_{DP}(r_{\ww}^\ast)}{r_{\uu}^\ast}|+|\frac{\sigma_{DP}(r_{\ww}^\ast)}{r_{\uu}^\ast}-\frac{\sigma_{DP}(r_{\ww}^\ast)}{r_{\ww}^\ast}|\\
 \leq& |r_{\uu}^\ast|^{-1}|\sigma_{DP}(r_{\uu}^\ast)-\sigma_{DP}(r_{\ww}^\ast)|+|\frac{\sigma_{DP}(r_{\ww}^\ast)}{r_{\ww}^\ast}||\frac{r_{\ww}^\ast-r_{\uu}^\ast}{r_{\uu}^\ast}|\nonumber\\
  \leq& |r_{\uu}^\ast|^{-1}C|r_{\uu}^\ast-r^\ast_{\ww}|+|\frac{\sigma_{DP}(r_{\ww}^\ast)}{r_{\ww}^\ast}||\frac{r_{\ww}^\ast-r_{\uu}^\ast}{r_{\uu}^\ast}|\nonumber\\
   \leq& |r^C|^{-1}C|r_{\uu}^\ast-r^\ast_{\ww}|+|\frac{\sigma_{DP}(r_{\ww}^\ast)}{r_{\ww}^\ast}||\frac{r_{\ww}^\ast-r_{\uu}^\ast}{r^C}|.\nonumber
 \end{align}
The third inequality follows from the fact that the function $\sigma_{DP}(r^\ast)$ is uniformly Lipschitz on $[0,\infty)$.
For $r^*_{\uu} \geq r^C$ have two cases $r^*_{\ww}<r^C$ and $r^*_{\ww}\geq r^C$. For the first case $\sigma_{DP}(r^\ast_{\ww})=r^\ast_{\ww}$ so $|(\gamma_{\uu} - \gamma_{\ww})|\leq C |r^\ast_\uu-r^\ast_\ww|$. For the second case we also see that $(\gamma_{\uu} - \gamma_{\ww})|\leq C |r^\ast_\uu-r^\ast_\ww|$. Now if $r_{\uu}^*<r^C$ swap $r^*_{\uu}$ with $r^*_{\ww}$ in 
 \eqref{lip1} and argue as before. Last if both $r^*_{\uu}$ and $r^*_{\ww}$ are less than $r^C$ then $|\gamma_{\uu}-\gamma_{\ww}|=0$ and we conclude that for all cases the phase field  $\gamma$ is
Lipschitz continuous in $r^*_{\uu}$, i.e.
 \begin{align}
 \label{lip2}
  |(\gamma_{\uu} - \gamma_{\ww})|\leq C|r^*_{\uu}-r^*_{\ww}|.
 \end{align}
 
 Next we  see from \eqref{rescale} that $r^\ast(t)=r^\ast(t,\xx,\bbxi,\uu)=\max_{0\leq \tau\leq t}q(\mathbb{C}(\xx-\bbxi^n)\bS(t,\xx-\bbxi^n,\bbxi,\uu))$ so
  \begin{align}
 \label{lip3}
  |r^*_{\uu}-r^*_{\ww}|&=\frac{\sqrt{|\bbxi|}}{L}|\max_{0\leq\tau\leq t}q(\mathbb{C}(\xx-\bbxi^n)\EE^{\bbxi}\uu(t,\xx-\bbxi^n))-\max_{0\leq\tau\leq t}q(\mathbb{C}(\xx-\bbxi^n)\EE^{\bbxi}(t,\xx-\bbxi^n))|\nonumber\\
  &\leq \frac{\sqrt{|\bbxi|}}{L}\max_{0\leq\tau\leq t}|q(\mathbb{C}(\xx-\bbxi^n)\EE^{\bbxi}\uu(t,\xx-\bbxi^n))-q(\mathbb{C}(\xx-\bbxi^n)\EE^{\bbxi}(\ww(t,\xx-\bbxi^n))|\nonumber\\
  &\leq \frac{\sqrt{|\bbxi|}}{L}\max_{0\leq\tau\leq t}C|\mathbb{C}(\xx-\bbxi^n)\EE^{\bbxi}\uu(t,\xx-\bbxi^n))-\mathbb{C}(\xx-\bbxi^n)\EE^{\bbxi}(\ww(t,\xx-\bbxi^n))|_H\nonumber\\
  &\leq\frac{\sqrt{|\bbxi|}}{L}\max_{0\leq\tau\leq t}C|\EE^{\bbxi}\uu(t,\xx-\bbxi^n))-\EE^{\bbxi}\ww(t,\xx-\bbxi^n))|_H\nonumber\\
  &\leq\frac{\sqrt{|\bbxi|}}{L}\max_{0\leq\tau\leq t}C|\EE^{\bbxi}\uu(t,\xx-\bbxi^n))-\EE^{\bbxi}\ww(t,\xx-\bbxi^n))|_H\nonumber\\
  &\leq\frac{\sqrt{|\bbxi|}}{L}C\frac{\Vert\uu(t,\xx)-\ww(t,\xx)\Vert_{CX}}{|\bbxi|}=C\frac{\Vert\uu(t,\xx)-\ww(t,\xx)\Vert_{CX}}{\sqrt{|\bbxi|}}.
 \end{align}
 where the third inequality follows from the Lipschitz continuity of $q$ and the fourth inequality follows from \eqref{coersivebounded} and in the last
 line $L$ is included in the constant. To complete the upper estimate on $I_1$ we note $0\leq r_{\uu}^*\leq r^F$,  and from \eqref{bounding ball} we see that 
 \begin{align*}
 |\mathbb{C}(\xx-\bbxi^n)\EE^{\bbxi}(\uu(t,\xx-\bbxi^n))|_H\leq R^F\frac{L}{\sqrt{|\bbxi|}},
 \end{align*}
 so $I_1\leq C|\bbxi|^{-1}\Vert\uu(t,\xx)-\ww(t,\xx)\Vert_{CX}$. The second term $I_2$ is estimated using the bound $0\leq \gamma_{\ww}\leq 1$ and estimating the strain as in $I_1$ gives
 $I_2 \leq C|\bbxi|^{-1}\Vert\uu(t,\xx)-\ww(t,\xx)\Vert_{CX}$ and the estimate \eqref{eq: third} follows.  Thus the integrand \eqref{eq: second}  admits a bound of the form \eqref{eq: third} multiplied by $\rho^{*,\epsilon}(\xx,\bbxi)$ and we  have for $\xx$ in $\Omega$ and $0\leq t\leq T$,
 \begin{align*}
|\mathcal{L}^\epsilon[\mathbf{u}](t, \mathbf{x}) - \mathcal{L}^\epsilon[\mathbf{w}](t, \mathbf{x})| \leq C \|\mathbf{u} - \mathbf{w}\|_{CX}
 \int_\Omega\,\rho^{*,\epsilon}(\xx,\bbxi)\frac{1}{|\bbxi|}\,d\bbxi.
 \end{align*}
 The integral is finite and this proves the lemma.
 \end{proof}
 
From the inequality \eqref{lipshitzg} and setting $\ww=0$ we get
\begin{align}\label{lipshitz2}
\Vert \mathcal{L}^\epsilon[\mathbf{u}]\Vert_{CX} \leq C \|\mathbf{u}\|_{CX}.
\end{align}
So  the operator $\LL^\epsilon[\uu]$ is a map from $CX$ into itself.\\

To conclude the proof of Theorem \ref{exist}  one uses Lemma \ref{linftcnttheta}   to  show the solution $\uu$ is the unique  fixed point of $\uu(t)=(I \uu)(t)$ where $I$ maps $CX$ into itself and is defined by
\begin{equation}\label{Fixedpointmap}
\begin{aligned}
(I\uu)(t)=&\uu_0+t\vv_0+\rho^{-1}\int_0^t(t-\tau)\LL^\epsilon[\uu](\tau)+\bb(\tau)\,d\tau.
\end{aligned}
\end{equation}
This problem is equivalent to finding the unique solution of the initial value problem given by
\eqref{eq: prescrib boundary displacement2} through \eqref{initialconditions2}  We absorb the factor $\rho^{-1}$ into $\mathcal{L}^\epsilon[\uu]+\bb$ and show that $I$ is a contraction map on $CX$. We proceed similarly to  \cite{coskun2025quasi}, \cite{bromleylipton}, to  see that $I$ is a contraction map with respect to an equivalent norm. The equivalent norm is
\begin{equation}\label{Equivnorm}
\begin{aligned}
|||\uu|||_{CX}=\max_{t\in[0,T_0]}\{e^{-2LTt}\Vert \uu(t)\Vert_{X}\},
\end{aligned}
\end{equation}
For $t\in [0,T]$ one has
\begin{equation}\label{Contrction}
\begin{aligned}
\Vert (I\uu)(t)-(I\ww)(t)\Vert_{X}\leq&\int_0^t(t-\tau)\Vert \LL[\uu](\tau)-\LL[\ww](\tau)\Vert_{X}\,d\tau\\
&\leq LT\int_0^t\Vert \uu-\ww\Vert_{C([0,\tau];X)}\,d\tau\\
&\leq LT\int_0^t\max_{s\in[0,\tau]}\{\Vert \uu(s)-\ww(s)\Vert_{X}e^{-2LTs}\}e^{2LT\tau}\,d\tau\\
&\leq\frac{e^{2LTt}-1}{2}|||\uu-\ww|||_{CX},
\end{aligned}
\end{equation}
hence
\begin{equation}\label{ContrctionD}
\begin{aligned}
|||(I\uu)(t)-(I\ww)(t)|||_{CX}\leq\frac{1}{2}|||\uu-\ww|||_{CX},
\end{aligned}
\end{equation}
and  $I$ is a contraction.  From the Banach fixed point  theorem there is a unique fixed point $\uu(t)$ belonging to $CX$ and it is evident from \eqref{Fixedpointmap} that $\uu(t)$ also belongs to $C^2X$.

\section{Power balance for quasibrittle fracture evolutions with damage}
\label{powerbalanceprocesszone2}

In this section, we establish the power balance for damage and fracture evolutions and then discuss the inelastic energy. Multiplying the linear momentum equation given in \eqref{eq: linearmomentumbal3} with $\dot{\uu}(t,\xx)$ and integrating over the domain $\Omega$ results in 
 \begin{equation}
	\int_{\Omega} \rho \ddot{\uu}(\xx, t) \cdot \dot{\uu}(\xx, t) \,d\xx + 
	\int_{\Omega} \LL^\epsilon [\uu](t,\xx) \cdot \dot{\uu}(\xx, t) \,d\xx = \int_{\Omega}\,\bb(t,\xx)\cdot\dot{\uu}(t,\xx)\,d\xx.
 \label{eq: udotlinearmomentumbalInt2}
\end{equation}
The first term is readily identified as
\begin{align}\label{eq:prebalancekenetic}
\int_{\Omega} \rho \ddot{\uu}(\xx, t) \cdot \dot{\uu}(\xx, t) \,d\xx=\partial_t\left(\int_\Omega\, \rho\,\frac{|\dot{\uu}(t)|^2}{2}\,d\xx \right)=\dot{\mathcal{K}}(t),
\end{align}
where $\mathcal{K}(t)$ is the Kinetic energy of the motion. 
The last term is identified as the external power supplied by the body force
\begin{align}
\label{forcepower}
\mathcal{P}_F(t)= \int_{\Omega}\,\bb(t,\xx)\cdot\dot{\uu}(t,\xx)\,d\xx.
\end{align}

To identify the power associated with the second term we expand it and interchange summation and integration to get
 \begin{align}\label{eq: force24}
 \int_{\Omega} \LL^\epsilon [\uu](t,\xx)\cdot\dot{\uu}(t,\xx)\,d\xx=&\sum_{n=1}^d-\int_{\Omega}\int_{\Omega}\left[ \rho^{\ast, \epsilon}(\xx,\bbxi)\, \ss(t,\xx,\bbxi)\ee^n\cdot\dot{\uu}(t,\xx)\right]\,d\bbxi\,d\xx\nonumber\\
&+\sum_{n=1}^d\int_{\Omega}\int_{\Omega}\left[\rho^{\ast, \epsilon}(\xx-\bbxi^n,\bbxi)\,\ss(t,\xx-\bbxi^n,\bbxi)\ee^n\cdot\dot{\uu}(t,\xx)\right] d\bbxi\,d\xx,
\end{align}
Next we examine the second double integral of \eqref{eq: force24} and  and applying Fubini's theorem we interchange order of integration to get
\begin{align}
\label{fubiniandmore}
&\int_{\Omega}\int_{\Omega}\left[\rho^{\ast, \epsilon}(\xx-\bbxi^n,\bbxi)\,\ss(t,\xx-\bbxi^n,\bbxi)\ee^n\cdot\dot{\uu}(t,\xx)\right] d\bbxi\,d\xx\nonumber\\
&= \int_{\HH_{\epsilon}(0)}\int_{\Omega}\left[\rho^{\ast, \epsilon}(\xx-\bbxi^n,\bbxi)\,\ss(t,\xx-\bbxi^n,\bbxi)\ee^n\cdot\dot{\uu}(t,\xx)\right]\,d\xx\, d\bbxi =A_n.
\end{align}
Now change variables in $A_n$ setting $\yy=\xx-\bbxi^n$, noting $d\yy=d\xx$ and $\xx=\yy+\bbxi^n$. For a frame to be admissible the shifted frame $\yy=\xx-\bbxi^n$ must lie in $\Omega$ so
\begin{align}
\label{shifted}
A_n&= \int_{\HH_{\epsilon}(0)}\int_{\Omega}\left[\rho^{\ast, \epsilon}(\yy,\bbxi)\,\ss(t,\yy,\bbxi)\ee^n\cdot\dot{\uu}(t,\yy+\bbxi^n)\right]\,d\yy\, d\bbxi\nonumber\\
&=\int_{\Omega}\int_{\Omega}\left[\rho^{\ast, \epsilon}(\xx,\bbxi)\,\ss(t,\xx,\bbxi)\ee^n\cdot\dot{\uu}(t,\xx+\bbxi^n)\right]\,d\bbxi\,d\xx,
\end{align}
where in the second line we applied Fubini's theorem and relabeled the outer variable of integration $\xx$. A subset of frame legs $x+\bbxi^n$ lie in $\Omega_D$ consequently $\uu(t,\xx+\bbxi^n)=\UU(t,\xx+\bbxi^n)$ there. The set of frames with at least one frame leg lying in $\Omega_D$ is denoted by $B$ and defined unambiguously as
$$B=\{\hbox{frames parametrized by }(\xx,\bbxi)\hbox{ in }\Omega\times\HH_\epsilon(0):\hbox{ having at least one leg }\xx+\bbxi^n\hbox{ in }\Omega_D\}.$$ 
Gathering results, denoting the characteristic function of the set $B$ by $\chi_B$, and applying \eqref{scale33} we  have 
\begin{align}
\label{parce}
A_n&= \int_{\Omega}\int_{\Omega}\left[\rho^ \epsilon(\xx,\bbxi)\,\ss(t,\xx,\bbxi)\ee^n\cdot\dot{\uu}(t,\xx+\bbxi^n)\right]\,d\xx\, d\bbxi-{\mathcal{P}}_D^n(t)
\end{align}
where ${\mathcal{P}}_D^n(t)$ is due to the externally applied power flowing into the body due to the prescribed displacement $\UU(t,\xx)$ on $\Omega_D$ given by
\begin{align}
\label{displacementpower}
\mathcal{P}_D^n(t)=\int\int\,\chi_B(\xx,\bbxi)\left[\rho^{\ast, \epsilon}(\xx,\bbxi)\,\ss(t,\xx,\bbxi)(-\ee^n)\cdot\dot{\UU}(t,\xx+\bbxi^n)\right]\,d\bbxi\,d\xx.
\end{align}
To conclude we apply  \eqref{parce} to \eqref{fubiniandmore} and substitute into \eqref{eq: force24} then swap integration and summation to rewrite \eqref{eq: force24} in terms of the adjusted non local kernel and total externally applied power ${\mathcal{P}}_D=\sum_{n=1}^d{\mathcal{P}}_D^n$ as 
\begin{align}\label{eq: force24adj}
\int_{\Omega} \LL^\epsilon [\uu](t,\xx)\cdot\dot{\uu}(t,\xx)\,d\xx=\int_{\Omega}\int_{\Omega}\,\rho^{\epsilon}(\xx,\bbxi)|\bbxi|\,\ss_{ij}(t,\xx,\bbxi)\EE_{ij}^{\bbxi}(\dot{\uu}(t,\xx))\,d\bbxi\,d\xx-{\mathcal{P}}_D
\end{align}

Equation \eqref{eq: udotlinearmomentumbalInt2}
is now written
\begin{align}\label{prepowerbalamce}
\dot{\mathcal{K}}(t)+ \int_{\Omega}\int_{\Omega}\,\rho^{\epsilon}(\xx,\bbxi)|\bbxi|\,\ss_{ij}(t,\xx,\bbxi)\EE_{ij}^{\bbxi}(\dot{\uu}(t,\xx))\,d\bbxi\,d\xx={\mathcal{P}}_D(t)+{\mathcal{P}}_F(t).
\end{align}
It is clear that the same ideas used to obtain \eqref{prepowerbalamce} can be used to calculate the first variation of the elastic energy given in Section \ref{Linear Elastic Materials}.

From \eqref{prepowerbalamce} we discover that:
\begin{proposition}
\label{conjugate}
The total internal power is
\begin{align}
\label{internal power}
\mathcal{P}_{int} = \int_{\Omega}\int_{\Omega}\,\rho^{\epsilon}(\xx,\bbxi)|\bbxi|\,\ss_{ij}(t,\xx,\bbxi)\EE_{ij}^{\bbxi}(\dot{\uu}(t,\xx))\,d\bbxi\,d\xx,
\end{align}
so  the nonlocal stress $\ss(t,\xx,\bbxi)$ and nonlocal strain rate $\EE^{\bbxi}(\dot{\uu}(t,\xx))$ are work conjugate fields.
\end{proposition}

Now we split the internal power into two components, one given by the elastic power component  and the other giving the power required for the inelastic process of elastic softening.
Observe that for all frames $(\xx,\bbxi)$  in $UZ^\epsilon(t)$  that  $\gamma(t,\xx,\bbxi,\uu))=1$ and  $\partial_t\gamma(t,\xx,\bbxi,\uu)=0$  so  the stress power is the elastic power  expressed by
\begin{align}
\label{identiy1}
\rho^{\epsilon}(\xx,\bbxi)|\bbxi|\,\mathbb{C}(\xx)(\xx)\EE^{\bbxi}(\uu(t,\xx)):\EE^{\bbxi}(\dot{\uu}(t,\xx))=\rho^{\epsilon}(\xx,\bbxi)|\bbxi|\,\partial_t\left(\frac{1}{2}\mathbb{C}(\xx)\EE^{\bbxi}(\uu(t,\xx)):\EE^{\bbxi}(\uu(t,\xx))\right).
\end{align}
where the first line is the stress power and second line  elastic  power.
Observe for all frames $(\xx,\bbxi)$ in $DZ^\epsilon(t)$ we  have $1>\gamma(t,\xx,\bbxi,\uu)\geq 0$ and $\partial_t\gamma(t,\xx,\bbxi,\uu)=0$ so the stress power is the elastic power given by
\begin{align}
\label{identiy2}
&\rho^{\epsilon}(\xx,\bbxi)|\bbxi|\,\gamma(t,\xx,\bbxi,\uu)\mathbb{C}(\xx)\EE^{\bbxi}(\uu(t,\xx)):\EE^{\bbxi}(\dot{\uu}(t,\xx))=\nonumber\\
&=\rho^{\epsilon}(\xx,\bbxi)|\bbxi|\,\partial_t\left(\frac{1}{2}\gamma(t,\xx,\bbxi,\uu)\mathbb{C}(\xx)\EE^{\bbxi}(\uu(t,\xx)):\EE^{\bbxi}(\uu(t,\xx))\right).
\end{align}

Next observe for all frames  $(\xx,\bbxi)$ in $PZ^\epsilon(t)$, the stress power  is a combination of elastic power and damage power. Here $ \partial_t\gamma(t,\xx,\bbxi,\uu)<0$ and
\begin{align}
\label{identiy3}
&\rho^{\epsilon}(\xx,\bbxi)|\bbxi|\gamma(t,\xx,\bbxi,\uu)\,\mathbb{C}(\xx)\EE^{\bbxi}(\uu(t,\xx)):\EE^{\bbxi}(\dot{\uu}(t,\xx))=\nonumber\\
&=\rho^{\epsilon}(\xx,\bbxi)|\bbxi|\,\partial_t\left(\frac{1}{2}\gamma(t,\xx,\bbxi,\uu)\mathbb{C}(\xx)\EE^{\bbxi}(\uu(t,\xx)):\EE^{\bbxi}(\uu(t,\xx))\right)\nonumber\\
&-\rho^{\epsilon}(\xx,\bbxi)|\bbxi|\,\partial_t\gamma(t,\xx,\bbxi,\uu)\frac{1}{2}\left(\mathbb{C}(\xx)\EE^{\bbxi}(\uu(t,\xx)):\EE^{\bbxi}(\uu(t,\xx))\right).
\end{align}
On adding all contributions the total stress power $\dot{\mathcal{S}}^\epsilon(t)$ in 
$$\{(\xx,\bbxi)\hbox{ in }\Omega\times\HH_{\epsilon}\}=UZ^\epsilon(t)\cup DZ^\epsilon(t)\cup PZ^\epsilon(t). $$
 is a combination of elastic and damage power given by
\begin{align}\label{totaldamagepower}
\dot{\mathcal{S}}^\epsilon(t)=\dot{\mathcal{E}}^\epsilon(t)+\dot{\mathcal{D}}^\epsilon(t),
\end{align}
where the elastic power is
\begin{align}\label{elasticpzpower2}
\dot{\mathcal{E}}^\epsilon(t)=\int_{\Omega}\int_{\HH_\epsilon(0)}\,\rho^{\epsilon}(\xx,\bbxi)|\bbxi|\,\partial_t\left(\frac{1}{2}\gamma(t,\xx,\bbxi,\uu)\mathbb{C}(\xx)\EE^{\bbxi}(\uu(t,\xx)):\EE^{\bbxi}(\uu(t,\xx))\right)\,d\bbxi\,d\xx.
\end{align}
and the damage power (energy dissipation rate) is confined to the process zone $PZ^\epsilon(t)$ and given by
\begin{align}\label{damagepower}
\dot{\mathcal{D}}^\epsilon(t)=-\int_{\Omega}\int_{\HH_\epsilon(0)}\chi_{PZ^\epsilon(t)}(\xx,\bbxi)\,\rho^{\epsilon}(\xx,\bbxi)|\bbxi|\,\partial_t\gamma(t,\xx,\bbxi,\uu)\frac{1}{2}\left(\mathbb{C}(\xx)\EE^{\bbxi}(\uu(t,\xx)):\EE^{\bbxi}(\uu(t,\xx))\right)\,d\bbxi\,d\xx,
\end{align}
where $\dot{\mathcal{D}}^\epsilon>0$ since $\partial_t\gamma(t,\xx,\bbxi,\uu)<0$ for frames in $PZ^\epsilon(t)$.
On collecting results, we have the power balance

\bigskip

\noindent{\bf Power Balance for damage and fracture evolution}
\begin{align}\label{eq:rateenergybalance}
\dot{\mathcal{K}}(t) + \dot{\mathcal{E}}^\epsilon(t) + \dot{\mathcal{D}}^\epsilon(t)={\mathcal{P}}_D(t)+{\mathcal{P}}_F(t).
\end{align}

It is clear that the damage energy only changes when  $\partial_t\gamma(t,\xx,\bbxi,\uu)<0$ and is determined by the evolving displacement field, through the change in elastic energy, kinetic energy and work done against the load. Rearranging terms these observations are summarized in the following Lemma.
\begin{lemma}[{\bf Growth of the Damage Energy and Process Zone}]
\label{damagezonegrowth}
We have the power balance:
\begin{align}\label{eq:processzonepowerbalance}
\dot{\mathcal{D}}^\epsilon(t) ={\mathcal{P}}_D(t)+{\mathcal{P}}_F(t)-\dot{\mathcal{K}}(t) - \dot{\mathcal{E}}^\epsilon(t)  \geq 0.
\end{align}
\end{lemma}
It is observed that conditions for which damage nucleates and propagates follows directly from \eqref{eq:processzonepowerbalance} and is given below.

\bigskip

{\bf  Condition for Damage}

\begin{remark}
\label{damagecondition}
If the rate of energy put into the system exceeds the material's capacity to generate kinetic and elastic energy through  displacement and velocity, then damage occurs.
\end {remark}

\bigskip

Note here that \eqref{eq:processzonepowerbalance} is  consistent with thermodynamics and this formulation follows directly  from the equations of motion \eqref{eq: linearmomentumbal3}.
In summary, power balance  shows that $\dot{\mathcal{D}}^\epsilon(t)>0$ only on  $PZ^\epsilon(t)$ and zero elsewhere. If a crack exists, the strain is greatest in the neighborhood of its tips and the location of the process zone is at the crack tips.

\section{Energy balance satisfied by quasibrittle fracture  evolutions.}% initial value problem for ductile fracture.}
\label{energybalanceprocesszone2}

Explicit formulas for the elastic energy and damage energy are obtained by exchanging time and space integrals in $\int_0^t\dot{\mathcal{E}}(\tau)+\dot{\mathcal{D}}(\tau)\,d\tau$. 
To obtain a more explicit representation we partition the domain into subdomains given by
$$\{(\xx,\bbxi)\hbox{ in }\Omega\times\HH_{\epsilon}\}=UZ^\epsilon(t)\cup DZ^\epsilon(t)\cup PZ^\epsilon(t). $$

For pairs $(\xx,\bbxi)$ in $UZ^\epsilon(t)$ the stress work is only elastic and given by
\begin{align}\label{stressworkundamaged}
w_{UD}(t,\xx,\bbxi)&=\int_0^t\chi_{UZ^\epsilon(t)}(\xx,\bbxi)\,\rho^{\epsilon}(\xx,\bbxi)|\bbxi|\,\mathbb{C}(\xx)\EE^{\bbxi}(\uu(\tau,\xx)):\EE^{\bbxi}(\dot{\uu}(\tau,\xx))\,d\tau\nonumber\\
&=\chi_{UZ^\epsilon(t)}(\xx,\bbxi)\,\rho^{\epsilon}(\xx,\bbxi)|\bbxi|\frac{1}{2}\left(\mathbb{C}(\xx)\EE^{\bbxi}(\uu(t,\xx)):\EE^{\bbxi}{\uu}(t,\xx))-\mathbb{C}(\xx)\EE^{\bbxi}(\uu(0,\xx)):\EE^{\bbxi}{\uu}(0,\xx))\right).
\end{align}
Next, for  pairs $(\xx,\bbxi)$  in $DZ^\epsilon(t)$ we define $t^C_{\xx,\bbxi}$  to be the instant $t$  when $q^\ast(t,\xx,\bbxi,\uu)=q^C$. For pairs $(\xx,\bbxi)$  in $DZ^\epsilon(t)$ we also define  $t^U_{\xx,\bbxi}\leq t$ to be the most recent time of unloading defined by $\dot{q}(t^U_{\xx,\bbxi},\xx,\bbxi,\uu)\leq 0$  with $ q(t^U_{\xx,\bbxi},\xx,\bbxi,\uu)\geq q^C$.
 For pairs $(\xx,\bbxi)$  in $DZ^\epsilon(t)$,  the elastic and inelastic stress work is given by
\begin{align}\label{stressworkdamaged}
w_{DZ}(t,\xx,\bbxi)&=\int_0^{t^C_{\xx,\bbxi}}\chi_{DZ^\epsilon(t)}(\xx,\bbxi)\rho^{\epsilon}(\xx,\bbxi)|\bbxi|\,\mathbb{C}(\xx)\EE^{\bbxi}(\uu(\tau,\xx)):\EE^{\bbxi}(\dot{\uu}(\tau,\xx))d\tau\nonumber\\
&+\int_{t^C_{\yy,\xx}}^{t^U_{\xx,\bbxi}}\chi_{DZ^\epsilon(t)}(\xx,\bbxi)|\bbxi| \rho^{\epsilon}(\xx,\bbxi)\gamma(\tau,\xx,\bbxi,\uu) \mathbb{C}(\xx)\EE^{\bbxi}(\uu(\tau,\xx)):\EE^{\bbxi}(\dot{\uu}(\tau,\xx))d\tau\nonumber\\
&+\int_{t^U_{\xx,\bbxi}}^t\chi_{DZ^\epsilon(t)}(\xx,\bbxi)|\bbxi| \rho^{\epsilon}(\xx,\bbxi)\gamma(t^U_{\xx,\bbxi},\xx,\bbxi,\uu) \mathbb{C}(\xx)\EE^{\bbxi}(\uu(\tau,\xx)):\EE^{\bbxi}(\dot{\uu}(\tau,\xx)) d\tau,
\end{align}
The middle term in \eqref{stressworkdamaged} is given by the non-negative quantity $w_{IDZ}(t,\yy,\xx)$ defined as
\begin{align}\label{product2}
w_{IDZ}(t,\xx,\bbxi)&=\int_{t^C_{\yy,\xx}}^{t^U_{\xx,\bbxi}}\chi_{DZ^\epsilon(t)}(\xx,\bbxi)|\bbxi| \rho^{\epsilon}(\xx,\bbxi)\gamma(\tau,\xx,\bbxi,\uu) \mathbb{C}(\xx)\EE^{\bbxi}(\uu(\tau,\xx)):\EE^{\bbxi}(\dot{\uu}(\tau,\xx))d\tau\nonumber\\
&=\int_{t^C_{\xx,\bbxi}}^{t^U_{\xx,\bbxi}} \chi_{DZ^\epsilon(t)}(\xx,\bbxi)|\bbxi| \rho^{\epsilon}(\xx,\bbxi)\partial_\tau\left(\gamma(\tau,\xx,\bbxi,\uu) \frac{1}{2}\mathbb{C}(\xx)\EE^{\bbxi}(\uu(\tau,\xx)):\EE^{\bbxi}(\uu(\tau,\xx))\right)d\tau\nonumber\\
&+d_{DZ}(t,\xx,\bbxi) \chi_{DZ^\epsilon(t)}(\xx,\bbxi)\rho^{\epsilon}(\xx,\bbxi).
\end{align}
where
\begin{align}\label{product4}
d_{DZ}(t,\xx,\bbxi)=-\int_{t^C_{\xx,\bbxi}}^{t^U_{\xx,\bbxi}}|\bbxi| \partial_\tau\gamma(\tau,\xx,\bbxi,\uu) \frac{1}{2}\left(\mathbb{C}(\xx)\EE^{\bbxi}(\uu(\tau,\xx)):\EE^{\bbxi}(\uu(\tau,\xx))\right)d\tau.
\end{align}
Integrating the first term on the right hand side of \eqref{product2} gives
\begin{align}\label{product10}
&\int_{t^C_{\xx,\bbxi}}^{t^U_{\xx,\bbxi}} \chi_{DZ^\epsilon(t)}(\xx,\bbxi)|\bbxi| \rho^{\epsilon}(\xx,\bbxi)\partial_\tau\left(\gamma(\tau,\xx,\bbxi,\uu) \frac{1}{2}\mathbb{C}(\xx)\EE^{\bbxi}(\uu(\tau,\xx)):\EE^{\bbxi}(\uu(\tau,\xx))\right)d\tau=\nonumber\\
&= \chi_{DZ^\epsilon(t)}(\xx,\bbxi)|\bbxi| \rho^{\epsilon}(\xx,\bbxi)\left(\gamma(t^U_{\xx,\bbxi},\xx,\bbxi,\uu) \frac{1}{2}\mathbb{C}(\xx)\EE^{\bbxi}(\uu(t,\xx)):\EE^{\bbxi}(\uu(t,\xx))\right)\nonumber\\
&- \chi_{DZ^\epsilon(t)}(\xx,\bbxi)|\bbxi| \rho^{\epsilon}(\xx,\bbxi)\left(\gamma(t^C_{\xx,\bbxi},\xx,\bbxi,\uu) \frac{1}{2}\mathbb{C}(\xx)\EE^{\bbxi}(\uu(t^C_{\xx,\bbxi},\xx)):\EE^{\bbxi}(\uu(t^C_{\xx,\bbxi},\xx))\right),
\end{align}
so, integrating the first and third terms of \eqref{stressworkdamaged} and applying \eqref{product2}  gives the elastic and inelastic stress work
\begin{align}\label{stressworkdamagedintegrated}
w_{DZ}(t,\xx,\bbxi)&= \chi_{DZ^\epsilon(t)}(\xx,\bbxi)\frac{1}{2} d_{DZ}(t,\xx,\bbxi)\nonumber\\
&+\chi_{DZ^\epsilon(t)}(\xx,\bbxi)|\bbxi|\rho^{\epsilon}(\xx,\bbxi)\left(\gamma(t^U_{\xx,\bbxi},\xx,\bbxi,\uu) \frac{1}{2}\mathbb{C}(\xx)\EE^{\bbxi}(\uu(t,\xx)):\EE^{\bbxi}(\uu(t,\xx))\right),\nonumber\\
&-\chi_{DZ^\epsilon(t)}(\xx,\bbxi)|\bbxi|\rho^{\epsilon}(\xx,\bbxi)\left(\frac{1}{2}\mathbb{C}(\xx)\EE^{\bbxi}(\uu(0,\xx)):\EE^{\bbxi}(\uu(0,\xx))\right).
\end{align}

 In $PZ^\epsilon(t)$,  the total elastic and inelastic stress work is given by
\begin{align}\label{stressworkundamaging}
w_{PZ}(t,\xx,\bbxi)&=\int_0^{t^C_{\xx,\bbxi}}\chi_{PZ^\epsilon(t)}(\xx,\bbxi)\rho^{\epsilon}(\xx,\bbxi)|\bbxi|\,\mathbb{C}(\xx)\EE^{\bbxi}(\uu(\tau,\xx)):\EE^{\bbxi}(\dot{\uu}(\tau,\xx))d\tau\nonumber\\
&+\int_{t^C_{\xx,\bbxi}}^{t}\chi_{PZ^\epsilon(t)}(\xx,\bbxi)|\bbxi| \rho^{\epsilon}(\xx,\bbxi)\gamma(\tau,\xx,\bbxi,\uu) \mathbb{C}(\xx)\EE^{\bbxi}(\uu(\tau,\xx)):\EE^{\bbxi}(\dot{\uu}(\tau,\xx))d\tau.
\end{align}

Proceeding using the analogous arguments developed for the total stress work for pairs $(\xx,\bbxi)$ in $DZ(t)$ we get that the total elastic and inelastic stress work done in $PZ(t)$ is given by
\begin{align}\label{stressworkundamagingintegrated}
w_{PZ}(t,\xx,\bbxi)&=\chi_{PZ^\epsilon(t)}(\xx,\bbxi)\rho^{\epsilon}(\xx,\bbxi)d_{PZ}(t,\xx,\bbxi)\nonumber\\
&+\chi_{PZ^\epsilon(t)}(\xx,\bbxi)|\bbxi|\rho^{\epsilon}(\xx,\bbxi)\left(\gamma(t,\xx,\bbxi,\uu) \frac{1}{2}\mathbb{C}(\xx)\EE^{\bbxi}(\uu(t,\xx)):\EE^{\bbxi}(\uu(t,\xx))\right),\nonumber\\
&-\chi_{PZ^\epsilon(t)}(\xx,\bbxi)|\bbxi|\rho^{\epsilon}(\xx,\bbxi)\left(\frac{1}{2}\mathbb{C}(\xx)\EE^{\bbxi}(\uu(0,\xx)):\EE^{\bbxi}(\uu(0,\xx))\right).
\end{align}
where
\begin{align}\label{product80}
d_{PZ}(t,\xx,\bbxi)=-\int_{t^C_{\xx,\bbxi}}^{t}|\bbxi| \partial_\tau\gamma(\tau,\xx,\bbxi,\uu) \frac{1}{2}\left(\mathbb{C}(\xx)\EE^{\bbxi}(\uu(\tau,\xx)):\EE^{\bbxi}(\uu(\tau,\xx))\right)d\tau.
\end{align}

Having integrated over time for each fixed pair $(\xx,\bbxi)$, we now integrate \eqref{stressworkundamaged} \eqref{stressworkdamagedintegrated} and \eqref{stressworkundamagingintegrated} over $\bbxi$ and $\xx$ variables to obtain
the elastic energy at the present time $t$:
\begin{align}\label{elastict}
\mathcal{E}^\epsilon(t)&=\int_{\Omega}\int_{\HH_\epsilon(0)} \chi_{UZ^\epsilon(t)}(\xx,\bbxi)\,\rho^{\epsilon}(\xx,\bbxi)|\bbxi|\frac{1}{2}\left(\mathbb{C}(\xx)\EE^{\bbxi}(\uu(t,\xx)):\EE^{\bbxi}{\uu}(t,\xx))\right)d\bbxi d\xx\nonumber\\
&+\int_{\Omega}\int_{\HH_\epsilon(0)}\chi_{PZ^\epsilon(t)}(\xx,\bbxi))|\bbxi|\rho^{\epsilon}(\xx,\bbxi)\frac{1}{2}\left(\gamma(t,\xx,\bbxi,\uu) \mathbb{C}(\xx)\EE^{\bbxi}(\uu(t,\xx)):\EE^{\bbxi}(\uu(t,\xx))\right) \,d\bbxi d\xx\nonumber\\
&+\int_{\Omega}\int_{\HH_\epsilon(0)} \chi_{DZ^\epsilon(t)}(\xx,\bbxi)|\bbxi|\rho^{\epsilon}(\xx,\bbxi)\frac{1}{2}\left(\gamma(t^U_{\xx,\bbxi},\xx,\bbxi,\uu)\mathbb{C}(\xx)\EE^{\bbxi}(\uu(t,\xx)):\EE^{\bbxi}(\uu(t,\xx))\right)\,d\yy d\xx,
\end{align}
where the initial elastic energy is
\begin{align}\label{elasticti}
{\mathcal{E}}^\epsilon(0)&=\int_{\Omega}\int_{\HH_\epsilon(0)}|\bbxi|\rho^{\epsilon}(\xx,\bbxi)\frac{1}{2}\left(\mathbb{C}(\xx)\EE^{\bbxi}(\uu(0,\xx)):\EE^{\bbxi}(\uu(0,\xx))\right)\,d\bbxi\,d\xx.
\end{align}
The non-negativity of the elastic energy ${\mathcal{E}}^\epsilon(t)$ follows from \eqref{elastict}.
Recalling that $\mathcal{D}(0)=0$ and collecting results shows the damage energy expended from $0$ to $t$ is given by the non-negative quantity
\begin{align}\label{damageenergy}
{\mathcal{D}}^\epsilon(t)&=\frac{1}{2}\int_{\Omega}\int_{\HH_\epsilon(0)}\rho^{\epsilon}(\xx,\bbxi)(\chi_{DZ^\epsilon(t)}(\xx,\bbxi)d_{DZ}(t,\xx,\bbxi)+\chi_{PZ^\epsilon(t)}(\xx,\bbxi)d_{PZ}(t,\xx,\bbxi))\,d\bbxi\,d\xx
\end{align}

The kinetic energy at time $t$ is
\begin{align}\label{eq: balancek}
\mathcal{K}(t)=&\frac{\rho}{2}\int_\Omega\, |\dot{\uu}(t)|^2\,d\xx.
\end{align}
The corresponding energy balance  is given by:

\bigskip

\noindent{\bf Energy balance for simultaneous load and displacement controlled fracture evolution}
\begin{align}\label{eq:processzoneenergybalance}
{\mathcal{D}}^\epsilon(t) =\int_0^t{\mathcal{P}}_D(\tau)d\tau+\int_0^t{\mathcal{P}}_F(\tau)d \tau-({\mathcal{K}}(t) + {\mathcal{E}}^\epsilon(t)  - {\mathcal{K}}(0) - {\mathcal{E}}^\epsilon(0) ),
\end{align}
where
 $\mathcal{E}^\epsilon(0)$, $\mathcal{E}^\epsilon(t)$, $\mathcal{D^\epsilon}(t)$,  $\mathcal{K}(t)$,   are given by \eqref{elasticti},  \eqref{elastict}, \eqref{damageenergy},   \eqref{eq: balancek} and  $\mathcal{D}^\epsilon(0)=0$,

\section{Calibration of the nonlocal model using measured elastic properties, strength and fracture toughness of the material} %and \textcolor{red}{Size effect} }
\label{sec.Straight Cracks}
In this section we return to the inelastic IBVP of Section \ref{sec.Constiutivelaws BVP} and calibrate the elastic tensor and the parameters responsible for the strength and fracture toughness of the frame gradient  model  using quantities obtained directly from the dynamics and equating them to the material properties of the specimen. %The elastic moduli and critical energy release rate follow from $\Gamma$-convergence as $\epsilon\rightarrow 0$, see \cite{lipton2016cohesive}, \cite{lipton2024energy}. But they are obtained here by Taylor series and geometric measure theory arguments presented in \cite{lipton2024energy}. 
\subsection{Elastic properties}
\label{elasticproperties}
The horizon length scale $\epsilon$ is chosen small enough to resolve the process zone hence smaller than the length scale of the process zone of the material. Assume $\mathbb{C}(\xx)$ is a constant  $\mathbb{C}$ and undamaged inside $\mathcal{H}_\epsilon({\bf x})$.
For this case the energy is given by
the nonlocal elastic energy
\begin{align}\label{elastictundamaged}
\mathcal{E}(t,\xx,\uu)=\frac{1}{2}\int_{\Omega}\,\rho^{\epsilon}(\xx,\bbxi)|\bbxi|\,\ss_{ij}\EE_{ij}^{\bbxi}(\uu(t,\xx))\,d\bbxi.
\end{align}
For a neighborhood ${\mathcal{H}}_\epsilon(\xx)$ chosen in undamaged material we assume the displacement is elastic and the horizon is small enough so that the displacement is linear.
For this case $\uu=M{\bf x}$ where $M$  is a constant matrix. Next change variables of integration according to $\bbxi=\epsilon\bzz$ where $\bzz$ is any point in the unit ball $\mathcal{H}_1(0)$.  So the elastic energy density over a neighborhood of $\xx$ at  time $t$ is now written
 \begin{align}
\mathcal{E}(t,{\bf x,\uu})&=\frac{1}{2\omega_d}\int_{\mathcal{H}_1(0)}|\bzz| J(|\bzz|)\mathbb{C}_{ijkl}\EE_{kl}^{\epsilon\bzz}(\uu(t,\xx))\EE_{ij}^{\epsilon\bzz}(\uu(t,\xx))\,d\bzz
\label{LEFMequality}
\end{align}
Note we have the resolution of the $I^{d\times d}$ identity given by $I^{d\times d}=\sum_{m=1}^d \ee^m\otimes\ee^m$ and
\begin{eqnarray}
\nabla^{\epsilon\bzz}\uu(t,\xx,\epsilon\bzz)=\sum_{m=1}^d M\ee^m\otimes\ee^m=M\sum_{m=1}^d \ee^m\otimes\ee^m=M,
\label{resident}
\end{eqnarray}
so $\EE^{\epsilon\bzz}(\uu(t,\xx,\epsilon\bzz))=F$ where $F=(M^T+M)/2$.
Hence
\begin{eqnarray}
\mathcal{E}(t,{\bf x})=\frac{1}{2}\sum_{ijkl}p\mathbb{C}_{ijkl}F_{ij}F_{kl}
\label{leading order}
\end{eqnarray}
with 
\begin{eqnarray}
p=\frac{1}{\omega_d}\int_{\mathcal{H}_1(0)}|\bzz| J(|\bzz|)\,d\bzz.
\label{elasticpart3}
\end{eqnarray}
On the other hand, the potential elastic energy per unit volume of a material characterized by a constant elasticity tensor $\overline{\mathbb{C}}$ in $\mathcal{H}_\epsilon(\xx)$  is given by
\begin{align}
U&=\frac{1}{2}\sum_{ijkl}\mathbb{\overline{C}}_{ijkl}F_{ij}F_{kl}
\label{LEFMequality13}
\end{align}
where $\mathbb{\overline{C}}$ can be any prescribed  elasticity tensor. 
 Setting $U=\mathcal{E}(t,{\bf x})$, we get $p^{-1}\mathbb{\overline{C}}=\mathbb{C}$. 
 It is now clear that for the frame gradient method  there is no constraint placed on on using any  elasticity tensor hence isotropic materials exhibiting the complete intervals of Poisson ratios can be considered.

 Note in $\mathbb{R}^d$ and for the choice ${\overline{\mathbb{C}}}=p\mathbb{C}$ where
 \begin{eqnarray}
\mathbb{C}=2\mu \mathbb{K} + 3\kappa\mathbb{J},%\mathbb{\overline{C}}_{ijkl}=G (\delta_{ik} \delta_{jl} +\delta_{il}\delta_{kj})+ \lambda\delta_{ij}\delta_{kl}
\label{elasticpart4}
\end{eqnarray}
 we arrive at the calibration for determining the shear moduli $\overline{\mu}$ and bulk moduli $\overline{\kappa}$ of $\overline{\mathbb{C}}$. %Or equivalently we can use  values of $\overline{\mu}

\subsection{Critical energy release rate for brittle fracture and the dissipation energy required to form a crack}
\label{criticalenergyrelease}

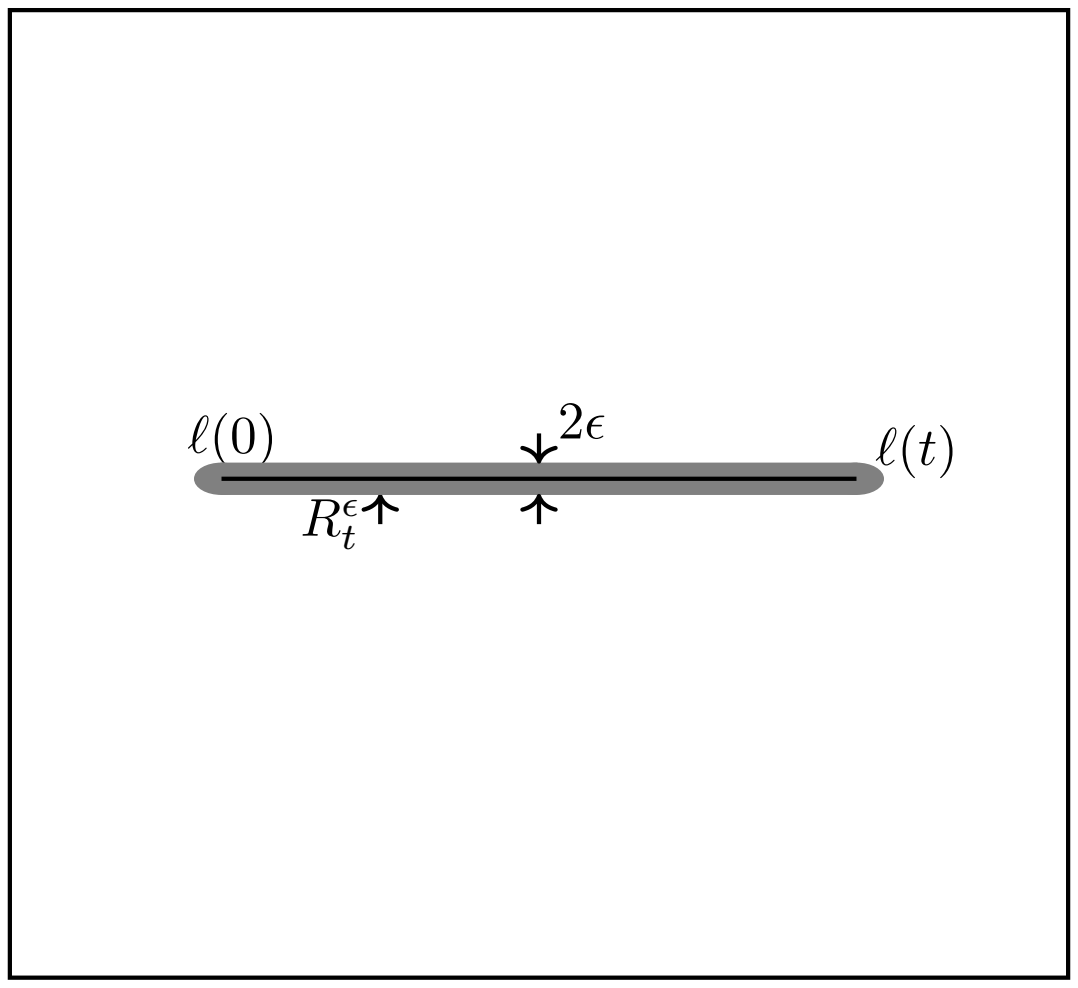
\begin{figure} 
\centering
\begin{tikzpicture}[xscale=0.70,yscale=0.40, rotate=90]

%\draw [fill=gray, gray] (-7.01,-0.5) rectangle (0.0,0.5);

\node[left] at (-0.5,1.5) {$R_t^\epsilon$};

\draw [->,thick] (-0.5,1.5) -- (-0.0,1.5);

%\draw[fill=gray, gray] (0,-0.5) arc (-90:90:0.5)  -- cycle;

%\draw[fill=gray, gray](-7.01,0.0) circle (0.5);

%\draw [-,thick] (-7.01,0) -- (-0.0,0);

%%% Specimen border
\draw [thick] (-8,-5) rectangle (8,5);

%%%%%%%%%%%%%%%%%%%%%%%%%%%%%%%%%%

%%%%%%%%%%%%%%%%%%%%%%%%%%%%%%%%%%

%\node [above] at (-1.6,0) {$\epsilon$};

%\node [below] at (-1.6,0) {$\epsilon$};

%%% ????????
%\draw [thick] (-1.3,.5) -- (-1.1,.5);

%\draw [<->,thick] (-1.215,.5) -- (-1.215,0);

%\draw [<->,thick] (-1.215,-.5) -- (-1.215,0);

%\draw [thick] (-1.3,-.5) -- (-1.1,-.5);
%%%%%%%%
%% Right end of previous crack
%%%%%\draw [->,thick] (0.0,-1.0) -- (0.0,-0.1);

%%%%%  top end of crack and annotate top of crack by $\ell(t)$
\node [right] at (0.9,3.5) { $\ell(0)$};

%%%%%%%%%% Fill rectangle around crack
\draw [fill=gray, gray] (0.5,-3.0) rectangle (0.0,3.0);
%fill circle at top of crack
\draw[fill=gray, gray] (0.25,3.0) circle (0.25);

%%%% arrow at left end of old crack
%\draw [->,thick] (-7,-1.0) -- (-7,-0.1);

%Annotate bottom of crack by $\ell(0)$
\node [right] at (0.7,-3.0) { $\ell(t)$};
% fill circle at bottom of crack
\draw[fill=gray, gray] (0.25,-3.0) circle (0.25);
 %%%%%% Vertical crack
\draw [thick] (0.25,-3.0) -- (0.25,3.0);
%%%%%%%%

\draw [->,thick] (-0.5,0.0) -- (-0.0,0.0);
\draw [<-,thick] (0.5,0.0) -- (1.0,0.0);
%Annotate crack width $2\epsilon$
\node [right] at (1.2,0.0) {$2\epsilon$};
\end{tikzpicture} 
\caption{{  A flat crack.  The crack $R_t^\epsilon=\{x_2=0;\,\ell(0)<x_1<\ell(t)\}$ is an interior surface formed by the evolution across which  traction forces above cracks do not influence the motion of points below the crack and vice versa. Consequently the displacement need not be continuous across the crack.}}
 \label{P}
\end{figure}

Brittle  fracture occurs under high stress triaxiality, so we assume all frames that intersect  the crack surface $R_t^\epsilon$ are in the triaxial stress state  $\mathcal{J}_2=0$, see \eqref{Jtwo} for $\mathcal{J}_2$ in $d=3$ and \eqref{psdp} for $\mathcal{J}_2$ in plane stress.  A nonlocal traction force originating from a point $\xx$ above the  crack surface should have no influence on points below the crack surface. 
This notion  is made precise when considering a flat crack  given by the flat surface $R_t^\epsilon$ on the $x_d=0$ plane, see Figure \ref{P}. The half space containing points with $x_d>0$ is denoted $\mathbb{R}^d_+$ and the half space  with $x_d<0$ is denoted $\mathbb{R}^d_-$. This neighborhood is shown in Figure \ref{P}.
The failure set $\Gamma^\epsilon(t)$ for a flat crack of area  $R_t^\epsilon$ is a  subset of  $\mathbb{R}^d$ is the collection of pairs $(\xx,\bbxi)$ in $\Omega\times H_\epsilon(0)$ such that  the specific  traction force per unit volume between $\xx-\bbxi$ in $\mathbb{R}^d_+$ and $\xx$ in $\mathbb{R}^d_-$ vanishes after some time $\tau$, $0<\tau\leq t$ i.e.,
\begin{align}
      \label{eq: force200}
\rho^{\epsilon}(\xx,\bbxi)\, \ss(t,\xx,\bbxi)\ee =\rho^{\epsilon}(\xx-\bbxi,\bbxi)\,\ss(t,\xx-\bbxi,\bbxi)\ee=0.
 \end{align}
 This is the gray shaded region in Figure \ref{P}. 
 Thus $\ee$ is an eigenvector of $ \ss(t,\xx,\bbxi)$ and $\ss(t,\xx-\bbxi,\bbxi)$ and since these are symmetric  stress tensors the vectors $\ee^\perp$ and   $\ee^\times:=\ee\times\ee^\perp$  are also eigenvectors. 
 Since the stress state is assumed triaxial, i.e., $\mathcal{J}_2(\ss)=0$ within an $\epsilon$ neighborhood of the crack the eigenvector  $\ee^\perp$ is associated with a zero eigenvalue, i.e., 
 \begin{align}
      \label{eq: force2002}
\rho^{\epsilon}(\xx,\bbxi)\, \ss(t,\xx,\bbxi)\ee^\perp =\rho^{\epsilon}(\xx-\bbxi,\bbxi)\,\ss(t,\xx-\bbxi,\bbxi)\ee^\perp=0,
 \end{align}
 and  $\ee^\times$ is also associated with a zero eigenvalue
 \begin{align}
      \label{eq: force2002}
\rho^{\epsilon}(\xx,\bbxi)\, \ss(t,\xx,\bbxi)\ee^\times =\rho^{\epsilon}(\xx-\bbxi,\bbxi)\,\ss(t,\xx-\bbxi,\bbxi)\ee^\times=0.
 \end{align}
Thus a frame is failed  for $d=2,3$ when one of its legs crosses the crack.

 In dimension $d=2,3$ counting shows every single interval $(\xx,\bbxi)$ of length $|\bbxi|$ crossing the crossing the crack corresponds to  $d$ different frame legs of length $|\bbxi|$ also crossing the crack.
\begin{align}\label{brokenset}
\Gamma^\epsilon(t)=\{(\xx,\bbxi)\hbox{ in }\Omega\times H_\epsilon(0); \, \hbox{ any interval $(\xx,\xx+\bbxi^n)$ of a frame based at $\xx$ that intersects $A$} \}
\end{align}
with indicator function
\begin{equation}\label{busted}
\chi_{\Gamma^\epsilon(t)}(\xx,\bbxi)=\left\{\begin{array}{ll}1,&\hbox{if  $(\xx,\xx+\bbxi^n)$ belongs to $\Gamma^\epsilon(t)(\xx,\bbxi)$ }\\
0,& \hbox{otherwise}.\end{array}\right.
\end{equation}

The dissipation energy necessary to form a crack of a given area (length) is seen to follow from the evolution equations.  The calculation given below shows that the dissipation energy  is given by the crack area multiplied by a scalar. This scalar is  precisely the critical energy release rate (or equivalently the fracture toughness)  and is calibrated to that of the material. In this way  the Griffith fracture energy for brittle cleavage fracture is recovered  directly from the dissipation  energy.  Calculation and theory show the fracture energy is independent of the horizon $\epsilon$ used in the frame gradient model. This procedure is explicitly carried out in the paragraphs below.

From Section  \ref{energybalanceprocesszone2} and the definition of  $\chi_{\Gamma^\epsilon(t)}(\xx,\bbxi)$ it follows that the dissipation energy necessary to form a crack of area $A$ at time $t_F$ is given by  the non-negative quantity
\begin{align}\label{failenergy}
{\mathcal{D}}^\epsilon(t_F)&=\int_{\Omega}\int_{\HH_{\epsilon}(0)}\chi_{\Gamma^\epsilon(t)}(\xx,\bbxi)\rho^\epsilon(\xx,\bbxi)\,d_{F}(t_F,\xx,\bbxi)\,d\bbxi\,d\xx,
\end{align}
where the stress work  per unit area  necessary to eliminate traction forces across the crack due to one frame of length $|\bbxi|$ is given by 
\begin{align}\label{product4}
d_{F}(t_F,\xx,\bbxi)=-\int_{{t^C}_{\xx,\bbxi}}^{{t^F}_{\xx,\bbxi}} |\bbxi| \partial_{\tau}\gamma(\tau,\xx,\bbxi,\uu)\frac{1}{2}\mathbb{C}\EE^{\bbxi}(\tau,\xx):\EE^{\bbxi}(\tau,\xx)d\tau,
\end{align}
with $t^C_{\xx,\bbxi}$ equal to the time when the stress leaves the strength domain and $t^F_{\xx,\bbxi}$ equal to the time when the material connecting $\xx-\bbxi$ to $\xx$   is stressed to failure. 

From \eqref{product4} we see the stress work  per unit frame length $|\bbxi|$ to fully soften (fail) a frame is given by
\begin{align}\label{product5}
a_{F}(t_F,\xx,\bbxi)=-\int_{{t^C}_{\xx,\bbxi}}^{{t^F}_{\xx,\bbxi}}  \partial_{\tau}\gamma(\tau,\xx,\bbxi,\uu)\frac{1}{2}\mathbb{C}\EE^{\bbxi}(\tau,\xx):\EE^{\bbxi}(\tau,\xx)d\tau,
\end{align}
and we calculate it explicitly for a cleavage crack where the stress state is assumed triaxial within an $\epsilon$ neighborhood of the crack.
Since $\mathcal{J}_2(\ss)=0$ and we write the elastic energy density in terms of the stress $\mathbb{C}\EE^{\bbxi}(\uu(\tau,\xx)):\EE^{\bbxi}(\uu(\tau,\xx))=\mathbb{C}^{-1}\ss(\tau,\xx,\bbxi):\ss(\tau,\xx,\bbxi)$. Hence 
$$(1/2)\mathbb{C}^{-1}\ss(\tau,\xx,\bbxi):\ss(\tau,\xx,\bbxi)=(1/2)\mathbb{C}^{-1}I:I(tr\{\ss(\tau,\xx,\bbxi)\}/3)^2,$$
and $tr\{\ss(\tau,\xx,\bbxi)\}=\frac{1}{\beta^C}q(\ss(\tau,\xx,\bbxi)):=\frac{1}{\beta^C}q(\tau)$.

Thus applying the definition of $\gamma$ and $\sigma(r^*)$, noting $r^\ast=r$ on the loading curve and $r=\sqrt{|\bbxi|/L}q$  we get
\begin{align}\label{product6}
a_{F}(t_F,\xx,\bbxi)=&-\frac{1}{2}\mathbb{C}^{-1}I:I\int_{{t^C}_{\xx,\bbxi}}^{{t^F}_{\xx,\bbxi}}  \partial_{\tau}\left(\frac{\partial_{r}g(r(\tau))}{r(\tau)} \right)\left(\frac{q(\tau)}{3\beta^C}\right)^2d\tau,\nonumber\\
=&-\frac{\mathbb{C}^{-1}I:I}{18(\beta^C)^2}\int_{{t^C}_{\xx,\bbxi}}^{{t^F}_{\xx,\bbxi}}  \left(\frac{\partial_{r}^2g(r(\tau))r(\tau)\dot{r}(\tau)}{r^2(\tau)}- \frac{\partial_{r}g(r(\tau))\dot{r}(\tau)}{r^2(\tau)}\right)(q(\tau))^2d\tau\nonumber\\
=&-\frac{L}{|\bbxi|}\frac{\mathbb{C}^{-1}I:I}{18(\beta^C)^2}\int_{{t^C}_{\xx,\bbxi}}^{{t^F}_{\xx,\bbxi}}  \left(\partial_{r}^2g(r(\tau))r(\tau)\dot{r}(\tau)- \partial_{r}g(r(\tau))\dot{r}(\tau)\right)d\tau\nonumber\\
=&-\frac{L}{|\bbxi|}\frac{\mathbb{C}^{-1}I:I}{18(\beta^C)^2}\int_{r^C}^{r^F}  \left(\partial_{r}^2g(r)r- \partial_{r}g(r)\right)d r\nonumber\\
=&\,\,\, \,\,\,\,\frac{L}{|\bbxi|}\frac{\mathbb{C}^{-1}I:I}{9(\beta^C)^2}g(r^F).
\end{align}
and $d_F$ is independent of $\xx,\bbxi$ and given by $d_F(t_F,\xx,\bbxi)=d_F=Lg(r^F)\frac{\mathbb{C}^{-1}I:I}{9(\beta^C)^2}$.
Here the scaling $\sqrt{|\bbxi|/L}$ is chosen such  that the stress work  to fail a frame of length $|\bbxi|$ is the same for all frames and we write
\begin{align}\label{failenergy2}
{\mathcal{D}}^\epsilon(t_F)&=d_F\int_{\Omega}\int_{\HH_{\epsilon}(0)}\chi_{\Gamma^\epsilon(t)}(\xx,\bbxi)\rho^\epsilon(\xx,\bbxi)\,d\bbxi\,d\xx,
\end{align}
and explicitly calculate the geometric integral
\begin{align}\label{geometric}
K_{frac}=\int_{\Omega}\int_{\HH_{\epsilon}(0)}\chi_{\Gamma^\epsilon(t)}(\xx,\bbxi)\rho^\epsilon(\xx,\bbxi)\,d\bbxi\,d\xx,
\end{align}
when the stress state is assumed triaxial.
%The critical energy release rate for a cleavage crack is $G_c$ and we use it to calibrate $g(r^F)$.
We change variables  in \eqref{geometric} and write $\boldsymbol{\zeta}=\bbxi/\epsilon$ and switch the order of integration to get the $d-1$ dimensional integral
\begin{align}
\label{1stenergy}
K_{frac}=\frac{1}{\omega_d}\int_{H_1(0)}\,J(|\bzz|)\frac{1}{\epsilon}\left(\int_\Omega\chi_{\Gamma^\epsilon(t)}(\xx,\epsilon\zeta)\,d\xx\right)\,d\bzz.
\end{align}
We change coordinates in $\xx$ using the slicing variables and write $x=\zz+\ee s$ Where $s\in\mathbb{R}$,  $\ee=\bzz/|\bzz|$ is the direction along $\bzz$ and $\zz$ is the $d$ dimensional subspace of points $\zz\perp\ee$. For $(\zz,\ee)$ fixed, consider the line $\Omega_\zz^\ee=\{s\in\mathbb{R};\,\zz+\ee s\}$ and $ \Omega^e=\{\zz;\,\Omega_\zz^\ee\cap\Omega\not=\emptyset\}$ and 
\begin{align}
\label{2ndtenergy}
K_{frac}=\frac{1}{\omega_d}\int_{H_1(0)}\,J(|\bzz|)\frac{1}{\epsilon}\int_{\Omega^\ee}\int_{\Omega^\ee_\zz}(d)\chi^+(\zz+\ee(s+\epsilon|\bzz|),\zz+\ee s,{t})\,ds\,d\zz\,d\bzz,
\end{align}
where from \eqref{brokenset}  we have $\chi_{\Gamma^\epsilon(t)}(\xx,\epsilon\bzz)=\chi^+(\zz+\ee(s+\epsilon|\bzz|),\zz+\ee s,{t})$ and
\begin{align}
\label{kernelkernel}
\chi^+(\zz+\ee(s+\epsilon|\bzz|),\zz+\ee s,{t})=\left\{\begin{array}{ll}1,&\hbox{if the pair $(\zz+\ee(s+\epsilon|\bzz|),\zz+\ee s)$ is in $\Gamma^\epsilon({t})$}\\
0,& \hbox{otherwise}.\end{array}\right.
\end{align}
Lastly $(d)=2,3$ is the number of frame legs of length $\epsilon|\bzz|$ on any interval crossing the crack.
We change to polar coordinates and $d\bzz=|\bzz|^{d-1}d|\bzz|\,d\ee$, where $d\ee$ is the surface area element of $\mathbb{S}^{d-1}$. After a change in the order of integration we arrive at the desired formula.
 \vspace*{1\baselineskip}

\noindent{\bf  Geometric integral representation}
\label{geometricintegral}

 \vspace*{1\baselineskip}
\begin{align}
\label{3rdtenergy}
K_{frac}=\frac{d}{\omega_d}\int_{\mathbb{S}^{d-1}}\int_{\Omega^\ee}\int_0^1\,J(|\bzz|)|\bzz|^d m({t},\ee,\zz,|\bzz|)\,d|\bzz|\,d\zz\,d\ee,
\end{align}
with
\begin{align}
\label{1stkernel}
m({t},\ee,\zz,|\bzz|)=\frac{1}{\epsilon|\bzz|}\int_{\Omega^\ee_\zz}\chi^+(\zz+\ee(s+\epsilon|\bzz|),\zz+\ee s,{t})\,ds.
\end{align}

Here
\begin{align}
\label{ml}
m({t},\ee,\zz,|\bzz|)=\left\{\begin{array}{ll}1,&\hbox{if the line $\xx=\zz+\ee s$ pierces $R_t$}\\
0,& \hbox{otherwise}.\end{array}\right.
\end{align}
for all $|\bzz|<1$ and we write $m({t},\ee,\zz,|\bzz|)=N(t,\ee,\zz)$, where $N(t,\ee,\zz)$ is the multiplicity function of any line $\Omega_\zz^\ee$ counting the number of times it pierces the surface (line segment). For this case, $R_t$ is flat so $N(t,\ee,\zz)$ is either one or zero. The failure energy ${\mathcal{D}}^\epsilon(t_F)$ becomes
\begin{align}
\label{crofton}
{\mathcal{D}}^\epsilon(t_F)=\mathcal{G}_c\frac{1}{2\omega_{d-1}}\int_{\mathbb{S}^{d-1}}\int_{\Omega^\ee}N(t,\ee,\zz)\,d\zz\,d\ee.
\end{align}
where the first factor of \eqref{crofton} is
\begin{align}
\label{G}
\mathcal{G}_c=d Lg(r^F)\frac{2\omega_{d-1}}{\omega_d}\frac{\mathbb{C}^{-1}I:I}{9(\beta^C)^2}\int_0^1\,J(|\bzz|)|\bzz|^d\,d|\bzz|.
\end{align}
The second factor of  \eqref{crofton}  is Crofton's formula and delivers the surface area  given by $\mathcal{H}^{d-1}(R_t^\epsilon)$  (for $d=3$) (or length $d=2$) of the internal boundary $R^\epsilon_t$. If the boundary of the flat crack is  rough for $d=3$, then the $2$ dimensional Hausdorff measure of $R^\epsilon_t$ is also written $\mathcal{H}^{2}(R^\epsilon_t)$ \cite{Morgan,Federer}, and for all cases
\begin{align}
\label{last}
{\mathcal{D}}^\epsilon(t_F)=\mathcal{G}_c\mathcal{H}^{d-1}(R_t).
\end{align}
This is the well known formula for Griffith fracture energy but now derived directly from the IBVP. %We believe this is the first such derivation
 What is distinctive is  that  the Griffith fracture energy  found here  follows directly from the nonlocal model without sending any parameter such as  $\epsilon$ to zero as in the phase field method.  Of course we can extend the formula \eqref{last} to a system of dispersed flat cracks separated by the distance $\epsilon$ with different orientations.  
 Here
 \begin{align}
    \label{gcofnonlocalparameters}
   \mathcal{G}_c= d Lg(r^F)\frac{2\omega_{d-1}}{\omega_d}\frac{\mathbb{C}^{-1}I:I}{9(\beta^C)^2}\int_0^1\,J(|\bzz|)|\bzz|^d\,d|\bzz|.
\end{align}

 The fracture toughness for the material specimen is given by $G_c$ and $L$ and $r^F$ lie on the locus
\begin{align}
    \label{locus}
    \frac{G_c}{L}= dg(r^F)\frac{2\omega_{d-1}}{\omega_d}\frac{\mathbb{C}^{-1}I:I}{9(\beta^C)^2}\int_0^1\,J(|\bzz|)|\bzz|^d\,d|\bzz|.
\end{align}

In this treatment we illustrate and validate the ideas numerically. In all plane stress simulations we apply the bilinear model and choose influence function $J(r)$ given by $1$ for $0<r<1$ and zero for $1\leq r$ and 

\begin{align}
    \label{lenthscale}
    L=\frac{18 (\beta^C)^2 G_c}{d\ r^F r^C (\mathbb{C}^{-1}I:I)},\,\,\,r^F\geq r^C=\sigma_{DP}^C.
\end{align}

\section{Numerical Solution and Results}
\label{sec.numerics}
In this section, we discuss the numerical implementation and results of the proposed frame gradient model. We first perform a mesh convergence study to ensure numerical accuracy. Then we  validate our numerical model on different sets of materials under different loading conditions. Throughout this section, the contour plots of the local damage variable are used to present the crack path resulting from the numerical analysis. The damage phase field variable $\gamma$, defined in \eqref{softening3d} can take any value between zero and one, representing completely damaged and virgin material points, respectively. A simple bilinear strength curve \eqref{bilinearloading} is used to define the constitutive law throughout the implementation. The details of the calibration of the material constants are provided in section \ref{sec.Calibration}. 

The dynamic relaxation method \cite{underwood1983} is used for the quasistatic test, with the damping coefficient fixed as in \cite{coskun2025quasi}. However, damage is not updated during the relaxation. Rather, a staggered solution scheme is adopted, where damage is updated after the mechanical equilibrium is reached. The algorithm for the quasi static test is summarized in Algorithm \ref{alg:algorithm}.

\begin{algorithm}[H]
\caption{\textbf{Dynamic relaxation for the quasi-static solution} (dynamic relaxation of \cite{littlewood2016roadmap, coskun2025quasi} is modified with staggered damage update)}
\label{alg:algorithm}
\begin{algorithmic}[1]
\State Initialization: $n = 0$, $\mathbf{u} = \mathbf{0}$, $\dot{\mathbf{u}}^{-\frac12} = \mathbf{0}$, $\ddot{\mathbf{u}} = \mathbf{0}$, $\gamma_f = 1$ and $r_f^* = 0$ for all frames, where $n$ is the load step.
\State Update load step and apply the prescribed displacement $\bar{\mathbf{u}}^{n}$ on the loading layer.
\State Iterative solution starts, initialize the iteration counter $i = 0$, $\mathbf{u}^i \gets \mathbf{u}^n$ and $\ddot{\mathbf{u}}^i \gets \ddot{\mathbf{u}}^n$.
\While{$e \geq$ tolerance}
  \State $\dot{\mathbf{u}}^{i+\frac12} \gets \dot{\mathbf{u}}^{i-\frac12} + \Delta t\,\ddot{\mathbf{u}}^{i}$ \Comment{half-step velocity update}
  \State Apply velocity boundary conditions, $\dot{\mathbf{u}}^{i+\frac12} = \mathbf{0}$ on constrained DOFs.
  \State $\mathbf{u}^{i+1} \gets \mathbf{u}^{i} + \Delta t\,\dot{\mathbf{u}}^{i+\frac12}$, and reset constrained DOFs to $\bar{\mathbf{u}}^{n}$ \Comment{update displacement}
  \State $\dot{\mathbf{u}}^{i+1} \gets \dot{\mathbf{u}}^{i+\frac12} + \tfrac12\Delta t\,\ddot{\mathbf{u}}^{i}$ \Comment{velocity for the damping term}
  \State Evaluate internal forces $\mathbf{f}_{int}^{i+1}$ with $\mathbf{u}^{i+1}$ and $\gamma_f^{n}$ held fixed
  \State $\ddot{\mathbf{u}}^{i+1} \gets \mathbf{M}^{-1}\big(\mathbf{f}_{int}^{i+1} - C\,\dot{\mathbf{u}}^{i+1}\big)$ \Comment{$C$ is the damping coefficient}
  \State $e = \dfrac{\|\mathbf{f}_{int}^{i+1}\|}{|R^{i+1}|}$ \Comment{$R$ is the reaction on the loaded layer}
  \State Update: $\mathbf{u}^{i} \gets \mathbf{u}^{i+1}$ and $\ddot{\mathbf{u}}^{i} \gets \ddot{\mathbf{u}}^{i+1}$, and $i \gets i + 1$.
\EndWhile
\State Solution is accepted: $\mathbf{u}^{n} \gets \mathbf{u}^{i}$ and $\ddot{\mathbf{u}}^{n} \gets \ddot{\mathbf{u}}^{i}$.
\For{each damageable frame $f$} \Comment{staggered damage update}
  \State $q_f \gets \sqrt{J_2(\sigma_f)} + \beta_C I_1(\sigma_f)$ from $\mathbf{u}^{n}$
  \State $r_f \gets \sqrt{|\boldsymbol{\xi}_f|/L}\; q_f$, and $r_f^* \gets \max(r_f^*, r_f)$
\If{$r_f^* \le r_C$}
  \State $\gamma_f^{n+1} \gets 1$
\ElsIf{$r_f^* < r_F$}
  \State $\gamma_f^{n+1} \gets \sigma_{DP}(r_f^*)\,/\,r_f^*$
\Else
  \State $\gamma_f^{n+1} \gets 0$
\EndIf
\EndFor
\State If $\min_f \gamma_f^{n+1} < 1$, damage onset is reached; stop.
\State If simulation is not complete, go to \textit{step 2}.
\end{algorithmic}
\end{algorithm}

For the dynamic simulation, $\epsilon \leq L$ is chosen and the local damping coefficient in the dynamic relaxation technique used in the static case is set to zero. The critical time step is calculated using the approach of \cite{silling2005meshfree}. For all simulation, a safety factor of 0.85 is chosen to calculate the final time step. Finally, for the comparison purposes, experimental and numerical data from the literature are extracted using WebPlotDigitizer \cite{WebPlotDigitizer}. 

\subsection{Mesh Convergence}
\label{sec.convergence}
The dynamic crack branching problem in glass like material was selected to perform the numerical convergence analysis of the proposed frame gradient model. The specimen has dimensions of $100 \times 40$ mm with a prenotch of $50$ mm at the middle of the left edge, subjected to a tensile stress of $\sigma = 1$ MPa. The material properties are provided in Table \ref{tab:glass}.

\begin{figure}[H]
    \centering
    \subfloat[]{\includegraphics[width=3in]{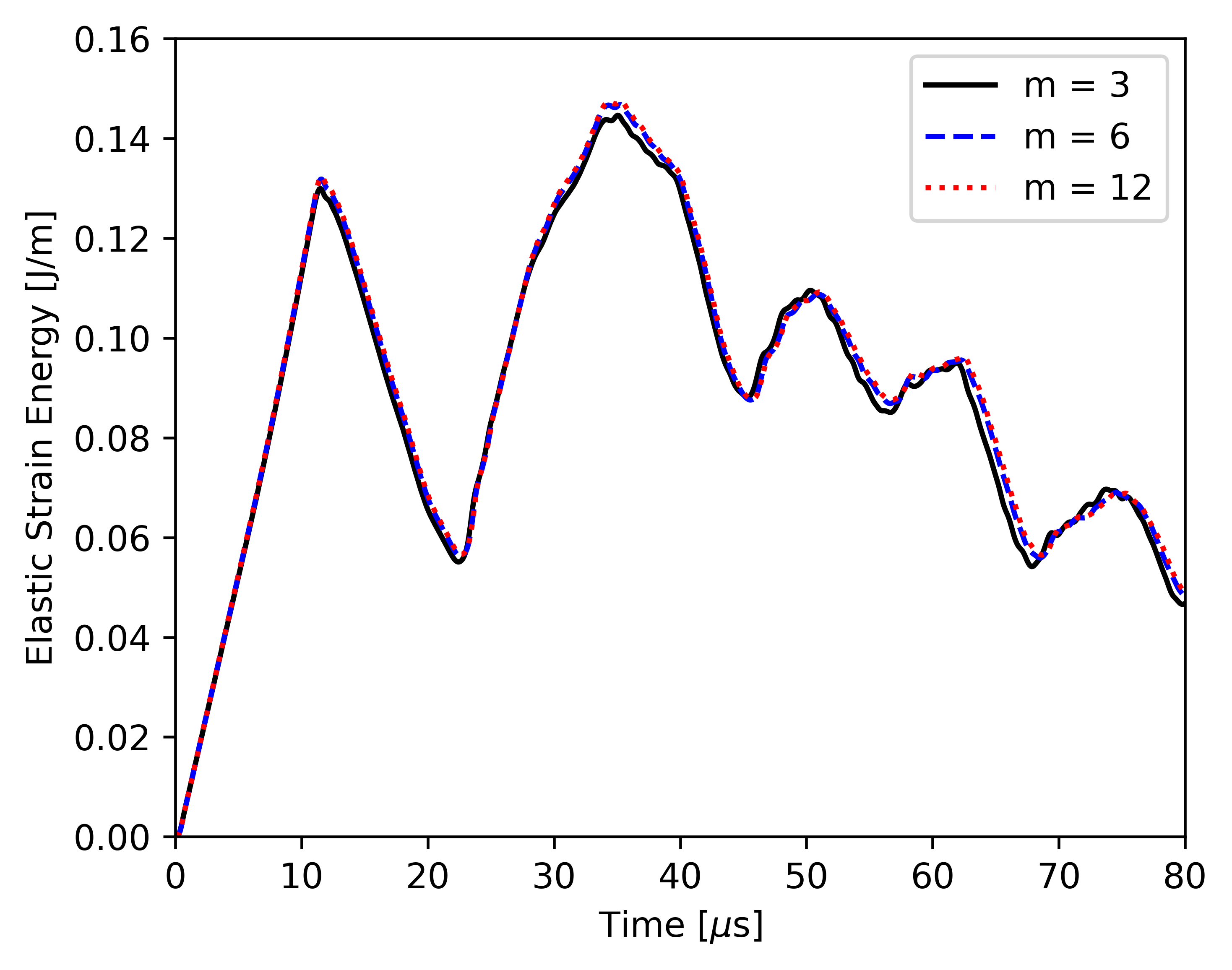}}
    \subfloat[]{\includegraphics[width=3in]{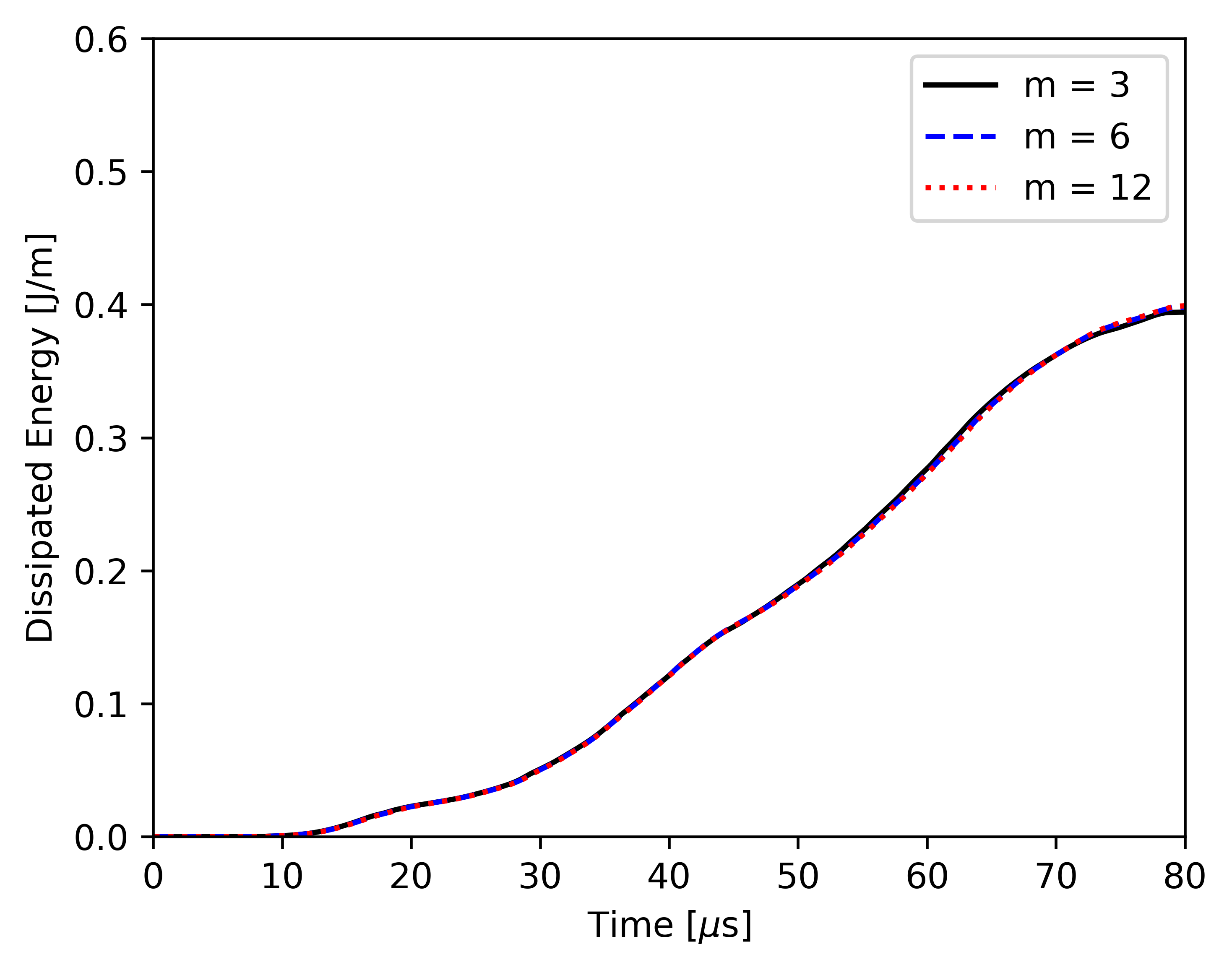}}
 \caption{%
    $m$-convergence: (a) elastic strain energy and (b) dissipated energy for Mesh~I ($m = 3$), Mesh~II ($m = 6$), and Mesh~III ($m = 12$) at fixed
    $\delta = 0.75$~mm.}
  \label{fig:convergence_mesh}
\end{figure}

\begin{table}[h]
  \centering
  \caption{Material properties of the specimen.}
  \label{tab:glass}
  \begin{tabular}{lll}
    \hline
    Symbol & Value & Units \\
    \hline
    $\rho$    & 2450   & kg/m$^3$ \\
    $E$       & 32     & GPa \\
    $\sigma_{tf}$ & 4.5   & MPa \\
    $\mathcal{G}_c$ & 3 & N/m \\
    \hline
  \end{tabular}
\end{table}
Two types of convergence were considered: $ m $ - convergence and $\delta$ - convergence, where $\epsilon = m h$. For $m$ convergence, the horizon was fixed at $ 0.75$, and the grid size $h$ was changed, where Mesh I = $ 0.25$, Mesh II = $ 0.125$, and Mesh III = $0.0625$ mm. The characteristic length $L$ was chosen to be close to the horizon, and it was not treated as a different mesh parameter. 
Figure \ref{fig:convergence_mesh} shows that the elastic strain energy and dissipated energy are nearly identical for three meshes, with the $m$ = 6 and $m$ = 12 curves, indicating that convergence has been reached at $m$ = 6.

\begin{figure}[H]
    \centering
    \subfloat[]{\includegraphics[width=3in]{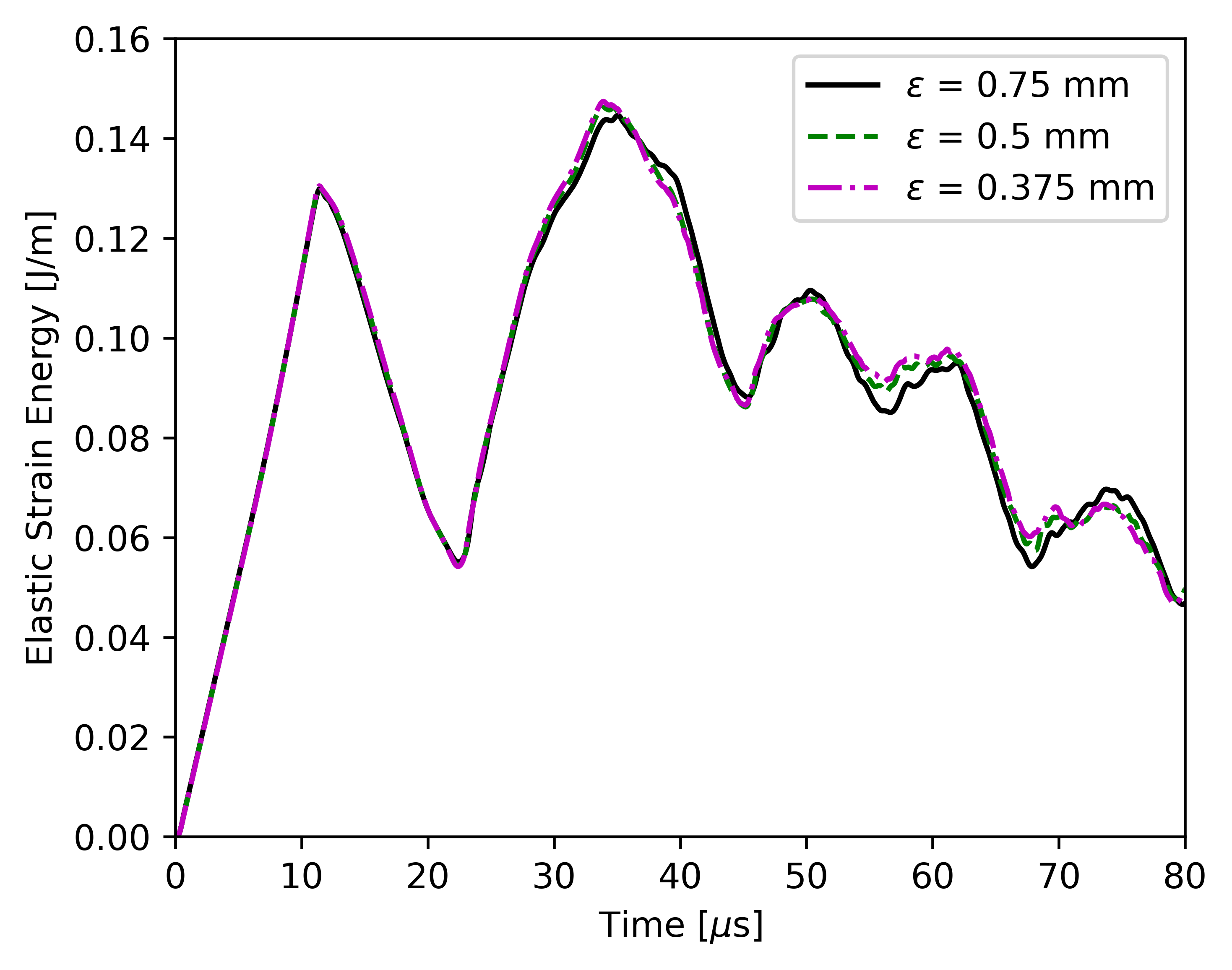}}
    \subfloat[]{\includegraphics[width=3in]{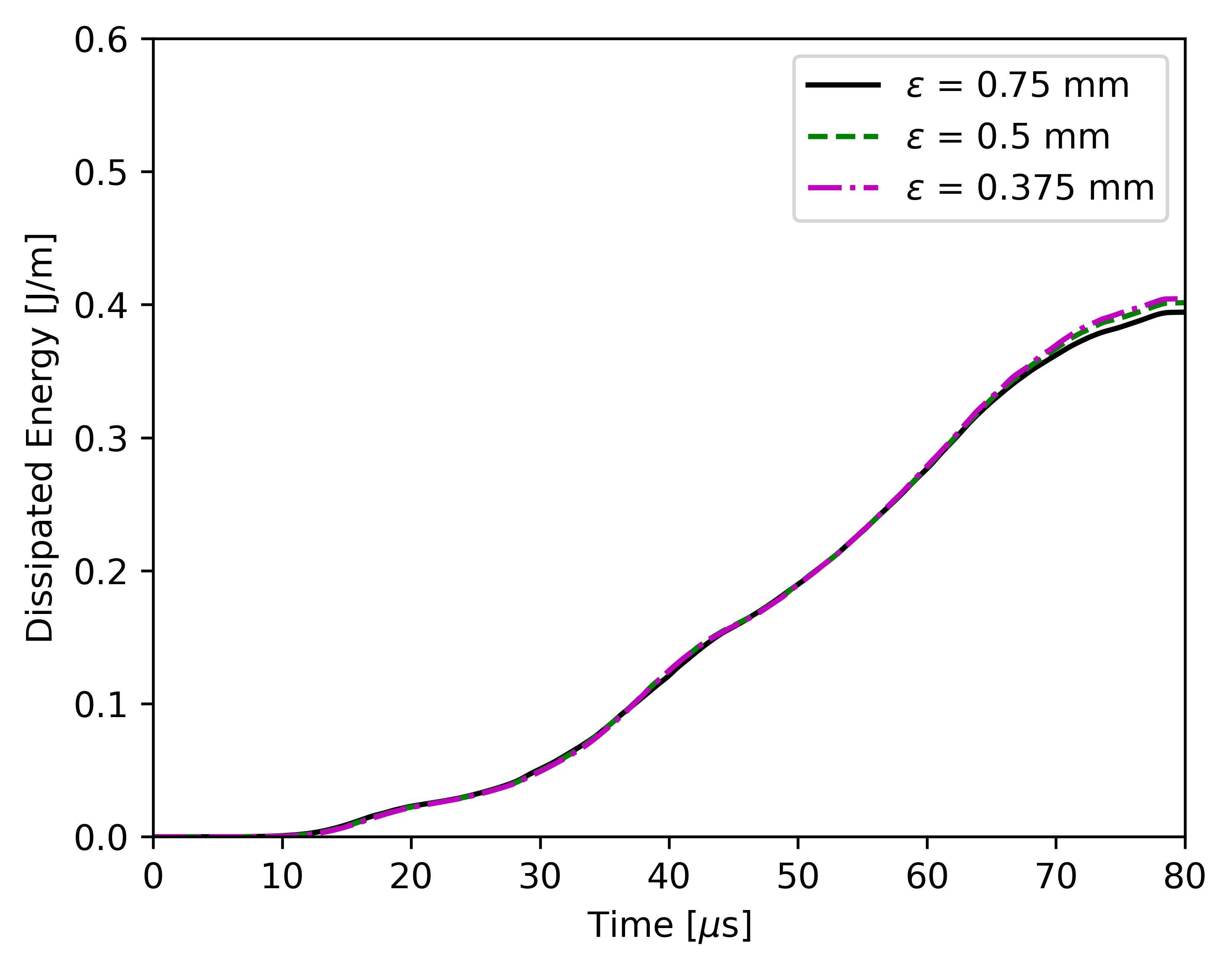}}
 \caption{%
    $\epsilon$-convergence: (a) elastic strain energy and (b) dissipated energy for $\epsilon = 0.75$, $\epsilon = 0.5$, and $\epsilon = 0.375$ at fixed
    $m = 3$.}
  \label{fig:convergence_delta}
\end{figure}

For $\epsilon$ - convergence, $m$ was fixed at 3 and the $\epsilon$ was changed. Three different values of $\epsilon$ were chosen: $0.75$, $0.5$, and $0.375$ mm. Figure \ref{fig:convergence_delta} shows the close agreement among the three curves in both quantities. After branching, the dissipated energy increases with decreasing $\epsilon$ which indicates consistency with the small horizon limit theory. 

\subsection{Biaxial Strength of titania}
\label{sec: Ely}
%Ely (1971)
\cite{ely1972strength} reported axial and hoop strength for titania and aluminum silicate in various biaxial stress states at room temperature. Here, the numerical tests are considered for Tensile - Tensile (TT) and Tensile -  Compression (TC) stress states for titania (AlSiMag 192). Only limitated information was provided regarding the fracture modes, so this section compares the predicted damage onset stresses with Ely's data using the proposed model. 

\begin{figure}[h]
  \centering
  \includegraphics[width=1.5in]{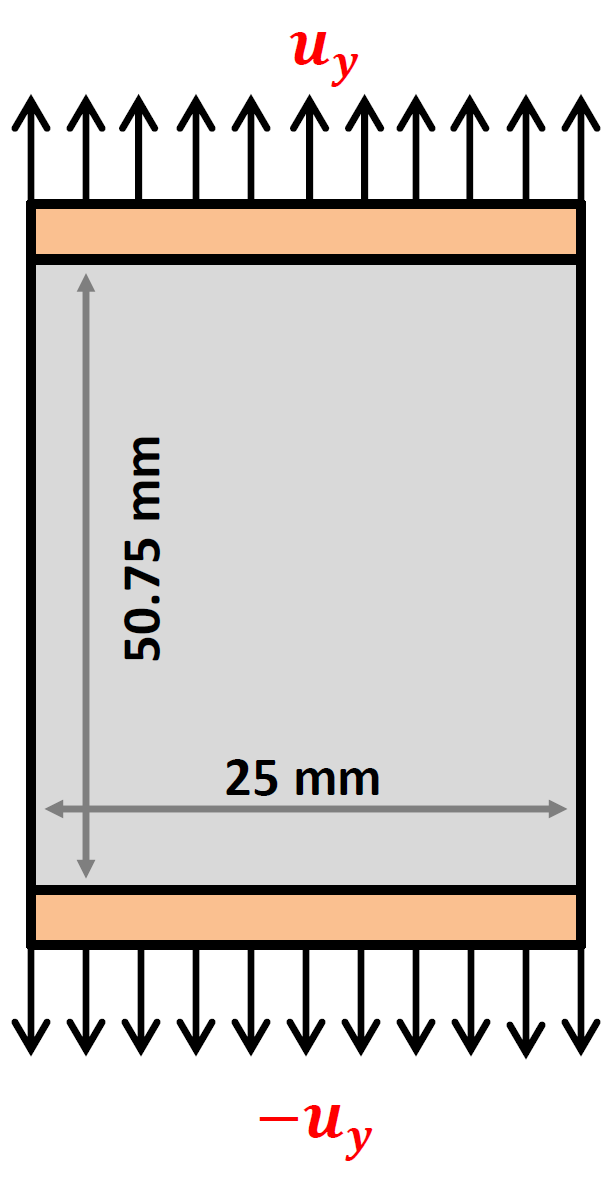}
  \caption{%
    Dimensions and boundary conditions of thin tube wall.}
  \label{fig:ely}
\end{figure}

The specimens in the actual test were thin walled tube. However, we modeled it as a flat patch of a thin tube wall in plane stress. The dimensions of the domain and loading conditions for uniaxial tension are provided in Figure \ref{fig:ely}. The dimensions were consistent throughout  the TT cases; however, the gage length was reduced to $38$ mm ($1.5$ in) for the TC cases. The loading conditions were changed for different biaxial stress states. An elastic layer surrounds the domain in the loading regions of $1$ mm in all cases.
The material properties are selected to be consistent with \cite{lopez2025classical}. However, the Poisson's ratio and compressive strength were chosen to be 0.20 and 1236 MPa, respectively, consistent with Ely's experiment. All the material properties are listed in Table \ref{tab:titania_properties}.
\begin{table}[h]
  \centering
  \caption{Material constants for titania.}
  \label{tab:titania_properties}
  \begin{tabular}{lll}
    \hline
    Symbol & Value & Units \\
    \hline
    $E$          & 250     & GPa \\
    $\nu$        & 0.20    &  \\
    $\sigma_{tf}$   & 100     & MPa \\
    $\sigma_{cf}$   & 1236    & MPa \\
    \hline
  \end{tabular}
\end{table}
The experiments were conducted in a force (pressure) controlled manner. However, before damage, every frame is elastic, so we prescribed displacement controlled quasi static loading using the staggered dynamic relaxation algorithm in \ref{alg:algorithm}. Each simulation is run until the first damage, i.e. $\gamma_{min} < 1$, and the onset stress is interpolated between the last fully elastic step and the first damage step.

In the four TT cases, the hoop and axial strains were prescribed proportionally, where the strain ratio was computed from Ely's stresses following the plane stress Hooke's law. The TC cases followed three loading paths: proportional (P), compression, then tension (CT), and proportional, then compression (PC) with tension in the hoop direction and compression in the axial direction. Under P loading, axial compression and hoop tension were applied at a fixed strain ratio from zero to $1.25$ times Ely's stress. Under CT loading, uniaxial compression was applied in 50 step size until Ely's axial stress and then hoop stress was applied in 125 increments at fixed axial stress. Under PC loading, both axial and hoop loads were increased proportionally over 50 increments until hoop tension reached Ely's hoop stress and axial compression reached one-half of Ely's axial stress. Then, the hoop stress was fixed and the axial compression was increased to 1.25 times Ely's axial stress.

\begin{figure}[H]
    \centering
    \subfloat[]{\includegraphics[width=3in]{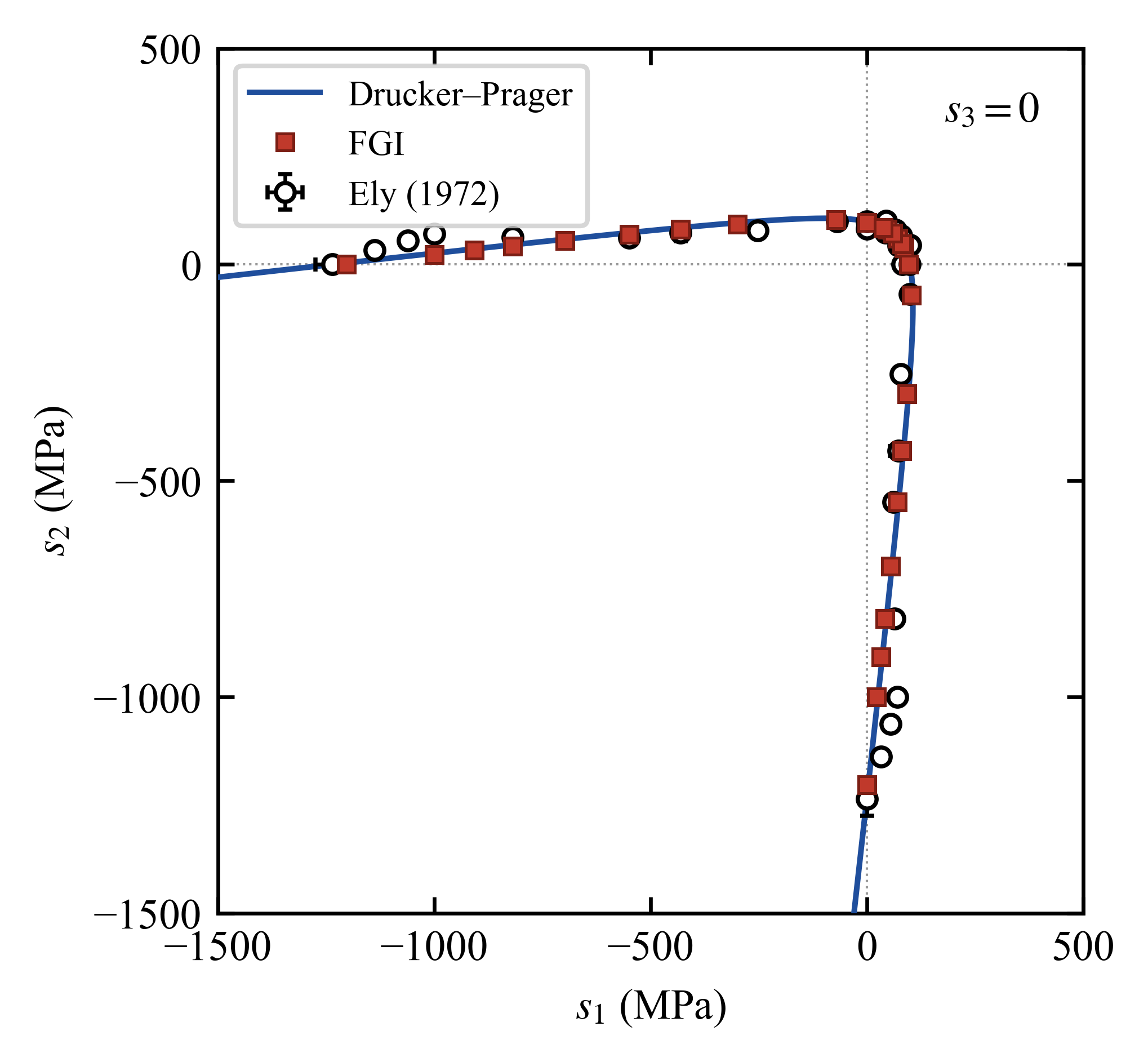}}
    \subfloat[]{\includegraphics[width=3in]{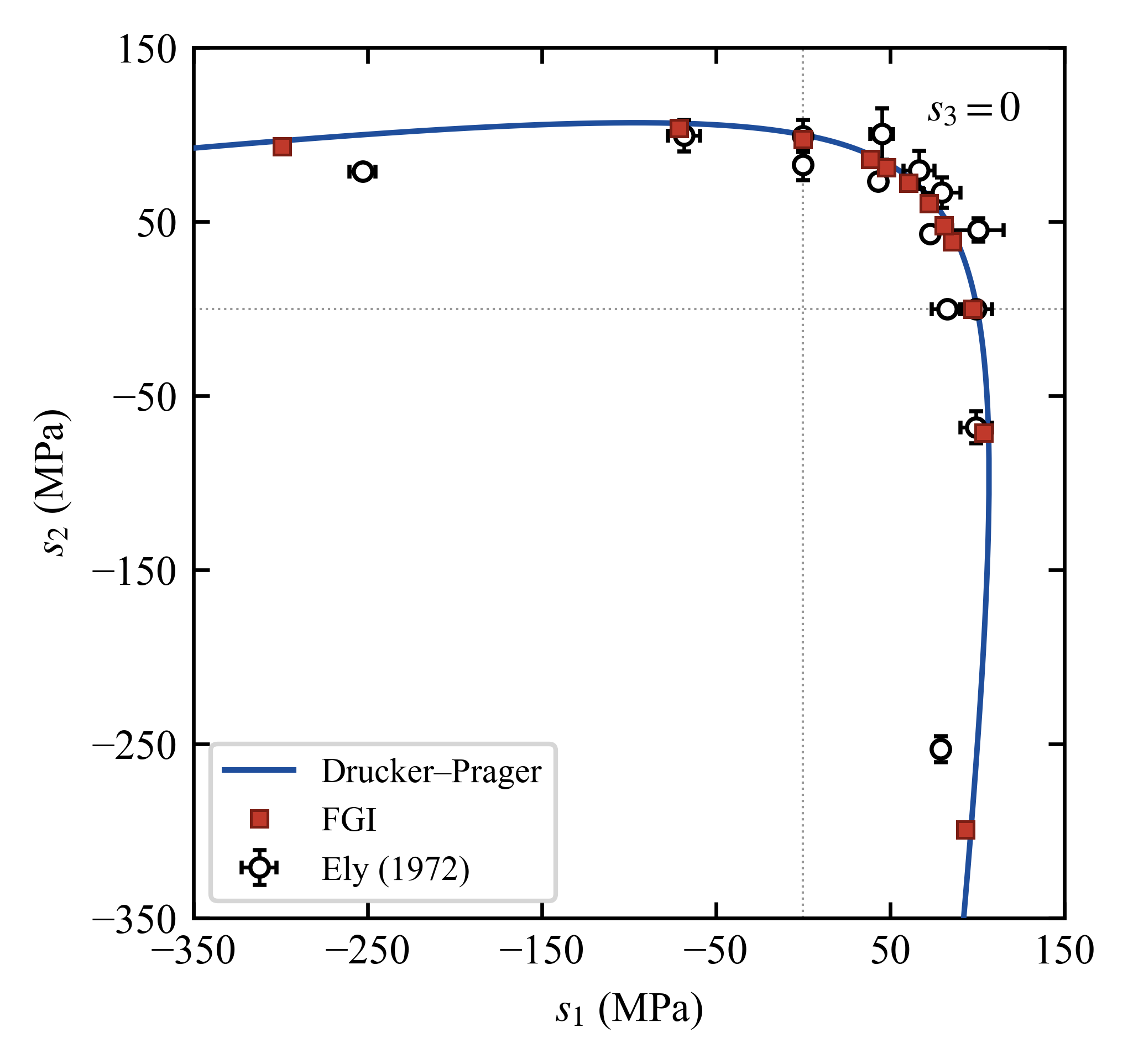}}
 \caption{%
  Frame gradient inelasticity (FGI)  predicted failure stresses compared with the DP failure surface and Ely's experimental data. (a) TT-TC regions (b) TT region}
  \label{fig:Ely_principle}
\end{figure}

Figure \ref{fig:Ely_principle} shows the predicted onset stresses compared with the Drucker - Prager (DP) failure surface and Ely's experimental data in the principal stress plane (from \cite{kumar2020revisiting}). The prediction of  frame gradient inelasticity  falls on the DP curve in all cases, confirming that the frame discretization does not introduce any numerical artifacts in the material strength.

Although Ely's stresses are near the DP curve in the TT region, the model as expected was not able to reproduce the asymmetry present in Ely's data where the hoop dominated points were above the curve and axial dominated points were below the curve. In the TC region, exactly under higher compression in the axial direction, Ely's data deviated from the DP curve. For cases 5 and 6, Ely's stresses were 1.4 and 2.7 times higher than the Drucker - Prager strength. Similarly, in cases 7 and 8, the difference was 1.45 and 1.21 times the DP value at the same hoop stress. Ely considered his uniaxial compression result unreliable because of the end constraints in the short tubes. The DP surface used here is calibrated using the lower uniaxial compressive strength which is 1236 MPa, which might be the reason why it underestimates the strength under high axial compression. 

\subsubsection{Yield strength versus Griffith fracture criterion}
\begin{figure}[h]
  \centering
  \includegraphics[width=3.5in]{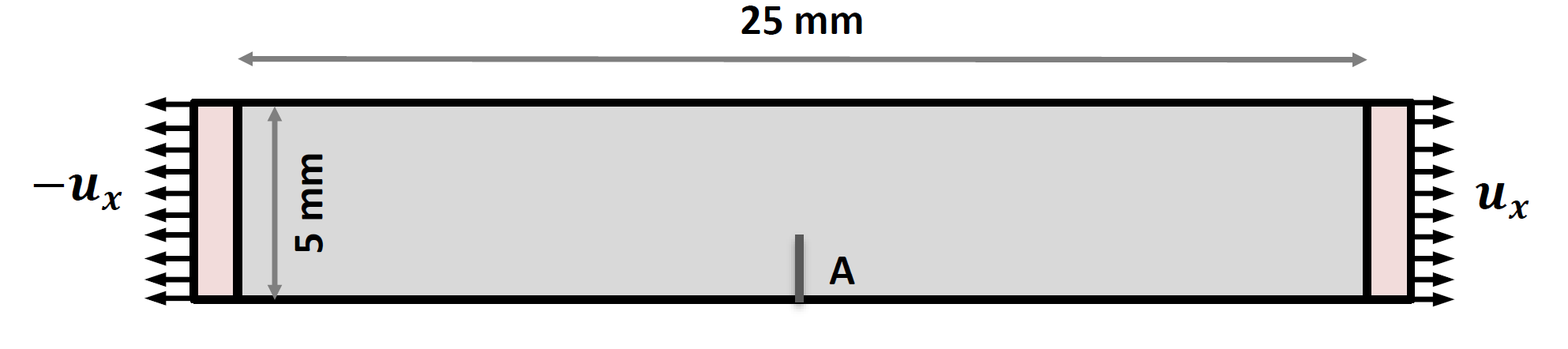}
  \caption{%
    Dimensions of single notch specimen.}
  \label{fig:LEFM_Geo}
\end{figure}

We further extend the numerical study to validate the fracture criterion based on yield strength and Griffith energy. We considered a single-edge notched tension test in \cite{kamarei2026nine}, \cite{bromleylipton}. The specimen has a length (L) of 25 mm, height (H) of 5 mm, and thickness (t) of 0.25 mm that contains a pre-existing crack with a length of A = 0.05, 0.1, 0.25, 0.5, 1, 1.5, and 2 mm.  The strips are clamped at the left and right boundaries and pulled apart by an applied displacement as shown in Figure \ref{fig:LEFM_Geo}. 
The reaction force is denoted by $F$ and the stress is calculated as $\sigma = \frac{F}{L\times H}$. As the applied displacement is increased, the strips deform elastically until a critical value at which the crack starts to propagate. The corresponding critical stress at this stage is bounded by the material tensile strength $\sigma_{tf}$ and critical stress $\sigma_{GC}$ when the Griffith condition is satisfied.  Here, the objective of this numerical analysis is to examine the dependence of the critical stress on the initial crack length such that for sufficiently small flaws, fracture is expected to be governed primarily by the material yield strength and as the crack length increases, the critical stress of the specimen begins to decrease.  We first perform the analysis on soda lime glass following the approach in \cite{kamarei2026nine} as a validation. Then we extend it to titania. Since titania is modeled here as isotropic, linear-elastic, and brittle using the same geometric setup, we chose the same closed form bound given by the equation (15) in the reference study. The material properties used in the simulations are listed in Table \ref{tab:titania_properties}. A grid size of 0.05 mm and a horizon of 0.15 mm were used for the both analyses. 

\begin{figure}[H]
    \centering
    \subfloat[]{\includegraphics[width=3in]{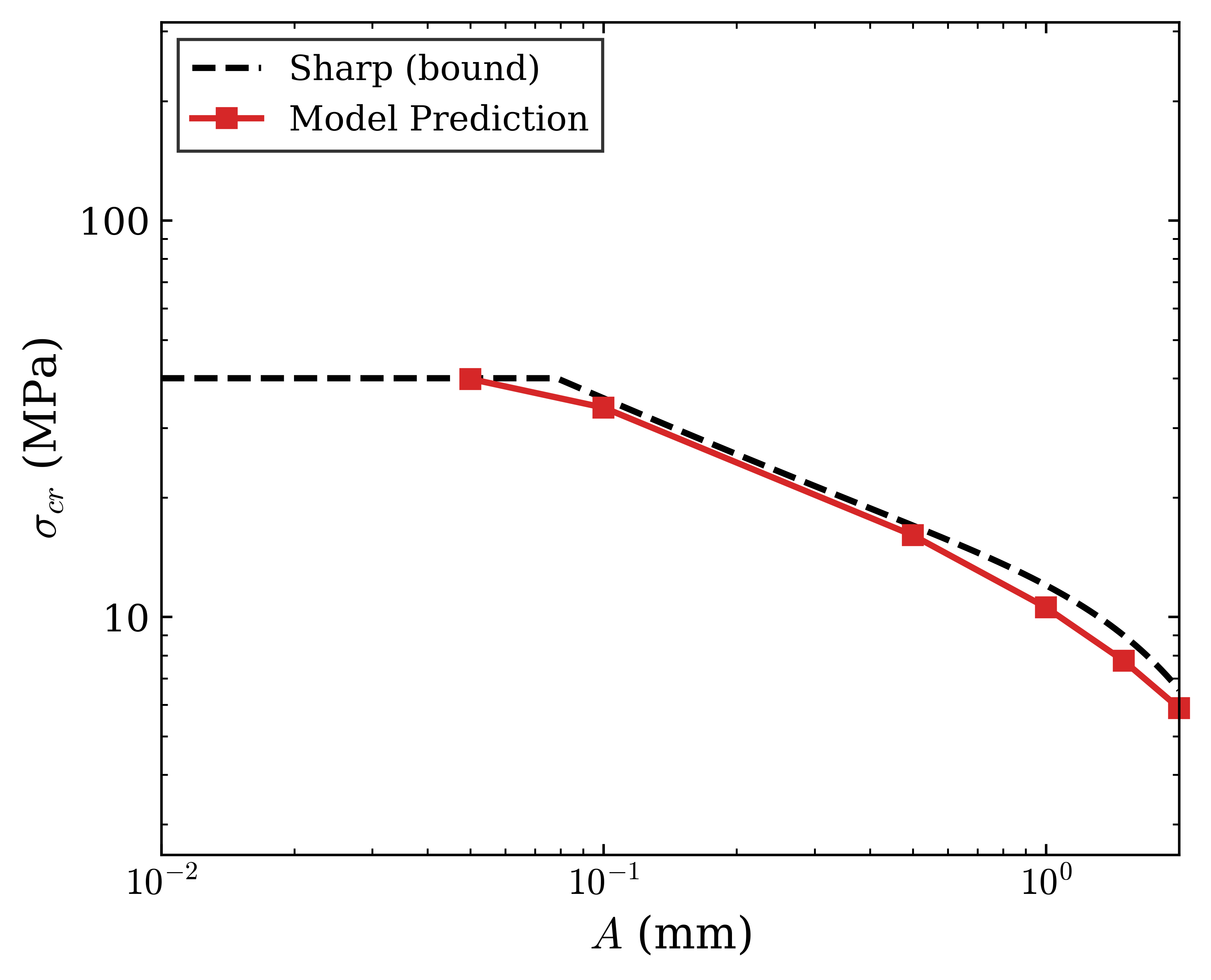}}
    \subfloat[]{\includegraphics[width=3in]{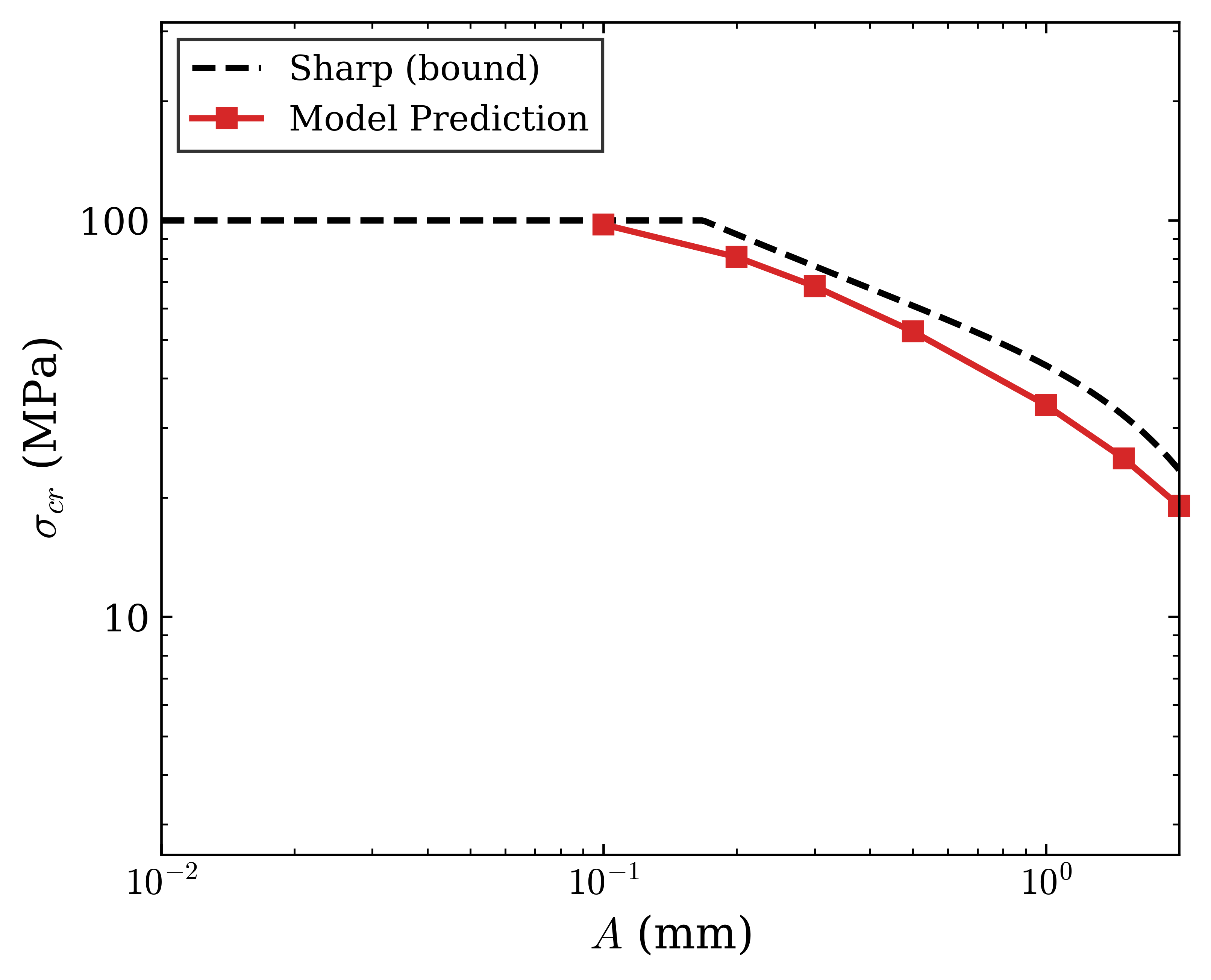}}
 \caption{%
    Comparison between the critical stress for the Sharp bound and model prediction. (a) Soda Lime Glass (b) Titania}
  \label{fig:LEFM}
\end{figure}

Figure \ref{fig:LEFM} shows that, in both cases, for small prenotch lengths, the predicted critical stress remains close to $\sigma_{tf}$. As the prenotch length increases, the critical stress values approach $\sigma_{GC}$.  This indicates that the model captures the expected transition from tensile-strength-controlled fracture for small flaws to LEFM-controlled behavior as the flaw size increases.

\subsection{Dynamic crack branching in glass sheet}
Numerous studies have been conducted on dynamic crack branching in brittle materials to understand the failure mechanisms \cite{borden2012phase, song2008comparative, bobaru2015cracks,bleyer2017dynamic,rakici2023discrete,coskun2025quasi,Partmann}. In this section, we apply the proposed frame gradient model to crack branching in glass and compare the results with those of \cite{borden2012phase}.

\begin{figure}[h]
  \centering
  \includegraphics[width=2.5in]{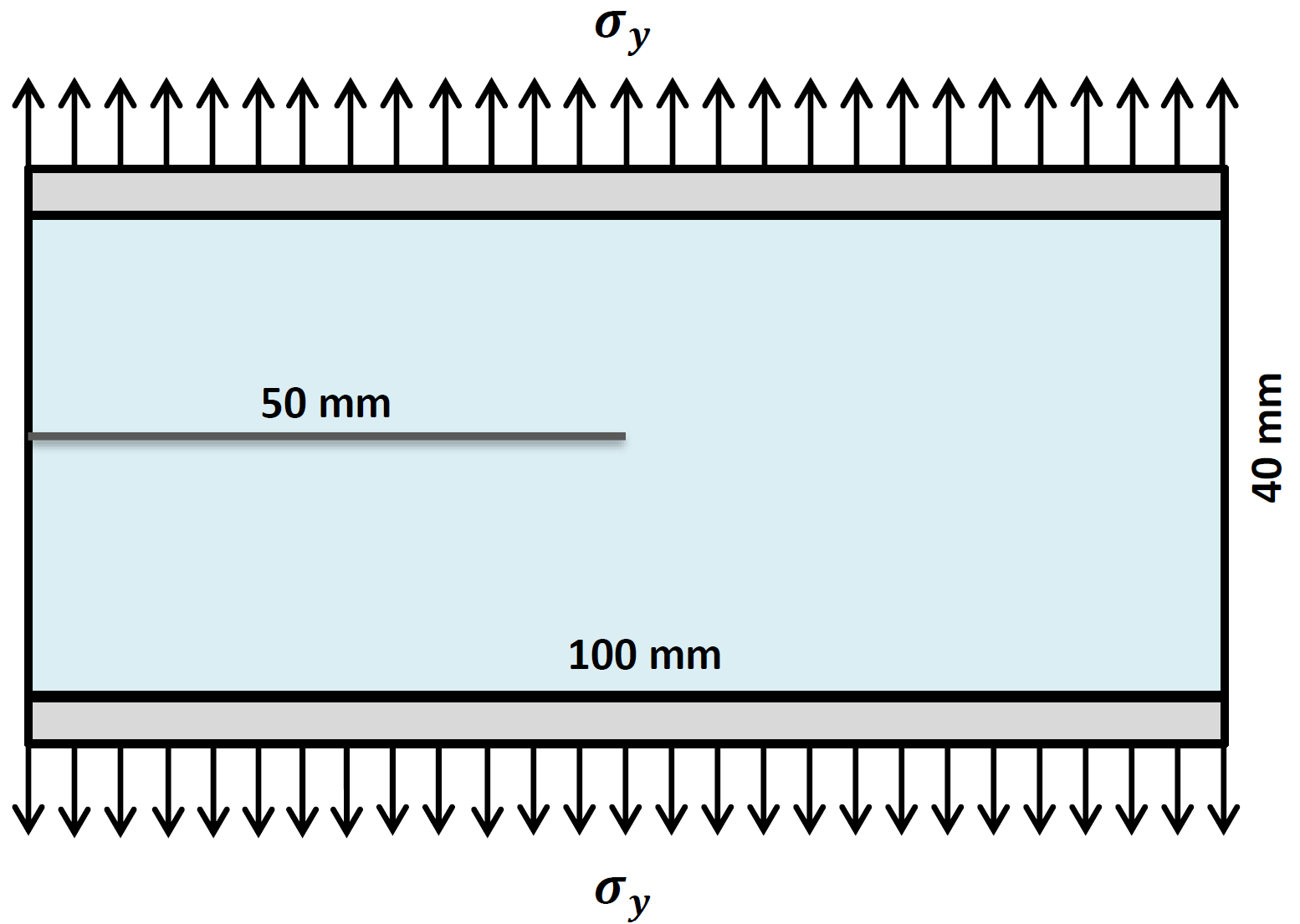}
  \caption{%
    Dimensions and boundary conditions of glass sheet.}
  \label{fig:glass}
\end{figure}

For the numerical simulation, a rectangular specimen with dimensions $100$ mm $\times$  $40$ mm is considered, consistent with the literature. The material domain has a pre-notch of length $50$ mm at the mid-height on the left edge, as shown in Figure \ref{fig:glass}.
A uniform tensile stress of 1 MPa is applied along the top and bottom edges in a 1 mm elastic loading region. The material properties are shown in Table \ref{tab:glass_properties}. These material properties are used in the studies by \cite{song2008comparative, borden2012phase, coskun2025quasi}. However, the tensile strength used in all these studies was different. In this case, we chose the critical strength of the material 4.5 MPa by using equation (27) in \cite{borden2012phase} with the 0.25 mm length scale used in their study. 
\begin{table}[h]
  \centering
  \caption{Material constants for Glass Sheet.}
  \label{tab:glass_properties}
  \begin{tabular}{lll}
    \hline
    Symbol & Value & Units \\
    \hline
    $\rho$          & 2450     & $\text{kg}\cdot\text{m}^{-3}$ \\
    $E$          & 32     & GPa \\
    $\nu$        & 0.20    &  \\
    $\sigma_{tf}$   & 4.5     & MPa \\
    $\sigma_{cf}$   & 1000    & MPa \\
    \hline
  \end{tabular}
\end{table}

Similar to the studies by \cite{borden2012phase, coskun2025quasi}, three different uniform grid spacings, $h$, are used: 0.25, 0.125, and 0.0625 mm. The horizon $\epsilon$ and the length $L$ are fixed at 0.75 mm for each grid spacing. Explicit Velocity Verlet is used for time integration with a time step of 0.85 times the critical time step where the final time steps, $\Delta t$, are 95, 89, and 87 $ns$.

The crack region was determined using a damage threshold criterion such that the damage at node I is the lost fraction of its weighted neighborhood, given by:
\begin{equation}
  \gamma_i = 1 - \frac{\sum \gamma_f V_f}{\sum V_f},
\end{equation}

\noindent which was selected as 0.7. Nodes with damage values exceeding this threshold limit were identified as a part of the crack region. However, in the bond based model, a node loses half of its neighbors, which makes the damage about 0.5, therefore, a threshold of 0.38 was used in \cite{coskun2025quasi}. In the proposed model, each frame has orthogonal legs and if any of the legs crosses the crack, the frame gets damaged. Thus, a node close to straight crack loses about 3/4 of its frames instead of half, resulting in a damage value of 0.75. Similar to \cite{coskun2025quasi}, a local damage threshold of 0.7 is chosen, which is lower than 0.75. Since the coordinates and the corresponding damage values were available as output of the solver, the crack tip was identified as the rightmost x-coordinate where damage exceeded this threshold. Because a frame reaches across a full horizon with both legs, a single crack produces a wider damage band than in the bond-based model. Before branching, the spread in our runs reached about 1.5 mm; therefore we used 3 mm to mark branching. The crack tip velocity was computed by linear regression of the tip position over a 2 $\mu s$ window to reduce noise. After branching, the velocity was computed along the upper branch, in the same way as in \cite{coskun2025quasi,borden2012phase}.

\begin{figure}[H]
    \centering
    \subfloat[]{\includegraphics[width=4.5in]{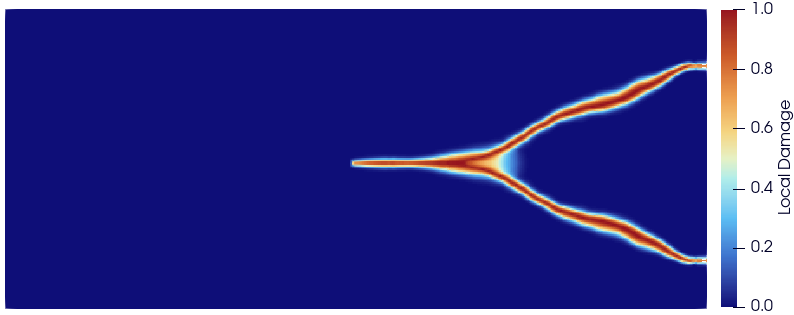}}
    \hfill
    \subfloat[]{\includegraphics[width=4.5in]{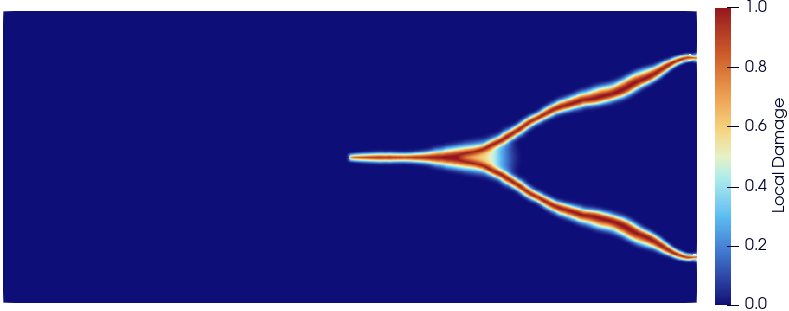}}
    \hfill
    \subfloat[]{\includegraphics[width=4.5in]{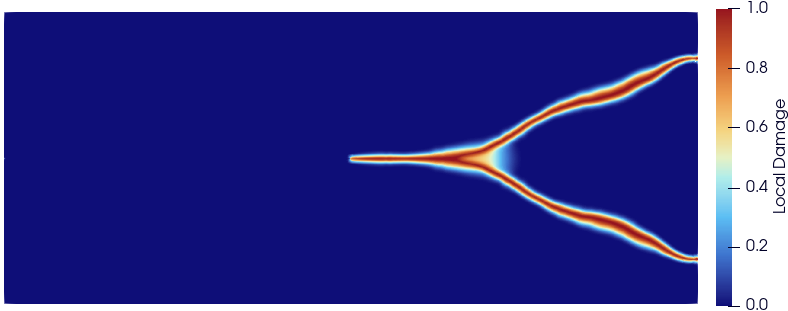}}
 \caption{%
    Damage variable contour plots for (a) Mesh I, and (b) Mesh II, and (c) Mesh III.}
  \label{fig:contour_glass}
\end{figure}

Figure \ref{fig:contour_glass} presents the local damage variable contour plots at the final time step for all three meshes.  Initially, the crack runs straight, but at approximately 30 $\mu s$, it widens and starts to bifurcate forming two symmetric branches at 70 mm at approximately 37 $\mu s$. After branching, each branch runs with an average speed of 855 m/s. Both branches reach the right boundary by $80 \mu s$ for all three meshes. 

\begin{figure}[H]
    \centering
    \subfloat[]{\includegraphics[width=2in]{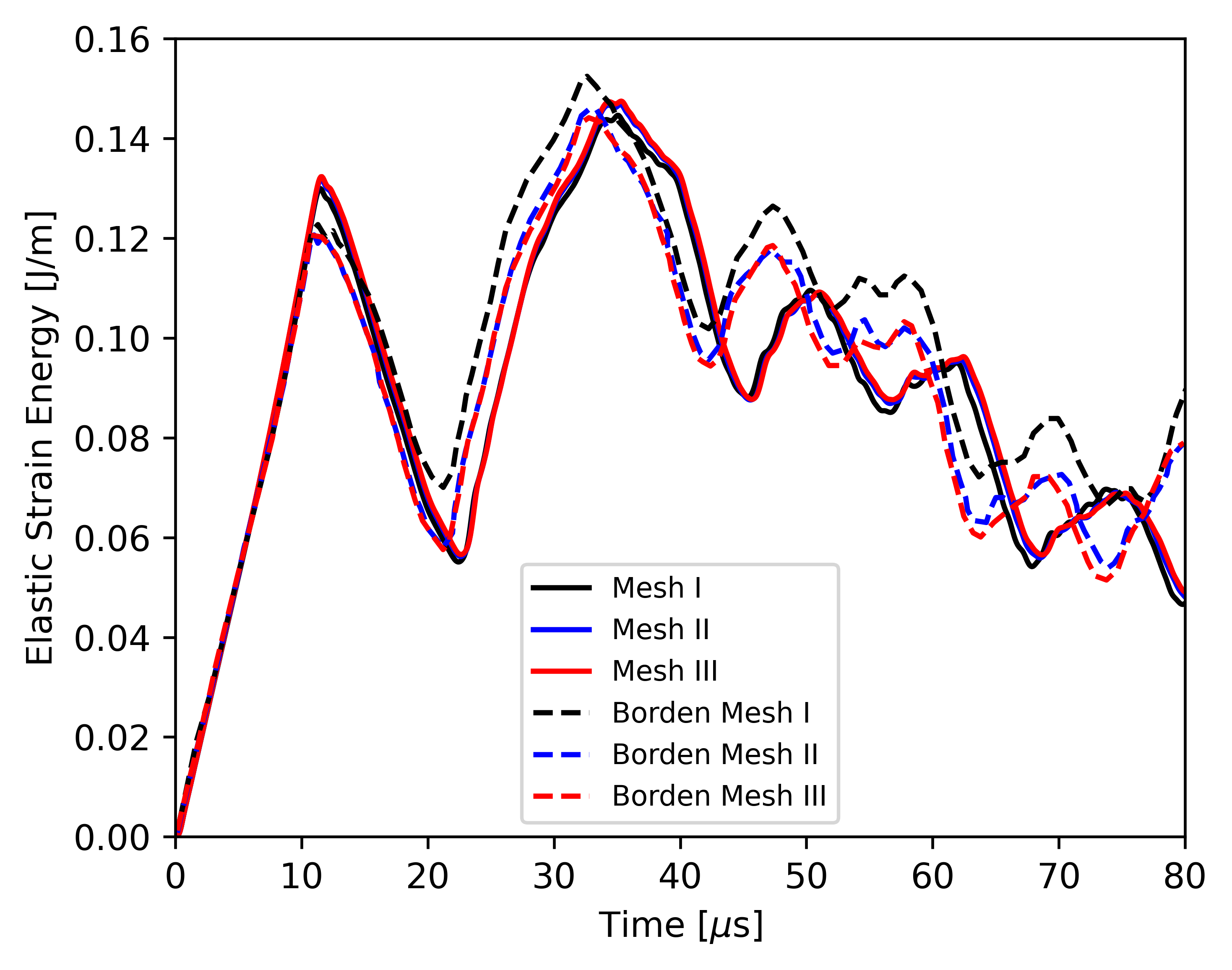}}
    \subfloat[]{\includegraphics[width=2in]{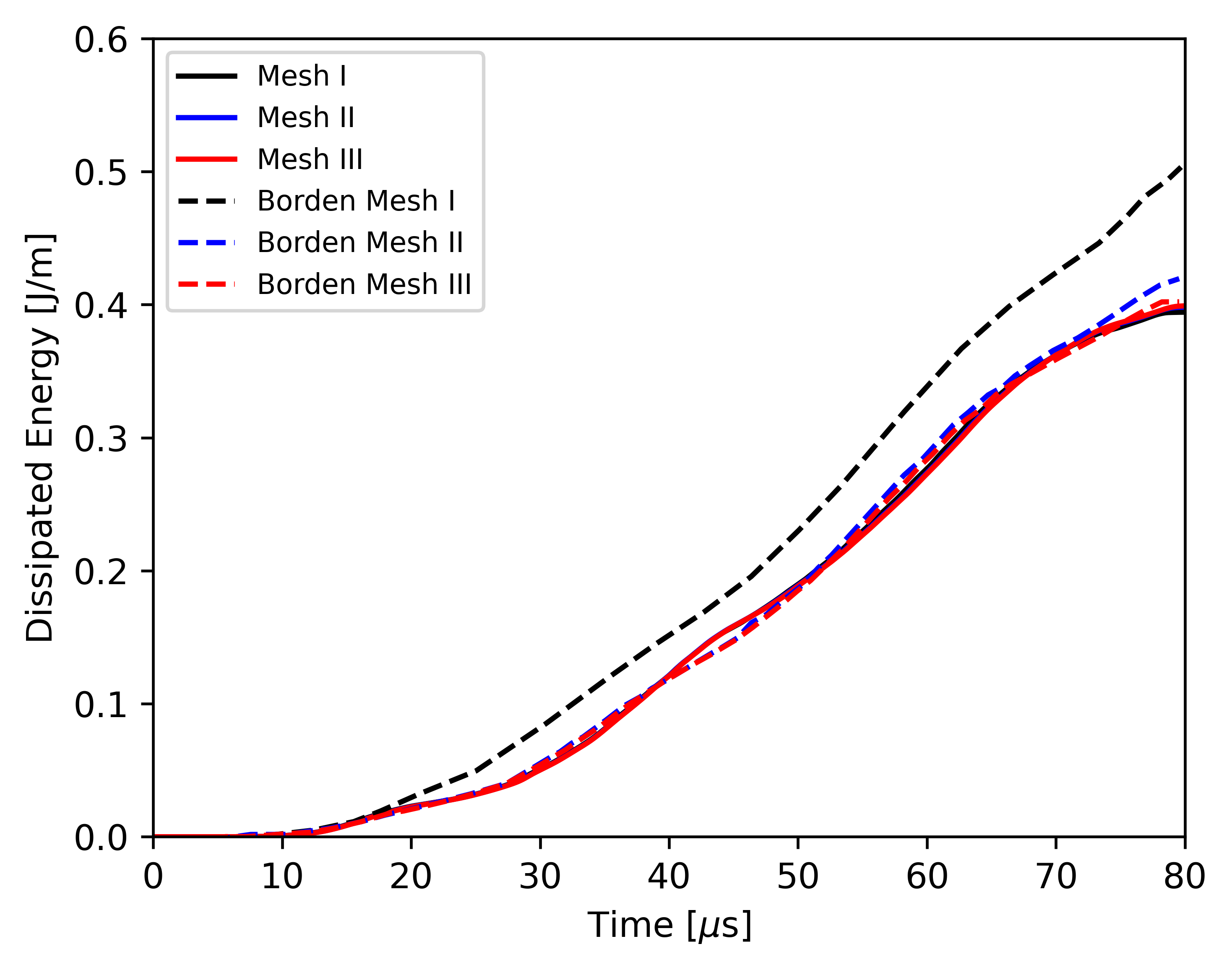}}
    \subfloat[]{\includegraphics[width=2in]{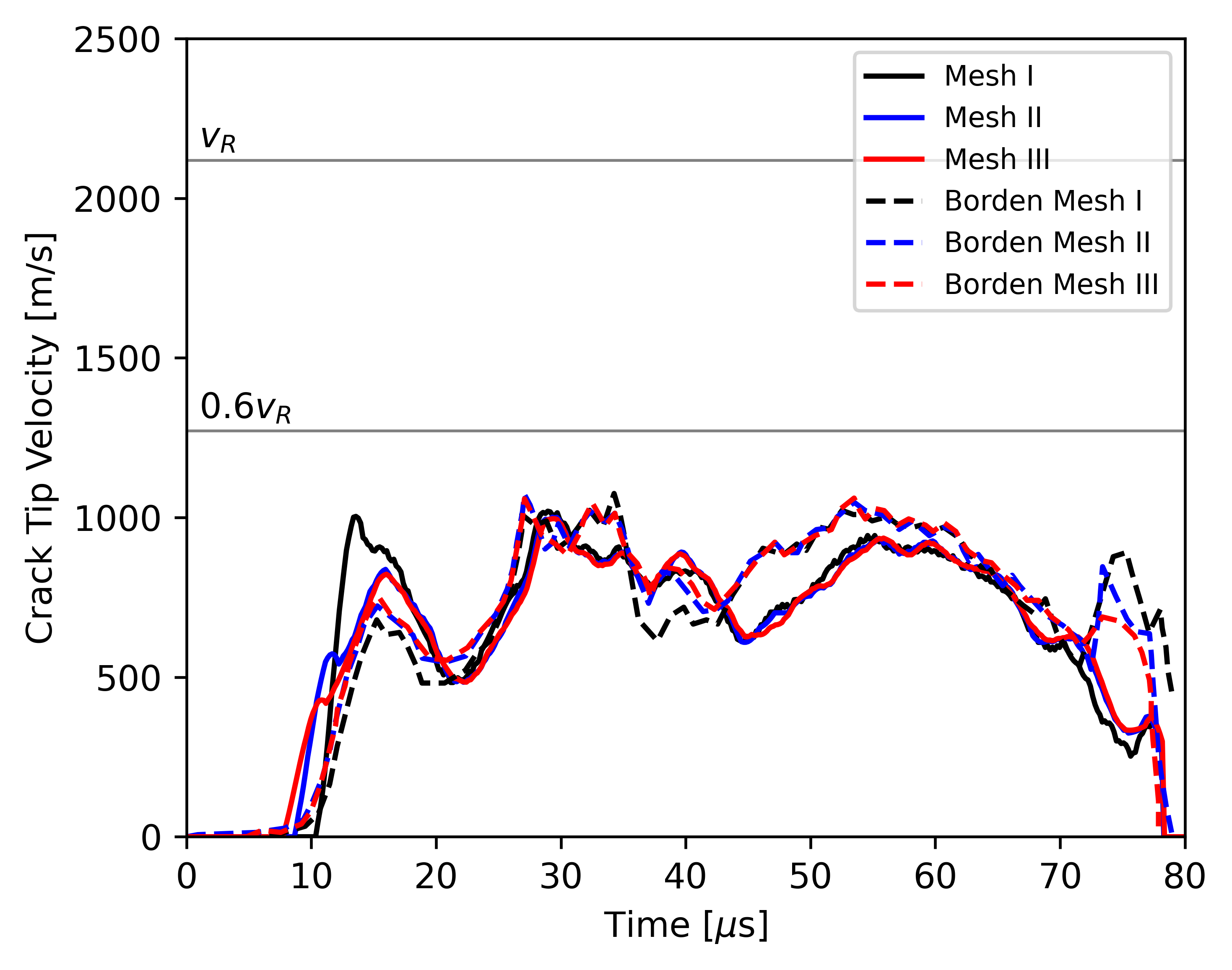}}
 \caption{%
    Comparison of (a) Elastic Strain Eneregy, (b) Dissipated Energy, (c) Crack Tip Velocity with \cite{borden2012phase}}
  \label{fig:comp_bordern}
\end{figure}

Figure \ref{fig:comp_bordern} presents a comparison of the elastic strain energy, dissipated energy, and crack tip velocity between the results of the current study and those of \cite{borden2012phase} for the three different mesh configurations. The elastic energy per unit thickness rises linearly and peaks when the crack starts, as shown in Figure \ref{fig:comp_bordern} (a). The first peak is about 7 \% above the reference study. However, unlike the reference study's mesh I, our coarse mesh does not over predict the elastic strain energy.  The Figure \ref{fig:comp_bordern} (b) shows the dissipated energy for all three meshes are within 2.5 \% after 20 $\mu s$. Additionally, all results lie in around 0.394 - 0.399 J/m energy at 80 $\mu s$, close to the converged value of 0.40 $\mu s$ in the reference study. 

In Figure \ref{fig:comp_bordern} (c), the crack tip velocities in all three meshes are almost identical, indicating that reasonable convergence has been achieved. The convergence study has been performed separately in Section \ref{sec.convergence}.  The crack speed remains below 60 \% of the Rayleigh wave speed (0.6 $v_R$) at all times, consistent with theoretical limit reported in \cite{freund1998dynamic}. Although the widening and branching happen 2 - 5 $\mu s$ later than in \cite{borden2012phase, coskun2025quasi}, these are defined using different  thresholds.

The numerical studies show that the frame gradient model can capture the dynamic crack branching phenomenon in a glass sheet. Dynamic fracture experiments in glass can exhibit complex and unstable crack propagation and micro-branching. However, our numerical simulation showed a perfectly straight crack propagation from the prenotch followed by symmetric crack branching. This is  attributed to the perfectly symmetric geometry, loading, and initial boundary conditions, as well as the absence of material imperfections.

\subsection{Dynamic fracture of concrete}
\begin{figure}[H]
    \centering
    \subfloat[]{\includegraphics[width=2in]{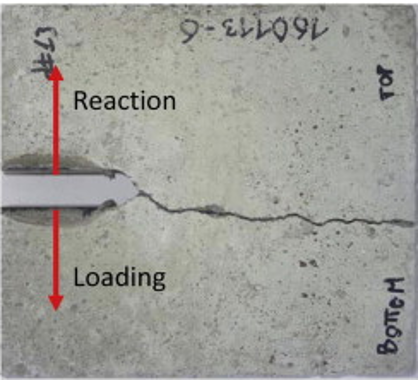}}
    \hfill
    \subfloat[]{\includegraphics[width=2in]{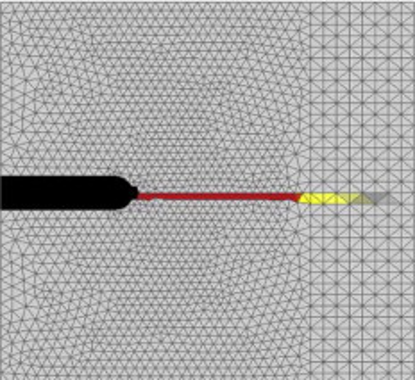}}
    \hfill
    \subfloat[]{\includegraphics[width=2in]{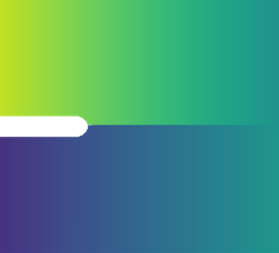}}

    \caption{%
     Local damage contour plots showing the crack propagation in (a) Experimental Test in \cite{ovzbolt2013dynamic}, (b) Numerical Test in \cite{ovzbolt2013dynamic}, (c) Frame Gradient Inelasticity (Current Study) for Test 2.(Figures (a) and (b) reprinted from \cite{ovzbolt2013dynamic} with permission from Elsevier.)}
    \label{fig:test_2}
\end{figure}

\begin{figure}[H]
    \centering
    \subfloat[]{\includegraphics[width=2in]{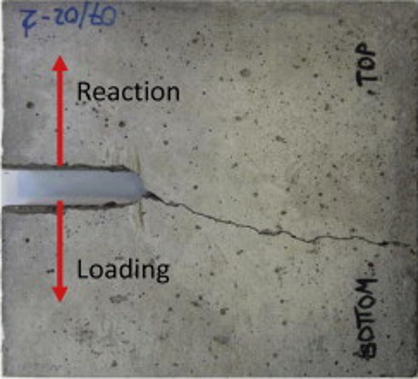}}
    \hfill
    \subfloat[]{\includegraphics[width=2in]{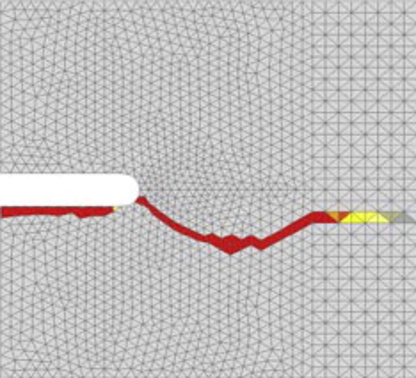}}
    \hfill
    \subfloat[]{\includegraphics[width=2in]{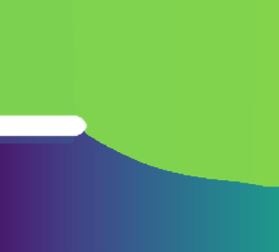}}

    \caption{%
     Local damage contour plots showing the crack propagation in (a) Experimental Test in \cite{ovzbolt2013dynamic}, (b) Numerical Test in \cite{ovzbolt2013dynamic}, (c) Frame Gradient Inelasticity  (Current Study) for Test 5. (Figures (a) and (b) reprinted from \cite{ovzbolt2013dynamic} with permission from Elsevier.)}
    \label{fig:test_5}
\end{figure}

\begin{figure}[H]
    \centering
    \subfloat[]{\includegraphics[width=2in]{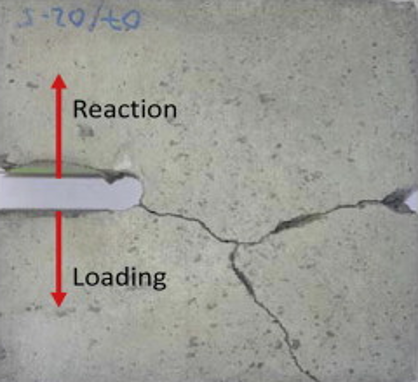}}
    \hfill
    \subfloat[]{\includegraphics[width=2in]{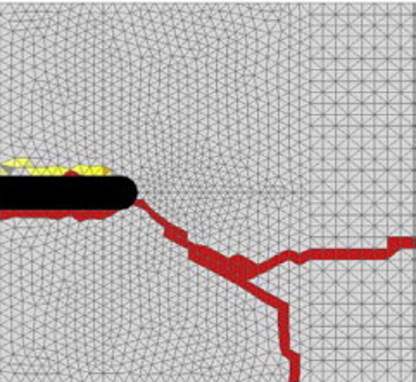}}
    \hfill
    \subfloat[]{\includegraphics[width=2in]{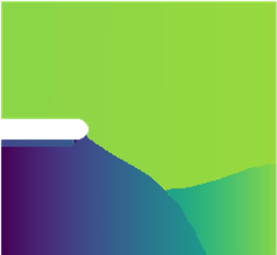}}
    \caption{%
     Local damage contour plots showing the crack propagation in (a) Experimental Test in \cite{ovzbolt2013dynamic}, (b) Numerical Test  in \cite{ovzbolt2013dynamic}, (c) Frame Gradient Inelasticity  (Current Study) for Test 12. (Figures (a) and (b) reprinted from \cite{ovzbolt2013dynamic} with permission from Elsevier.)}
    \label{fig:test_12}
\end{figure}

In the final numerical example, the proposed model is used to simulate the dynamic fracture in a compact tension specimen (CTS). Here as in \cite{ovzbolt2013dynamic}   the effect of loading rate is examined. The reference experiments were conducted by the authors of \cite{ovzbolt2013dynamic} to validate their numerical study in which a notched concrete specimen was loaded at three different displacement rates. The numerical study of \cite{ovzbolt2011dynamic} is based on a rate dependent microplane model. The crack pattern changed with the loading rate: a straight crack at low rates, a single inclined crack at intermediate rates, and a Y shaped branching crack at the highest rate (see Figure \ref{fig:test_2} - \ref{fig:test_12}). 

\begin{figure}[h]
  \centering
  \includegraphics[width=3in]{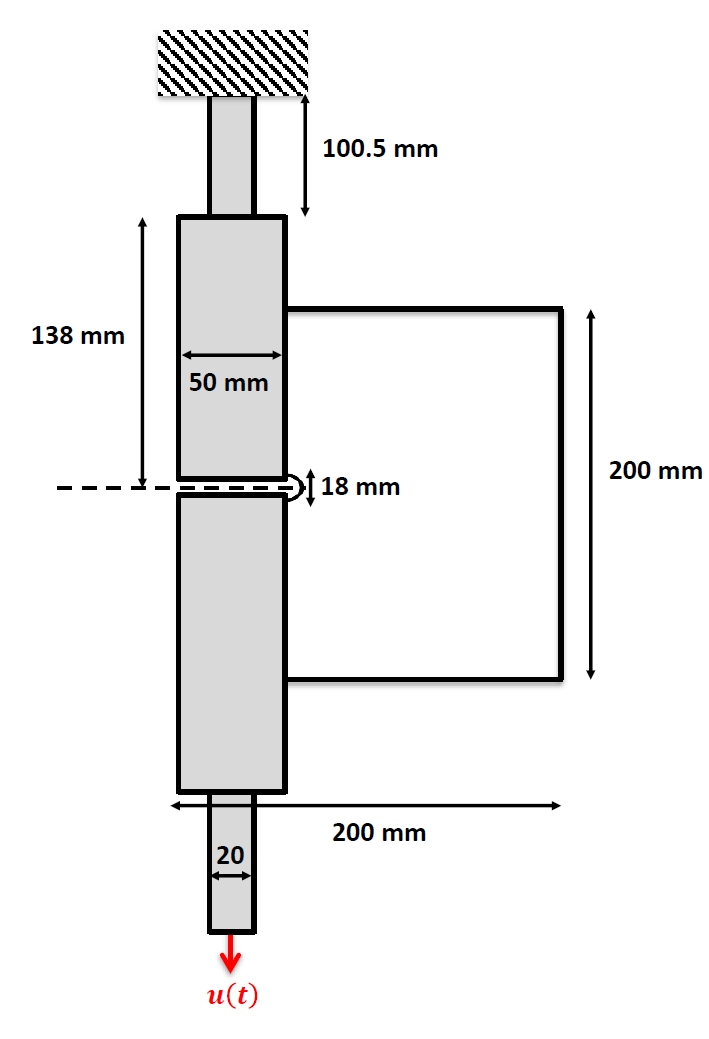}
  \caption{%
    Dimensions of compact tension specimen.}
  \label{fig:concrete}
\end{figure}

The specimen has dimensions of 200 $\times$ 200 $\times$ 25 mm with a prenotch 64 mm long and 18 mm wide with a rounded tip of 9 mm radius at the middle of the left edge. The geometry and boundary conditions are shown in Figure \ref{fig:concrete}. The experimental and numerical studies in \cite{ovzbolt2013dynamic, ovzbolt2011dynamic} were conducted in a three dimensional domain. However, in the present study, we consider a two dimensional plane stress case.
A simpler approach would be to prescribe the displacement directly on the notch surfaces without modeling the entire steel frame. However, in the experiment, the glued steel plates stiffen the notch faces and the rod transmits the load from a rod end far from the specimen. This means that the loading rate on the concrete surfaces differs from that at the rod end. The authors \cite{ovzbolt2013dynamic, ovzbolt2011dynamic} modeled the steel frames in their finite element simulations. Therefore, we modeled two steel loading frames and considered them fully elastic in this study. The upper frame is fixed and the reaction is measured there and the lower frame is pulled downward under displacement control. Only at the notch surfaces, the steel frames surfaces are attached to the concrete material where they are connected by frames through common grid points. Although the steel frames surround (as rings) the concrete arms, there were no frames connecting the steel nodes to the concrete nodes. Thus, the load goes from the end of the steel rod through the ring into the glued plate and finally into the concrete through the joint frames. The joint frames are assigned with concrete material properties and all the nodes in the steel remains elastic such that damage occurs only in the concrete. The material properties for both steel and concrete are provided in Table \ref{tab:cts_properties}

\begin{table}[h]
  \centering
  \caption{Material constants for the concrete specimen and the steel loading frames.}
  \label{tab:cts_properties}
  \begin{tabular}{lllllll}
    \hline
    Material & $\rho$ & $E$ & $\nu$ & $\sigma_{tf}$ & $\sigma_{cf}$ & $G_c$ \\
             & ($\text{kg}\cdot\text{m}^{-3}$) & (GPa) & & (MPa) & (MPa) & ($\text{N}\cdot\text{m}^{-1}$) \\
    \hline
    Concrete & 2400 & 36  & 0.18 & 3.8 & 53 & 65 \\
    Steel    & 7800 & 210 & 0.33 & --  & -- & -- \\
    \hline
  \end{tabular}
\end{table}

The uniform grid spacing is 1 mm, and the horizon is set to 3 mm. The characteristic length, L, is selected to be 12 mm. We consider three experimental cases: Test 2, Test 5, and Test 12 from \cite{ovzbolt2013dynamic}. All three cases are solved dynamically without damping using the explicit velocity Verlet method. As in the reference study, the displacement was applied at the end of the lower frame and the end layer of the upper rod was fixed. The rod end velocity increases linearly until a ramp time of 300 $\mu s$ and then stays at the same loading rate. A loading rate of 0.035, 1.40, and 3.60 m/s for the three tests was used, respectively, with the same ramp time of 300 $\mu s$. These load rates are in the range of the rates used in the experimental  study of  \cite{ovzbolt2013dynamic} which show crack branching for load rates greater than 3.30 m/s.

We measured the reaction force at the end point and the contact point of the steel frame and only the former one was compared with the experimental results. For all three cases, we considered the first peak reaction force, before the crack reached the far edge. The model predictions were 2.97 kN, 4.86 kN, and 5.69 kN, compared with experimental peaks 2.83 kN, 4.64 kN, and 5.66 kN, for Test 2, Test 5, and Test 12, respectively.  Here we compare crack path in  figures \ref{fig:test_2},  \ref{fig:test_5} and \ref{fig:test_12} given by experiment to the computed crack path for the frame based inelasticity method. The crack path can seen from the jump set of the second component $\uu_2$ of the displacement.   
 In Test 2, the crack starts at the notch and runs straight closely matching the numerical prediction of \cite{ovzbolt2013dynamic}. In Test 5, the crack starts in the lower region of the notch and follows a similar curve pattern as in the experimental result. Test 12 shows that the crack starts and branches with the two branches showing close agreement with the experimental result. 
\hfill
\hfill

The crack tip speed in Test 12 was evaluated based on the position of the crack tip with damage above 0.7 using an approach similar to that used in the glass branching test. We sampled the crack tip position every 1 $\mu s$ using a window of 20 samples. Before branching, the velocity was in the range of 700 - 890 m/s, close to the 835 m/s measured in the experiment. As mentioned before in  figures \ref{fig:test_2},  \ref{fig:test_5} and \ref{fig:test_12} the crack path for the frame based inelasticity method is rendered as the jump set  of the second component of the displacement $\uu_2.$  Simulations show that the crack is mode one for loading rate 0.035 and mixed mode one and two for the loading rates 1.40  m/s,  3.60 m/s . Here the jump set corresponds to the crack defined as the internal boundary where traction force on one side of the crack does not influence traction on the other side. Overall, the proposed model shows good agreement with the experimental results demonstrating its ability to capture rate dependent dynamic fracture in terms of peak reaction forces, crack pattern and crack tip velocities.

\section{Conclusion}
\label{sec:Conclusion}
The proposed frame based inelasticity is formulated as a mathematically well-posed initial boundary value problem for brittle and quasi-brittle fracture. The model inherently satisfies the principles of energy and power balance, exhibiting positive energy dissipation consistent with thermodynamics. The model calibration does not impose constraints on Poisson's ratio, or elastic symmetry and the formulation is theoretically applicable to predicting material strength and fracture behavior in anisotropic media. The numerical simulations demonstrate close agreement with the experimental results and other numerical methods for when predicting material strength, crack branching under dynamic loading, and crack patterns for different loading rates. These results support the model’s ability to predict real-world material behavior with minimal computational complexity. Although the numerical simulations presented in this study are carried out for isotropic material properties, the formulation is applicable in to general anisotropic materials. Future work will incorporate fracture in polycrystalline and heterogeneous media.

%%%%%%%%%%%%%%%%%%%%%%%%%%%%%%%%%%%%%%%%%%%%%%%%%%%%%%%%%%%%%%%%%%%%%%%%%%%%%%%%%%%%%%%%%%%%%%%
\section{Acknowledgment}
This material is based upon work supported by the U.S. Army Research Laboratory and the U.S. Army Research Office under Contract/Grant Number W911NF-19-1-0245 and W911NF-24- 2-0184 and NSF DMPS- 2607263.

\bigskip

%\noindent {\bf Data availability} The code used to generate the results in this study is available at \\ https://github.com/SemsiCoskun/PDBlend.

\appendix


%%       \citet{<label>} ==> Lamport (1994)
%%       \citep{<label>} ==> (Lamport, 1994)
%%
%%Example citation, See \citet{lamport94}.

\bibliographystyle{elsarticle-harv} 
\bibliography{references}



%\begin{thebibliography}{00}


%\bibitem[Lamport(1994)]{lamport94}
%  \textit{\LaTeX: a document preparation system},
%  Addison Wesley, Massachusetts,

%\end{thebibliography}
\end{document}